\documentclass[preprints,article,accept,pdftex,moreauthors]{Definitions/mdpi} 
\firstpage{1} 
\pubvolume{1}
\issuenum{1}
\articlenumber{0}
\pubyear{2025}
\copyrightyear{2025}
\datereceived{ } 
\daterevised{ } 
\dateaccepted{ } 
\datepublished{ } 
\hreflink{https://doi.org/} 

\usepackage{svg}
\usepackage{dcolumn}
\usepackage{tabularray}
\usepackage{longtable}
\usepackage[toc,page]{appendix}

\newcolumntype{Y}{>{\centering\arraybackslash}X}

\DeclareUnicodeCharacter{2212}{\ensuremath{-}}

\Title{A Systematic Literature Review of Machine Learning Techniques for Observational Constraints in Cosmology}

\TitleCitation{Title}

\Author{Luis Rojas $^{1,\ddagger}$*\orcidA{}, Sebasti\'an Espinoza $^{2,\ddagger}$\orcidB{}, Esteban González $^{3,\ddagger}$*\orcidC{}, Carlos Maldonado $^{4,\ddagger}$\orcidD{}, and Fei Luo $^5$\orcidE{}}

\AuthorNames{Firstname Lastname, Firstname Lastname and Firstname Lastname}

\isAPAStyle{%
       \AuthorCitation{Lastname, F., Lastname, F., \& Lastname, F.}
         }{%
        \isChicagoStyle{%
        \AuthorCitation{Lastname, Firstname, Firstname Lastname, and Firstname Lastname.}
        }{
        \AuthorCitation{Lastname, F.; Lastname, F.; Lastname, F.}
        }
}

\address{%
$^{1}$ \quad Facultad de Ingeniería, Universidad San Sebastián, Bellavista 7, Santiago 8420524, Chile; luis.rojasp@uss.cl\\
$^{2}$ \quad Departamento de Sistemas de Información, Universidad del Bío-Bío, Avenida Andrés Bello 720, Chillán 3800708, Chile; sespinoza@ubiobio.cl\\
$^{3}$ \quad Departamento de Física, Universidad Católica del Norte, Avenida Angamos 0610, Casilla 1280, Antofagasta 1270709, Chile; esteban.gonzalez@ucn.cl\\
$^{4}$ \quad Facultad de Ciencias, Universidad San Sebastián, Lago Panguipulli 1390, Puerto Montt 5501842, Chile; carlos.maldonado@uss.cl\\
$^{5}$ \quad School of Information Science and Engineering, East China University of Science and Technology, Shanghai 200237, China; luof@ecust.edu.cn}

\corres{Correspondence: luis.rojasp@uss.cl (L.R.); esteban.gonzalez@ucn.cl (E.G.)}

\secondnote{These authors contributed equally to this work.}

\abstract{This paper presents a systematic literature review focusing on the application of machine learning techniques for deriving observational constraints in cosmology. The goal is to evaluate and synthesize existing research to identify effective methodologies, highlight gaps, and propose future research directions. Our review identifies several key findings: (1) various machine learning techniques, including Bayesian neural networks, Gaussian processes, and deep learning models, have been applied to cosmological data analysis, improving parameter estimation and handling large datasets. However, models achieving significant computational speedups often exhibit worse confidence regions compared to traditional methods, emphasizing the need for future research to enhance both efficiency and measurement precision. (2) Traditional cosmological methods, such as those using Type Ia Supernovae, baryon acoustic oscillations, and cosmic microwave background data, remain fundamental, but most studies focus narrowly on specific datasets. We recommend broader dataset usage to fully validate alternative cosmological models. (3) The reviewed studies mainly address the $H_0$ tension, leaving other cosmological challenges—such as the cosmological constant problem, warm dark matter, phantom dark energy, and others—unexplored. (4) Hybrid methodologies combining machine learning with Markov chain Monte Carlo offer promising results, particularly when machine learning techniques are used to solve differential equations, such as Einstein Boltzmann solvers, as prior to Markov chain Monte Carlo models, accelerating computations while maintaining precision. (5) There is a significant need for standardized evaluation criteria and methodologies, as variability in training processes and experimental setups complicates result comparability and reproducibility. (6) Our findings confirm that deep learning models outperform traditional machine learning methods for complex, high-dimensional datasets, underscoring the importance of clear guidelines to determine when the added complexity of Learning models is warranted.}


\begin{document}

\section{\label{sec:Introduction}Introduction}
The field of cosmology has experienced significant growth thanks to advances in observational data collection, enabling a deeper understanding of the Universe's structure and evolution. However, traditional techniques for analyzing these data, such as Markov chain Monte Carlo (MCMC) or other methods for Bayesian inference, face challenges when applied to increasing volumes of complex data \citep{gelman1995bayesian, lewis2002cosmological}. In response, machine learning (ML) techniques have emerged as promising tools to enhance the efficiency and accuracy of cosmological parameter estimation \citep{ntampaka2015machine}.

While systematic literature reviews (SLRs) are widely employed in fields such as medicine, computing, and education to critically evaluate the state of the art and guide future research, they remain relatively unexplored in cosmology. For instance, SLRs have been used in medicine to assess the effectiveness of new treatments and technologies \citep{jpt2008cochrane}, in computing to analyze methodologies for applying ML techniques in electrical power forecasting \citep{lopez2023supervised}, and in education to synthesize pedagogical strategies \citep{bozkurt2020emergency}. The relevance of a SLR in cosmology lies in its ability to synthesize multiple research efforts applying ML to cosmological problems, identifying patterns, gaps, and promising directions for future investigations. These reviews can accelerate innovation by providing a comprehensive and critical view of accumulated knowledge, as seen in other scientific domains.

The aim of this paper is to contribute with a novel perspective by addressing the intersection between ML techniques and cosmological data constraints through a systematic review of the literature. Unlike previous surveys on ML in cosmology, which often focus on specific applications such as supernova detection \citep{lochner2016photometric} or galaxy classification \citep{dieleman2015rotation}, our work systematically categorizes and evaluates the diverse ML methodologies applied across a range of cosmological problems. We not only review the effectiveness of these techniques in improving parameter estimation but also highlight the limitations and propose future research directions. Our primary objectives are: 1) to systematically review and categorize ML techniques applied to cosmological parameter estimation, 2) to assess the effectiveness and limitations of these techniques, and 3) to identify gaps in the literature and suggest new research directions. This is the first comprehensive review to address the broad application of ML techniques for improving Bayesian inference and parameter constraints in cosmology, making it an important resource for researchers in both fields.

This paper is organized as follows: in Section \ref{sec:Background}, we present the theoretical background that underpins the methods and techniques analyzed in this SLR, whereas, in Subsection \ref{subsec:Cosmology}, we briefly describe some important theoretical aspect on cosmology, observational datasets, and the standard procedure for deriving cosmological constraints and, in Subsection \ref{subsec:MachineLearning}, we briefly describe some important theoretical aspects for ML and their changes when it is implemented to the Bayesian inference. In Section \ref{sec:RelatedWorks}, we discuss some previous SLRs related to our study. In Section \ref{sec:Methodology}, we describe the research methodology used to carry out our SLR, whereas, in Subsection \ref{subsec:Planning}, we present the research questions and objectives that will steer the review, in Subsection \ref{subsec:Executing}, we discuss the identification and selection of the pertinent studies, along with the application of filters to ensure data quality and relevance, and, in Subsection \ref{subsec:Reporting}, the findings amassed throughout the review process are synthesized and presented cohesively. In Section \ref{sec:Results}, we present the main results of our SLR according to the following structure: Subsection \ref{subsec:Relation} is devoted to presenting the thematic interconnection of the selected articles. In Subsection \ref{subsec:Databases}, we focus our analysis on the samples used according to the datasets considered in the reviewed papers. In Subsection \ref{subsec:MlDlModels}, we discuss the models of ML and deep learning (DL) considered in the reviewed papers. In Subsection \ref{subsec:Aim}, we present the main objectives that the reviewed papers aim to tackle with the ML techniques. In Subsection \ref{subsec:Metadata}, we discuss the metadata of the papers selected in our SLR. On the other hand, in Section \ref{sec:Findings}, we discuss the main findings obtained in the results of our review, whereas, in Subsection \ref{subsec:Outcomes}, we present the main results and, in Subsection \ref{subsec:Gaps}, we discuss the works and problems that can be addressed in the future. Subsection~\ref{sec:tech_benchmarking} provides a concise technique-level comparison and explains why cross-paper benchmarking is not methodologically sound. In Section \ref{sec:Threats}, we discuss the threats to the validity of our SLR, which is composed of the following subsections: Subsection \ref{subsec:External} focuses on the applicability of the results to domains outside of cosmology, Subsection \ref{subsec:Construct} addresses how variations in the implementation of models can affect the outcomes of our SLR, Subsection \ref{subsec:Internal} explores whether the conclusions drawn from the SLR accurately reflect the methods implemented in the reviewed papers, and Subsection \ref{subsec:Conclusion} presents the arguments supporting the validity of our conclusions. Finally, in Section \ref{sec:Conclusions}, we present some conclusions and a final discussion.

\section{\label{sec:Background}Theoretical Background}

\subsection{\label{subsec:Cosmology}Cosmology}
In cosmology, the Universe is ruled by the cosmological principle, which establishes homogeneity and isotropy at large scales ($>100\,Mpc$ \cite{Mukhanov:2005sc}), and it is dominated, in principle, by radiation and baryonic matter during their cosmic evolution. However, observations of some phenomena give us insights into two additional hypotheses to consider, namely dark matter and dark energy. The first is responsible for the formation of large-scale structure, with initial evidence coming from galaxy rotation curves \cite{Bertone:2016nfn,Salucci:2018eie}, and the second accounts for the late-time accelerated expansion of the Universe \cite{SupernovaSearchTeam:1998fmf,SupernovaCosmologyProject:1998vns}. These ingredients give us the standard cosmological model (also known as $\Lambda$CDM model), described by Friedmann's equations \cite{Fischer:2018zkr}
\begin{eqnarray}
H^{2}\equiv\left(\frac{\dot{a}}{a}\right)^{2}&=&\frac{8\pi G}{3}\rho+\frac{\Lambda c^2}{3}-\frac{k c^2}{a^2},
\label{hubble}\\
\frac{\ddot{a}}{a}&=&-\frac{4\pi G}{3}\left(\rho+\frac{3p}{c^2}\right)+\frac{\Lambda c^2}{3},
\end{eqnarray}
where $G$ is the Newtonian constant of gravitation, $c$ is the speed of light in vacuum, $k$ is the curvature of the Universe, dot ( $\dot{}$ ) accounts for the derivative with respect to the cosmic time $t$, $a$ is the scale factor (a quantity that describes the evolutionary/expansion history of the universe), $H$ is the Hubble parameter (the expansion rate of the Universe), and $\Lambda$ is the cosmological constant. The energy density and pressure of the Universe are $\rho=\rho_{r}+\rho_{m}$ and $p=p_{r}+p_{m}$, where the subscripts $r$ and $m$ account for radiation and matter, respectively. At the current time, the radiation component can be neglected and the matter density is composed of baryonic matter and cold dark matter (CDM), which represent approximately the $5\%$ and $25\%$ of the total energy budget of the Universe, respectively. The remaining $70\%$ corresponds to dark energy \cite{ParticleDataGroup:2022pth}, which is described in the model by the cosmological constant.

It is convenient to write the Hubble parameter in terms of the density parameter of each matter component \cite{Velten:2014nra}
\begin{equation}\label{defofOmega}
\Omega_{i}(z)=\Omega_{i,0}(1+z)^{3(1+\omega)},\;\; \text{with} \;\; \Omega_{i,0}=\frac{8\pi G\rho_{i,0}}{3H_{0}^{2}},
\end{equation}
where the redshift $z$ is related to the scale factor through the expression $1+z=a_{0}/a$, the subscript $0$ accounts for the values at the current time, and the subscript $i$ accounts for the radiation and matter components with the values $\omega=1/3$ and $\omega=0$, respectively. For the cosmological constant, the density parameter reads $\Omega_{\Lambda}=\Lambda c^{2}/3H_{0}^{2}$. Considering a flat Universe ($k=0$) \cite{Planck:2018vyg}, the current values of the density parameters are constrained through the Eq. \eqref{hubble} as
\begin{equation}\label{Friedmannconstraint}
\Omega_{r,0}+\Omega_{m,0}+\Omega_{\Lambda}=1.
\end{equation}
Therefore, the Hubble parameter for the $\Lambda$CDM model can be written in terms of the redshift as follows
\begin{equation}\label{LCDM}
H^2(z)=H_0^2\left[\Omega_{r,0}\left(1+z\right)^4+\Omega_{m,0}\left(1+z\right)^3+\Omega_{\Lambda}\right].
\end{equation}

The $\Lambda$CDM model is, up to date, the most successful cosmological model to describe the background cosmological data, from which we can highlight:
\begin{itemize}
    \item\textbf{Type Ia supernovae (SNe Ia):} Supernovae (SNe) are highly energetic explosions of some stars and play an important role in the fields of astrophysics and cosmology because they have been used as cosmic distance indicators. In particular, SNe Ia are considered standard candles to measure the geometry and the late-time dynamics of the Universe \cite{Liu:2023qmw}. In fact, between 1998 and 1999, the independent projects High-z Supernova Search Team \cite{SupernovaSearchTeam:1998fmf} and Supernova Cosmology Project \cite{SupernovaCosmologyProject:1998vns} showed results that suggested an acceleration in the Universe expansion using SNe Ia data. This behavior is now confirmed by several cosmological observations, establishing that the Universe is currently facing an accelerated expansion, which began recently in cosmic terms at a redshift of $z=0.64$ \cite{Moresco:2016mzx}; SNe Ia data are widely used to test the capability of alternative models to $\Lambda$CDM in describing the cosmological background. The sample used by the Supernova Search team consisted of 50 SNe Ia data points between $0.01<z<0.97$, while the sample of the Supernova Cosmology Project consisted of 60 SNe Ia data points between $0.014<z<0.83$. Nowadays, the samples of SNe Ia observations have grown in data points and redshift range, the most recent being the Pantheon sample \cite{Pan-STARRS1:2017jku}, consisting in 1048 SNe Ia data points between $0.01\leq z\leq2.3$; and the Pantheon+ sample \cite{Brout:2022vxf}, with 1701 SNe Ia data points between $0.001\leq z\leq 2.26$.
    \item\textbf{Observational Hubble parameter data (OHD):} Even though SNe Ia data provide consistent evidence about the existence of a transition epoch in cosmic history where the expansion rate of the Universe changes, it is important to highlight that this conclusion is obtained in a model-dependent way \cite{Moresco:2016mzx}. The study of the expansion rate of the Universe in a model-independent way can be carried out through observations of the Hubble parameter. Up to date, the most complete OHD sample was compiled by Magaña et al. \cite{Magana:2017nfs}, which consists of 51 data points in the redshift range of $0.07\leq z\leq 2.36$. In this sample, 31 data points are obtained using the Differential Age method \cite{Jimenez:2001gg}, while the remaining 20 data points come from baryon acoustic oscillations measurements \cite{Magana:2017nfs}.
    \item\textbf{Baryon acoustic oscillations (BAO):} BAO are the footprints of the interactions between baryons and the relativistic plasma in the epoch before recombination (the epoch in the early Universe when electrons and protons combined to form neutral hydrogen) \cite{DESI:2025zgx}. There is a significant fraction of baryons in the Universe and the cosmological theory predicts acoustic oscillations in the plasma that left ``imprints" at the current time in the power spectrum of non-relativistic matter \cite{Peebles:1970ag,Eisenstein:1997ik}. Many collaborations have provided BAO measurements like 6dFGS \cite{Beutler_2011}, SDSS-MGS \cite{Ross:2014qpa}, BOSS-DR12 \cite{BOSS:2016wmc}, and the Dark Energy Spectroscopic Instrument (DESI) \cite{DESI:2016fyo} to mention a few.    
    \item\textbf{Cosmic microwave background (CMB):} Since the discovery of the CMB in 1965 by Penzias and Wilson \cite{1965ApJ...142..419P}, the different acoustic peaks in the anisotropy power spectrum have become the most robust observational evidence to test cosmological models. In this sense, the different acoustic peaks provide information about the matter content and curvature of the Universe \cite{SDSS:2003eyi,Wang:2006ts}, and have been measured by different satellites like the Wilkinson Microwave Anisotropy Probe (WMAP) \cite{2013ApJS..208...19H} and Planck \cite{Planck:2018nkj}.
    \item\textbf{Large scale structure (LSS):} The LSS is the study of the distribution of the galaxies in the Universe at large scales (larger than the scale of a galaxies group) \cite{Springel:2006vs}. At small scales, gravity concentrates particles to give form to gas, then to stars, and finally to galaxies. At large scales the galaxies also group in different patterns called ``the cosmic web'', which is caused by fluctuations in the early Universe. This distribution has been quantified by various surveys, such as the 2-degree Field Galaxy Redshift Survey (2dFGRS) \cite{2DFGRS:2001zay} and the Sloan Digital Sky Survey (SDSS) \cite{SDSS:2000hjo}.
    \item\textbf{Gravitational Lensing (GL):} When a background object (the source) is lensed due to the gravitational force of an intervening massive body (the lens) it generates multiple images. Therefore, the light rays emitted from the source will take different paths through the space-time at the different image positions arriving at the observer at different times. This time delay depends on the mass distribution in the lensing and along the line of sight, and also of the cosmological parameters. For this data we can highlight the strong lensing measurements of the $H_{0}$ Lenses in COSMOGRAIL's Wellspring (H0LiCOW) collaboration \cite{Wong:2019kwg}, which consist of six gravitationally lensed quasars with measured time delays.
\end{itemize}

Although $\Lambda$CDM is the most robust cosmological model until now, some issues cannot be explained in this theory such as the nature of dark matter and dark energy, the asymmetry in baryons (the observed imbalance between matter and antimatter), the hierarchy problem (the large difference between the weak force and gravity), the neutrino mass (the small but nonzero mass of the neutrino), among others \cite{Turner:2022gvw, Abdalla:2022yfr}. Furthermore, as we entering in the so-called ``era of precision cosmology'' some observational tensions arise and become more problematic with the inclusion of new data. For example, local measurements of Cepheid for the Hubble constant $H_{0}$ (model-independent) present a discrepancy of $5\sigma$ with the value inferred by Planck CMB assuming the $\Lambda$CDM model \cite{Riess:2021jrx}. This tension is also supported by the H0LiCOW collaboration with a discrepancy of $5.3\sigma$ concerning the value inferred from Planck CMB \cite{Wong:2019kwg}.

The shortcomings exhibited by the $\Lambda$CDM model due to the inclusion of more observational data accentuates the importance of the parameter estimation (the best-fit values of the free parameter space of a certain cosmological model), not only to test the capability of the standard model to describe these new data but also to test the capability of alternative cosmological scenarios in the description of the cosmological background. To that end, Bayesian inference is commonly used in cosmology for parameter estimation (also referred to as cosmological constraints), which considers Bayes' theorem of the form
\begin{equation}\label{Bayes}
    P(\theta,D)=\frac{P(D,\theta)P(\theta)}{P(D)},
\end{equation}
where $P(\theta,D)$ is the posterior distribution and corresponds to the probability of obtaining the parameter space $\theta$ for a given observational data $D$, $P(D,\theta)$ is the likelihood and corresponds to the probability to obtain the observational data $D$ for a given parameter space $\theta$, $P(\theta)$ is the prior distribution and corresponds to the previous physical evidence about the parameter space, and $P(D)$ is the prior predictive probability which is extremely hard to calculate \cite{Hogg:2017akh}. To overcome this last problem, $P(D)$ is approximated using Monte Carlo methods, being one of the most used in cosmology the affine-invariant MCMC \cite{Goodman_Ensemble_2010}, implemented in algorithms like the pure-Python code \textit{emcee} \cite{Foreman-Mackey:2012any}. Nevertheless, this method is highly dependent on the initial conditions, requires exploring all the parameter space to obtain the best fits, has problems in the presence of multi-modal likelihoods, and the computing time grows exponentially for big datasets and free parameters \cite{Hajian:2006mt}.

Nowadays, in cosmology there are new surveys such as the Legacy Survey of Space and Time (LSST) \cite{LSST:2008ijt}, Euclid \cite{laureijs2011euclid}, the Spectro-Photometer for the History of the Universe, Epoch of Reionization, and Ices Explorer (SPHEREx) \cite{SPHEREx:2014bgr}, the Nancy G. Roman Space Telescope (NGRST) \cite{Spergel:2015sza}, the Dark Energy Spectroscopic Instrument (DESI) \cite{DESI:2016fyo}, and the Prime Focus Spectrograph (PFS6) \cite{Takada_2014}. These surveys will provide new data that, in addition to previous cosmological data, will increase the efficiency and computing time problems of the MCMC method, raising the ML as a powerful alternative to improve the cosmological constraints.

\subsection{\label{subsec:MachineLearning}Machine Learning}
It is often challenging for humans to explicitly define the logical rules underlying very complex tasks, such as image recognition or natural language understanding. In general, it can be said that ML is a paradigm shift in which rules can be defined by delivering the inputs and outputs of a particular task repeatedly to a model. In this way, the model “learns” the patterns present in the inputs that result in certain outputs using some performance measure that ensures the validity of those patterns \cite{janiesch_machine_2021}.

\begin{figure}
    \centering
    \includegraphics[scale=1]{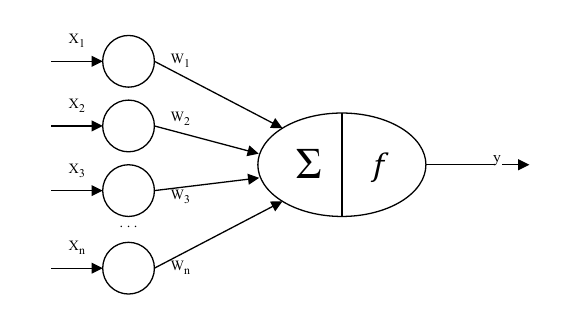}
    \caption{\label{img:perceptron} Example of basic structure of the Perceptron.}
\end{figure}

Deep learning (DL) is a subset of machine learning (ML) that uses neural networks with many layers. Neural networks are a component of ML, and when they consist of multiple layers, they are referred to as deep learning. The difference between ML and DL is that, in the latter, the models are capable of automatically learning representations of input data such as text, images, or videos, in much greater detail than ML models. This is due to the use of successive layers of data representation (hence the name Deep) \cite{janiesch_machine_2021}. NNs are one of the main and most popular DL techniques today. Their development has been going on for quite some time, having its origins in 1957 by Frank Rosenblatt with the Perceptron (see Fig. \ref{img:perceptron}) \cite{rosenblatt_perceptron_1958}. Despite its innovation, further research slowed down for a while due to several factors, including the criticism from Minsky and Papert in 1969 \cite{minsky_introduction_1969}, as well as limitations in hardware and the availability of sufficient data. It was not until the 1980s, with work such as that of Paul Werbos and David E. Rumelhart, which popularized the use of the Backpropagation algorithm in recurrent neural networks (RNNs), that renewed interest in the area. Since then, several advances have been made, as well as modifications and additions that allow the use of NNs for specific tasks, in particular, those related to the ML techniques applied for the cosmological constraints.

The construction of NNs varies depending on the use case, but their operation can be generalized through a structure called layers. This last one carries the information from the input to the output of the network, transforming it in the process and obtaining convenient representations for the task to be performed, commonly classifying or analyzing data. These layers receive as input the output of the previous layers or, failing that, the initial data input. The simplest form of these structures is the single-layer perceptron, which is one of the earliest examples of an NN \cite{rosenblatt_perceptron_1958}. For example, for binary classification as shown in Fig.~\ref{img:perceptron}, we have that \( \overline{X} = \left[ x_1, x_2, ..., x_n \right] \) contains the \( n \) feature values to be classified, and \( y \in [0, 1] \) is the resulting value. The model's output is calculated by applying an activation function \( f \) to the weighted linear combination of the inputs:

\begin{equation}
    z = \sum_{i=1}^{n} w_i x_i + b,
\end{equation}
\begin{equation}
    y = f(z),
\end{equation}
where \( w_i \) are the weights associated with each feature \( x_i \), \( b \) is the bias, and \( f(z) \) is the activation function. As an example, the step function can be used:

\begin{equation}
    f(z) = 
    \begin{cases}
        1 & \text{if } z \geq 0 \\
        0 & \text{if } z < 0
    \end{cases}.
\end{equation}
This function determines the output class based on whether the weighted sum of the inputs surpasses a threshold; in this example, the threshold is 0.

\begin{figure}
    \centering
    \includegraphics[scale=0.35]{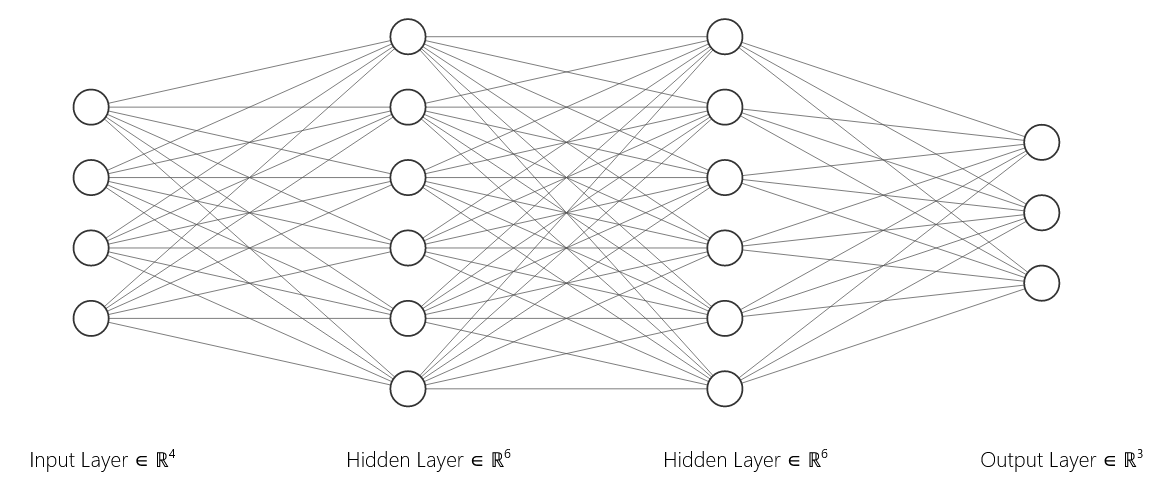}
    \caption{Example of multilayer NN (Inspired by the Figure 1.5 of Ref. \cite{chollet2021deep}).}
    \label{img:multilayer}
\end{figure}

In a multilayer neural network (see Fig. \ref{img:multilayer}), the outputs of each neuron feed into the subsequent neurons, allowing them to solve problems that are not linearly separable. This is achieved by a weighted sum of the inputs followed by an activation function \( f \), such as the sigmoid, which introduces non-linearity. Thus, the output of a neuron in a layer is:

\begin{equation}
    a_j = f\left(\sum_{i=1}^{n} w_{ji} x_i + b_j \right),
\end{equation}
where \( a_j \) is the output of the neuron, \( w_{ji} \) are the weights, \( x_i \) are the inputs, \( b_j \) is the bias, and \( f \) is the activation function, like the sigmoid. In this way, they are transformed into what is now known as a NN.

In general, the output values of each neuron are calculated based on the weights. In simple words, the layers of an NN are parameterized by the weights, allowing one to obtain different values depending on the context. Finding the ``ideal'' weights is fundamental to obtaining effective estimates. On the other hand, the activation function works as a filter or limiter that transforms the output values of each neuron. These transformations are usually nonlinear functions, which allow the NN to solve increasingly complex problems. Among the most relevant activation functions is the linear function defined by 
\begin{equation}
    f\left ( x \right ) = x, \label{nn:lineal}
\end{equation}
which does not limit the output of the neuron and is widely used in regression tasks. There is also sigmoidal function
\begin{equation}
    f\left ( x \right ) = \frac{1}{1+e^{-x}}, \label{nn:sigmoid}\\
\end{equation}
which limits the output of the network to a value between 0 and 1 and is often used as the activation function of the last neuron of a binary classification NN. Finally, one of the most commonly used functions is the ReLU function
\begin{equation}
    f\left ( x \right ) = max\left (0,x \right ), \label{nn:relu}
\end{equation}
which only allows the output of positive values of each neuron, suppressing negative values to zero, and it is mainly used to address the vanishing gradient problem in NNs, which results in faster convergence as a secondary effect. Finally, an optimizer is used to adjust the parameters. In general terms, the optimizer uses the loss function of the NN model to determine how much the weights should change to reduce the loss as much as possible. For this, some variant of gradient descent is usually used, with the logic of finding the parameters with which the derivative of the loss function finds convergence.

In summary, an artificial NN makes calculations on input data, which can be numbers, text, images, or other types of data. These calculations are propagated from the input to the output neurons using intermediate parameters called weights to adjust the subsequent calculations. Learning occurs when, given an input and the subsequent processing of the weights, the NN model can associate the input with an output, also called a label or prediction. It can be said that the model recognizes an input because it has learned patterns and generalizations from the data it was trained on. The goal is for the network to capture fundamental features of the data, enabling it to generalize to new, similar datasets. There are many types of artificial NNs, oriented to different types of tasks, but all maintain a similar structure and way of working \cite{aggarwal_introduction_2018}.

On the other hand, Bayesian machine learning (BML), although similar in operation to classical ML models, also includes techniques that use Bayesian principles to make predictions, allowing, for example, to calculate the uncertainty of these predictions \cite{bharadiya_review_2023}. Among these models are the Gaussian processes (GP) that are often used in complex ML problems due to their flexible and non-parametric nature \cite{seeger_gaussian_2004} and the Bayesian decision trees (BDT) that add Bayesian techniques to the classical decision trees, such as, for example, the uncertainty in the decisions of division of parent nodes to children \cite{nuti_bayesian_2019} \cite{dension_bayesian_1998}. In the context of NNs, Bayesian inference can be used to estimate the uncertainty of model predictions. This fusion is often referred to as Bayesian neural networks (BNNs), and replaces the classical network weights with fixed values by probability distributions, allowing the model to estimate uncertainty and fit the model with that approach \cite{blundell_weight_2015}.

\section{\label{sec:RelatedWorks}Related Works}
While there has been extensive research on the application of ML techniques in cosmology, most existing studies focus on specific applications rather than providing a comprehensive account of how these techniques are used to derive observational constraints, given that ML is still a relatively new and rapidly evolving field. For instance, ML methods have been widely used in classifying astronomical transients, such as supernovae, and in parameter estimation from large-scale structure surveys \citep{de2022observational, moriwaki2023machine, lahav2023deep}. These studies typically highlight the performance of ML models like convolutional neural networks (CNNs) and random forests in handling large and complex datasets.

Recent advances underscore the role of ML in large-scale surveys such as the LSST, which generates millions of transient alerts every night. These alerts far surpass the available spectroscopic resources for follow-up, making ML techniques indispensable for real-time classification and anomaly detection \citep{moriwaki2023machine, de2022observational, dvorkin2022machine}. ML has also been applied to challenges like photometric redshift estimation and galaxy clustering, with hybrid models combining NNs and support vector machines showing performance improvements \citep{han2015improving}. However, these efforts are often fragmented and focus on specific ML techniques or isolated aspects of observational constraints, without offering a comprehensive synthesis of methodologies specifically applied to observational data in cosmology. As of now, no systematic review has compiled and evaluated the full range of ML techniques applied to cosmological problems, such as those dealing with SNe Ia, BAO, and CMB data, among others  \citep{de2022observational, lahav2023deep}. This lack of synthesis reveals a critical gap in the literature.

Beyond application-specific studies, recent community-level and methodological advances further motivate our focus on observational constraints. The CosmoVerse White Paper synthesizes current observational tensions and the role of systematics, outlining a forward-looking agenda for robust inference pipelines \citep{CosmoVerse2025}. Methodologically, neural emulators such as \emph{CosmoPower} accelerate likelihood-based inference by replacing expensive Boltzmann solvers \citep{SpurioMancini2022CosmoPower}, while normalizing-flow approaches (e.g., \emph{emuflow}) enable efficient joint posteriors across heterogeneous datasets \citep{Mootoovaloo2025Emuflow}.

The present work aims to fill this gap by conducting a SLR that not only categorizes and evaluates the effectiveness of diverse ML techniques but also explores their integration with traditional cosmological methods. This study represents a novel contribution to the field, offering a valuable resource for researchers seeking to leverage ML for cosmological data analysis. The absence of similar reviews highlights the innovative nature of this work, positioning it as a critical step toward advancing ML applications in observational cosmology.

\section{\label{sec:Methodology}Research Methodology}
Our SLR is based on the principles for this kind of works in the area of software engineering \cite{kitchenham2007guidelines}, in combination with the methodologies outlined in Ref. \cite{kitchenham2004procedures}. Broadly, this approach encompasses three key phases:
   
\begin{itemize} 
    \item[\textbf{1.}] \textbf{Planning the Review}: This initial stage involves defining the research questions and objectives that will steer the review. This sets a clear framework for the study.
    \item[\textbf{2.}] \textbf{Executing the Review}: Here, the identification and selection of pertinent studies occur, along with the application of filters to ensure data quality and relevance. Information extraction is also pivotal during this phase.
    \item[\textbf{3.}] \textbf{Reporting the Review}: Finally, the findings amassed throughout the review process are synthesized and presented cohesively. This phase culminates in the succinct and organized presentation of research outcomes. 
\end{itemize}

\subsection{\label{subsec:Planning}Planning the Review}
In this subsection, a detailed overview of the key components comprising the review is provided. This includes outlining the research questions guiding the process, the selection of search engines, and the establishment of inclusion and exclusion criteria guiding the selection of materials.

The main objective of this study is to explore the current state of research related to the utilization of ML in assessing the feasibility of fitting observational data with cosmological models, focusing on the constraint of the free parameters of a certain cosmological model. The aim is to examine how ML approaches are being applied in cosmology to enhance the efficiency of Bayesian inference algorithms and other techniques used in model fitting, in particular, the MCMC method. For this purpose, the following research questions have been formulated:

\begin{itemize}
    \item\textbf{RQ1:} What ML approaches are most frequently used in the field of cosmology to adjust the free parameters of cosmological models to observational data?
    \item\textbf{RQ2:} To what extent does ML contribute to the field of fitting cosmological models to observational data, particularly in enhancing the efficiency of Bayesian inference algorithms and other fitting techniques?
    \item\textbf{RQ3:} What are the existing research gaps in the utilization of ML for fitting cosmological models and what are the opportunities for future research to address these gaps and enhance our understanding of observational cosmology?
    \item\textbf{RQ4:} What types of training data are commonly used in ML approaches applied to fitting cosmological models to observational data, and what methods are employed to obtain and process this data?
\end{itemize}

In terms of the research search, notable preprint repositories and digital databases such as arXiv, ScienceDirect, ACM Digital Library, Scopus, and Inspirehep were queried. Each search string was tailored to fit the specific formats of each database. These selections were made based on the reputable nature of these databases and their user-friendly search interfaces, which facilitate result filtering and exportation in convenient formats. arXiv was included because it is the primary repository for physics research, where all relevant studies, including those indexed in Web of Science (WoS), are conventionally shared, as well as works not published in journals. To ensure consistency in the selected articles, limit the amount of information under consideration, and maintain a clear focus on the main research themes, inclusion and exclusion criteria were established.
\\

\textbf{Inclusion Criteria:}
\begin{itemize}[topsep = 0pt]
    \item\textbf{IC1:} Research articles written in the English language, but may consider articles in other languages if relevant (with access to translation resources).
    \item\textbf{IC2:} Research works published from 2014 to 2024, selected based on relevance to the research topic.
    \item\textbf{IC3:} Articles published in conference/workshop proceedings, academic journals, and as thesis dissertations to encompass diverse scholarly sources.
    \item\textbf{IC4:} Complete (full-text) research articles to ensure comprehensive review.
\end{itemize}
\bigskip

\textbf{Exclusion Criteria:}
\begin{itemize}[topsep = 0pt]
    \item\textbf{EC1:} Exclude duplicate articles, ensuring data integrity.
    \item\textbf{EC2:} Exclude articles that are not focused on ML techniques applied to improve parameter estimation in cosmology.
    \item\textbf{EC3:} Exclude articles that are not aligned with the goals of the SLR, such as those describing ML techniques that do not improve Bayesian inference for parameter estimation.
\end{itemize}
\bigskip

\textbf{Protocol and reporting:} 
\begin{itemize}
    \item This SLR followed a pre-specified protocol aligned with PRISMA 2020; the full protocol is publicly archived on Zenodo~\cite{SLRProtocol2025}.
\end{itemize}

\subsection{\label{subsec:Executing}Executing the Review}
This Section outlines the methodology employed for conducting the review, including the processes of searching, filtering, and selecting the ultimate collection of research articles from which the necessary data were extracted. Subsequently, the data were synthesized and analyzed in preparation for the subsequent phase. This phase extended over a period of approximately eight months, from March 2024 to October 2024. In essence, this phase of the SLR encompasses the initial search and selection of review papers, defining the strategy for data extraction, and conducting data synthesis and analysis.

\subsubsection{Exploration and Concluding Selection of Reviewed Materials}
\begin{figure*}
\centering
\includegraphics[scale=1]{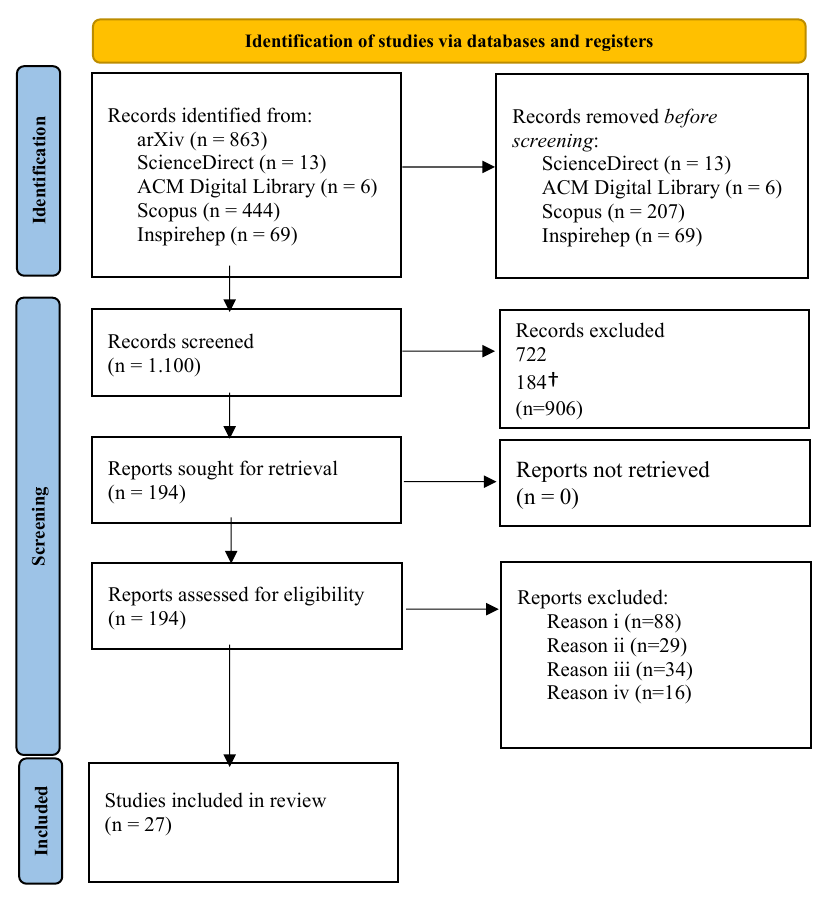}
\caption{\label{fig:prisma} PRISMA flow diagram illustrating the detailed process involved in selecting the final set of review materials.}
\end{figure*}

Initially, a thorough and comprehensive search was conducted across specified scholarly databases and the Google search engine. Various search strings were employed to retrieve research articles from diverse digital databases. Boolean operators, specifically AND and OR, were utilized in search syntax to refine the search results. Additionally, a wild card character (*) was incorporated into certain search queries to broaden the scope and capture matching results with one or more characters. Keywords were explored in different combinations within the title, abstract, and keywords sections of each article, as outlined in the search syntax format across various sources. The search string finally used for each source (digital database/library) is presented in Table~\ref{tab:search_details}. This systematic review was conducted according to the Preferred Reporting Items for Systematic Reviews and Meta-Analyses (PRISMA) guidelines. In Figure \ref{fig:prisma}, we present a PRISMA flow diagram that visually illustrates the detailed process involved in selecting the final set of review materials. The articles finally selected for the review are presented in detail in Appendix~A \ref{apendix:selectedPapers}.

\begin{table*}
\centering
\begin{tabularx}{\textwidth}{|c|Y|}
\hline
\textbf{Source} & \textbf{Search Syntax/String} \\
\hline
arXiv & Query: order: -announced\_date\_first; size: 200; include\_cross\_list: True; terms: AND abstract=COSMOLOGY OR "DARK ENERGY" OR "COSMOLOGICAL CONSTRAINTS" OR "OBSERVATIONAL CONSTRAINTS"; AND abstract="MACHINE LEARNING" OR "ARTIFICIAL INTELLIGENCE" OR "DEEP LEARNING" OR "NEURAL NETWORKS" \\
\hline
ScienceDirect & Title, abstract, keywords: (COSMOLOGY OR "DARK ENERGY" OR "COSMOLOGICAL CONSTRAINTS" OR "OBSERVATIONAL CONSTRAINTS") AND ("MACHINE LEARNING" OR "ARTIFICIAL INTELLIGENCE" OR "DEEP LEARNING" OR "NEURAL NETWORKS") \\
\hline
ACM Digital Library & [[Abstract: cosmology] OR [Abstract: "dark energy"] OR [Abstract: "cosmological constraints"] OR [Abstract: "observational constraints"]] AND [[Abstract: "machine learning"] OR [Abstract: "artificial intelligence"] OR [Abstract: "deep learning"] OR [Abstract: "neural networks"]] \\
\hline
Scopus & (COSMOLOGY OR "DARK ENERGY" OR "COSMOLOGICAL CONSTRAINTS" OR "OBSERVATIONAL CONSTRAINTS") AND ("MACHINE LEARNING" OR "ARTIFICIAL INTELLIGENCE" OR "DEEP LEARNING" OR "NEURAL NETWORKS") \\
\hline
Inspirehep & t(COSMOLOGY OR "DARK ENERGY" OR "COSMOLOGICAL CONSTRAINTS") AND t("MACHINE LEARNING" OR "ARTIFICIAL INTELLIGENCE" OR "DEEP LEARNING" OR "NEURAL NETWORKS") \\
\hline
\end{tabularx}
\caption{\label{tab:search_details} Search string used in the review for each digital database/library source.}
\end{table*}

\subsubsection{Data extraction strategy}
\begin{figure*}
\centering
\includegraphics[width=\textwidth]{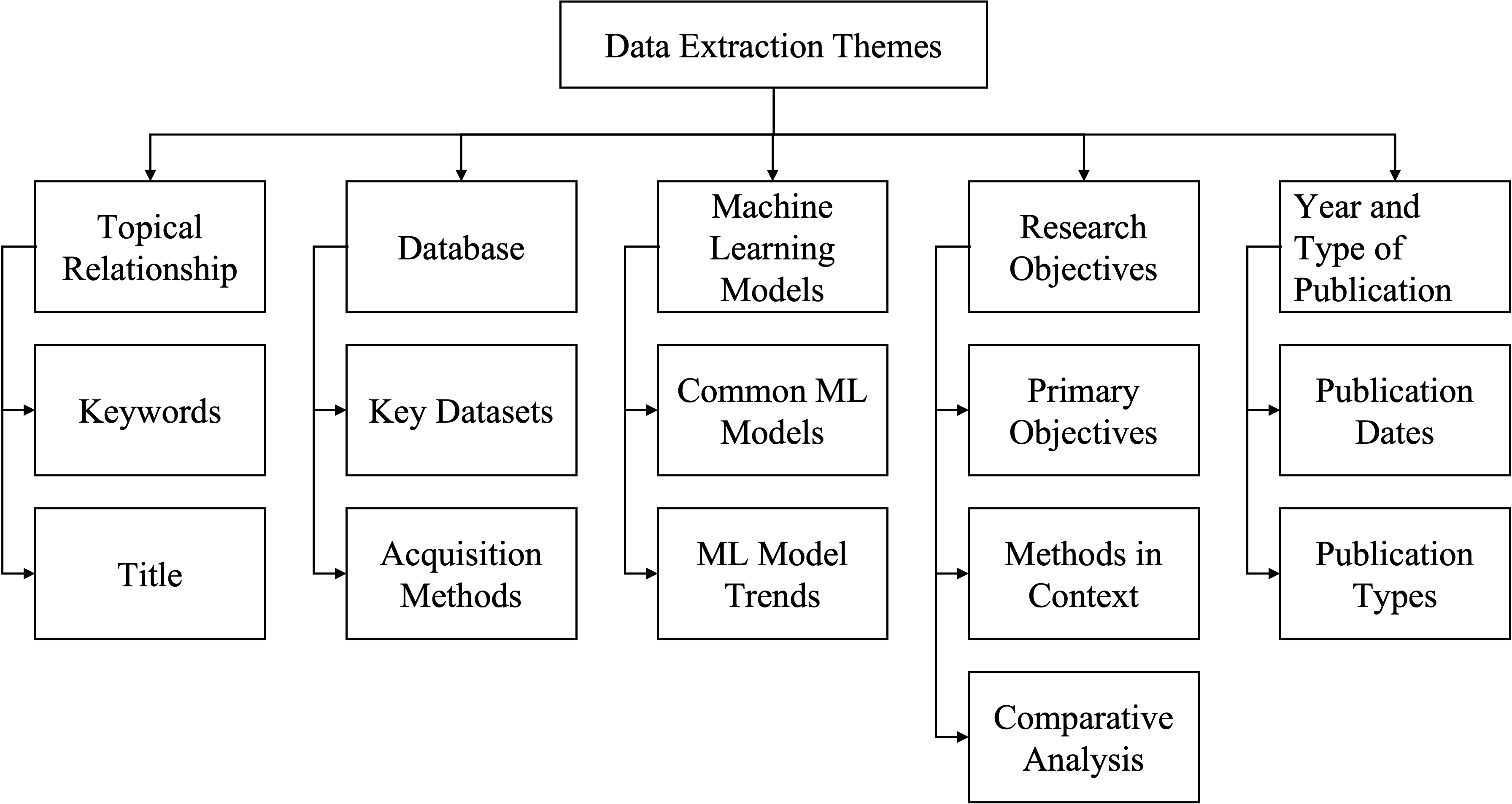}
\caption{\label{fig:EstrategiaD} Data extraction themes.}
\end{figure*}

A meticulous strategy is paramount for extracting data and conducting a structured literature review. This strategy is guided by five key themes, illustrated in Figure \ref{fig:EstrategiaD}. These themes have been expanded with specific attributes to structure data extraction effectively, ensuring alignment with the review's objectives. Below, each theme is described, supplemented by questions that elucidate the type of data extracted from the selected research articles.

\begin{itemize}
    \item\textbf{Topical Relationship}: This theme explores the thematic coherence among the reviewed articles. It is evaluated by the recurrence and relevance of keywords and the thematic correlation of titles, reflecting their collective contribution to the field of cosmology. This theme encompasses the following questions (i) How often do keywords appear across different articles? and (ii) Are the titles indicative of a common thematic focus?
    \item\textbf{Databases}: This theme investigates the datasets utilized in ML for cosmological model fitting. It examines the types of training data, the methodologies for data collection, and the techniques for processing this data to enhance model accuracy and reliability. This theme encompasses the following questions (i) What training datasets are prevalent in ML studies for cosmology? and (ii) What are the common methods for data acquisition and preprocessing?
    \item\textbf{Machine Learning Models:} This Section delves into the specific models and approaches that the field currently prioritizes, looking for patterns or trends in model selection and application. This theme encompasses the following questions (i) Which ML models are most commonly referenced in the literature? and (ii) Can we identify trends or preferences in the use of certain ML models for cosmological studies? 
    \item \textbf{Research Objectives:} The focus here is on understanding the primary goals of the research articles and how these align with the broader objectives of the field. It also assesses the structure and clarity with which these objectives are presented. This theme encompasses the questions (i) What are the primary objectives outlined in the articles?, (ii) How are the articles' methods and results situated within the broader context of cosmological research?, and (iii) Is there a comparative analysis between the presented research and other studies within the field? 
    \item \textbf{Year and Type of Publication}: This theme catalogs the articles based on their publication year and the medium of publication, which provides insight into the evolution of the field and the dissemination of findings. This theme encompasses the questions (i) When were the key articles in the domain published? and (ii) Are the articles predominantly from journals, conferences, workshops, or academic theses?
\end{itemize}

\subsection{\label{subsec:Reporting}Reporting the Review}
Data were extracted along five predefined themes and analyzed using both qualitative and quantitative approaches. The qualitative analysis is a narrative/thematic synthesis of methods, tasks, validation practices, and reported challenges, while the quantitative analysis consists of descriptive statistics (e.g., counts and proportions) of model families, cosmological probes, datasets, reported speedups, and uncertainty performance. Results are summarized and visualized in \ref{sec:Methodology}.

\section{\label{sec:Results}Results}
In this Section, we present the details of data synthesis and analysis of the reviewed articles concerning the five themes stated above.

\subsection{\label{subsec:Relation}Topical Relationship}
\begin{figure}
\centering
\includegraphics[scale=0.75]{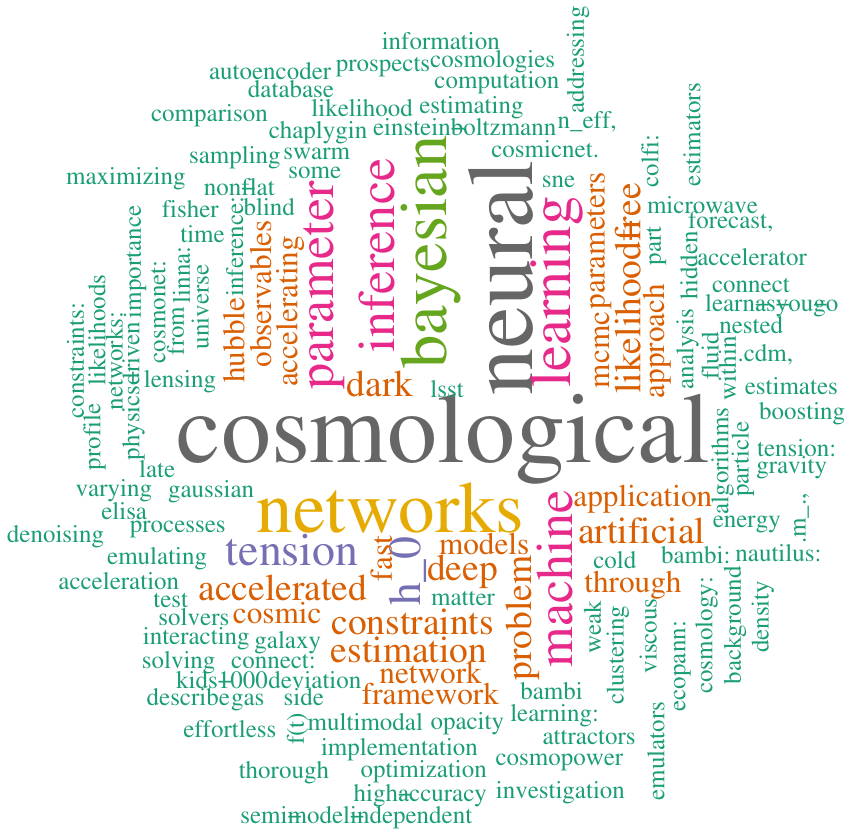}
\caption{\label{fig:wordcloud-titles} Word cloud for the titles of the twenty-seven reviewed articles.}
\end{figure}

\begin{figure}
\centering
\includegraphics[scale=0.95]{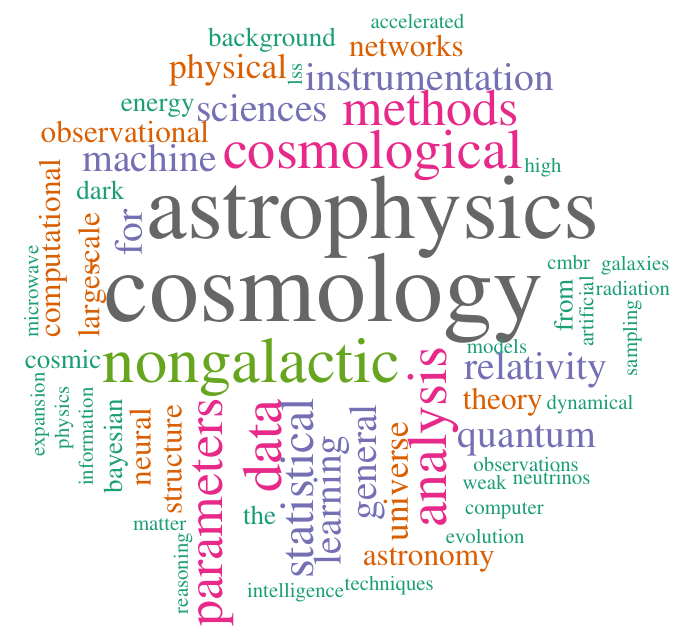}
\caption{\label{fig:wordcloud-tags} Word cloud for the keywords of the twenty-seven reviewed articles.}
\end{figure}

To visually illustrate the thematic interconnection of the selected articles in this SLR, we have adopted an approach akin to a ``word cloud". This graphical representation highlights the frequency of specific terms used in both the titles and keywords of the reviewed articles, reflecting their relevance and thematic focus. The word clouds for the titles and keywords of the articles are displayed in Figures \ref{fig:wordcloud-titles} and \ref{fig:wordcloud-tags}, respectively.

In Figure \ref{fig:wordcloud-titles}, it can be seen that some of the most frequent terms used in the titles include ``cosmological", ``neural", ``networks", ``bayesian", and ``learning". These terms are indicative of the convergence of the disciplines of ML and cosmology, reflecting the primary focus of this research on the effectiveness of ML techniques in fitting cosmological models with observational data.

On the other hand, in Figure \ref{fig:wordcloud-tags}, the most prominent keywords include “cosmology,” “astrophysics,” “nongalactic,” “analysis,” and “parameters.” The dominance of these terms highlights the strong thematic focus of the reviewed studies on cosmological and astrophysical domains, with particular attention to the analysis of nongalactic data and the estimation of cosmological parameters. Notably, other relevant keywords such as “neural,” “statistical,” “Bayesian,” and “machine” also emerge, reflecting the growing integration of machine learning techniques in addressing cosmological problems.

\subsection{\label{subsec:Databases}Databases}
\begin{figure*}
\centering
\includegraphics[width=\textwidth]{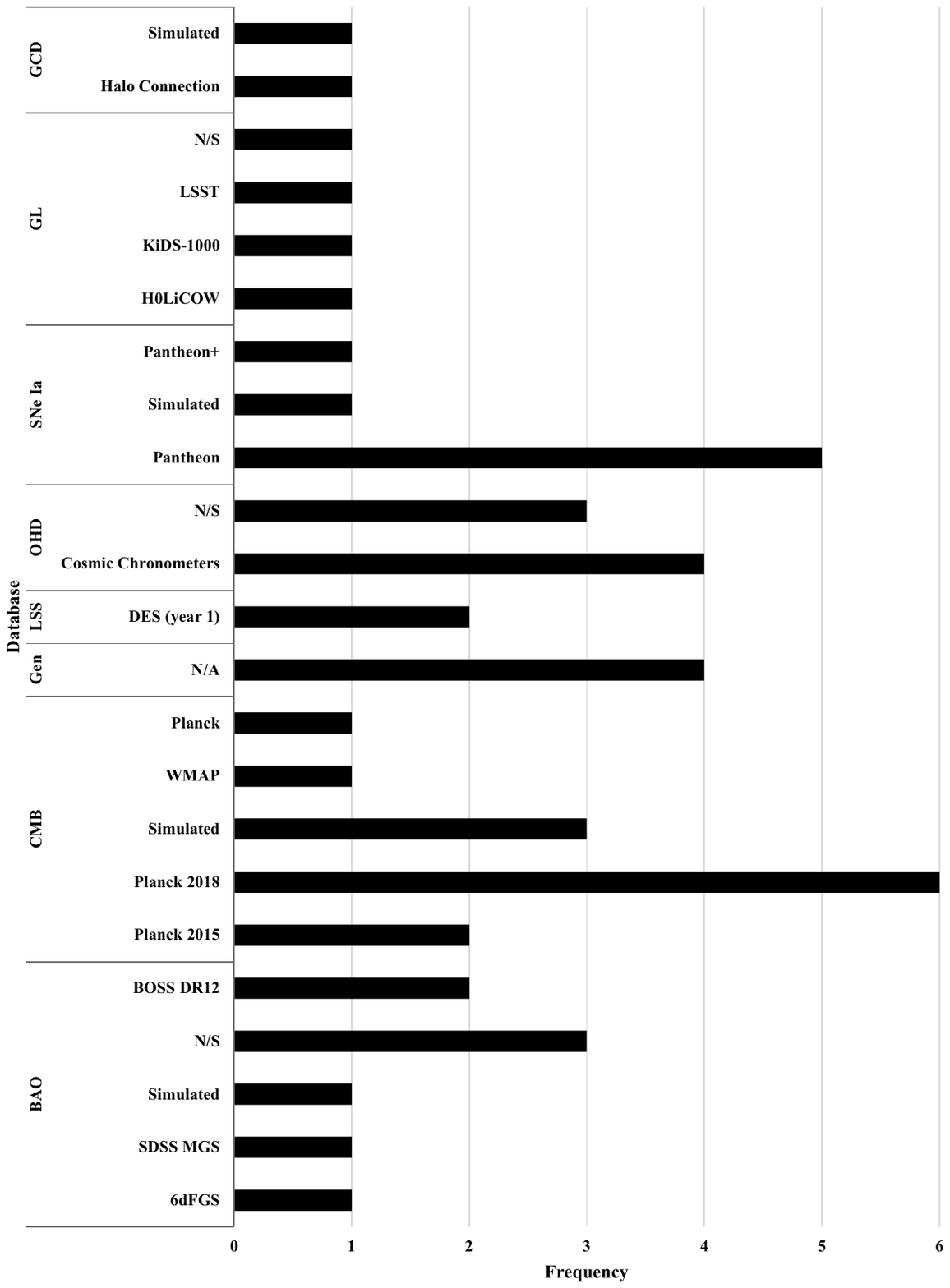}
\caption{\label{fig:databases}Frequency plot of the sample for each cosmological data considered for a total of twenty-seven reviewed papers. In the Figure, N/S and N/A stand for non-specified and not apply, respectively, simulated refers to a sample that comes from a particular database but is obtained through a ML technique, and generated corresponds to a sample that is obtained from a ML technique that does not have as a source a specific database.}
\end{figure*}

In this Section, we focus our analysis on the samples used according to the datasets considered in the reviewed papers. In particular, the main datasets are SNe Ia, OHD, BAO, CMB, LSS, GL, and galaxy clustering data (GCD). Some details about these are presented in Section \ref{subsec:Cosmology}. From the twenty-seven reviewed papers, we identify the data samples shown in Figure \ref{fig:databases}, where we depict a frequency plot of the sample for each cosmological data considered.  In the Figure, N/S and N/A stand for non-specified and not apply, respectively, simulated refers to a sample that comes from a particular database but is obtained through a ML technique, and generated corresponds to a sample that is obtained from a ML technique that does not have as a source a specific database. It is important to clarify that OHD is the name of the database and also of the data sample. In fact, cosmic chronometers are included in the OHD sample. In this sense, the N/S sample can be some specific points of the OHD database.

From Figure \ref{fig:databases}, we can see that, for the total of twenty-seven reviewed articles, CMB database is considered in twelve papers, one of which uses two catalogs from the same database, resulting in thirteen uses ($48.1\%$), BAO in six papers, and, as in the previous case, one of them uses three different catalogs, resulting in eight uses. ($29.6\%$), SNe Ia and OHD are considered in seven papers each one ($25.9\%$), GL in four papers ($14.8\%$), and LSS and GCD are considered in two papers for each one ($7.4\%$). Also, four papers ($14.8\%$) use a generated data sample from a ML technique without the use as a source of a specific database. The sum of the percentages presented above is no $100\%$ because, in general, the reviewed papers used more than one data set in their studies. From the latter, we can see that nearly $50\%$ of the twenty-seven reviewed papers use the CMB data, which is an expected result considering that is the most expensive data at a computational level. This is because CMB data can add to the constraint more than twenty free parameters (the computational time of the standard MCMC method increases with the number of free parameters), where most of them are nuisance parameters and do not correspond to free parameters of the cosmological model under study. Additionally, when deriving parameter constraints from observational data, the repeated evaluation of Einstein–Boltzmann solvers (e.g., CAMB/CLASS) is typically the principal computational bottleneck in the inference pipeline.

Focusing on the data samples presented in Figure \ref{fig:databases}, we can see that one paper ($50\%$) uses Halo Connection measurements \cite{id_25} and one paper ($50\%$) uses a simulated sample \cite{id_27} for the GCD; while for the GL data, we have the H0LiCOW \cite{id_19}, KiDS-1000 \cite{id_13}, and LSST \cite{id_02} samples, in one paper for each them ($25\%$), and one paper ($25\%$) does not specify the sample \cite{id_09}. In the case of SNe Ia data, five papers ($71.4\%$) use the Pantheon sample \cite{id_10,id_16,id_20,id_21,id_22}, one paper ($14.3\%$) uses the Pantheon+ sample \cite{id_06}, and one paper ($14.3\%$) uses a simulated sample from the future Wide Field Infrared Survey Telescope (WFIRST) experiment using ML techniques \cite{id_12}. On the other hand, for the OHD, we have four papers ($57.1\%$) that use the cosmic chronometers compilation \cite{id_10,id_19,id_20,id_06}, and three papers ($42.9\%$) do not specify the sample \cite{id_16,id_11,id_26}. For the LSS data, two papers ($100\%$) use DES (year 1) \cite{id_17,id_24}. The following samples are used for the CMB data: six papers ($46.1\%$) use the Planck 2018 release \cite{id_07,id_09,id_13,id_21,id_23,id_24}, two paper ($15.4\%$) use the Planck 2015 release \cite{id_01,id_22}, one paper ($7.7\%$) uses the WMAP sample without specifying the release year \cite{id_15}, one paper ($7.7\%$) uses the Planck sample without specify the release year \cite{id_15}, and three papers ($23.1\%$) use a simulated sample from CMB sky images using ML techniques \cite{id_03,id_12,id_18}. The BAO database is considered in two papers ($25\%$) through the BOSS DR12 release \cite{id_21,id_24}, SDSS-MGS and 6dFGS \cite{id_21} releases in one paper for each them ($12.5\%$), three papers ($37.5\%$) do not specify the sample \cite{id_09,id_10,id_20}, and one paper ($12.5\%$) uses a simulated sample from the future measurements of the SKA2 survey using ML techniques \cite{id_12}. Finally, five papers use a generated database \cite{id_04,id_05,id_08,id_14}.

\subsection{\label{subsec:MlDlModels}Machine Learning Models}

\begin{figure}
\centering
\includegraphics[scale=0.30]{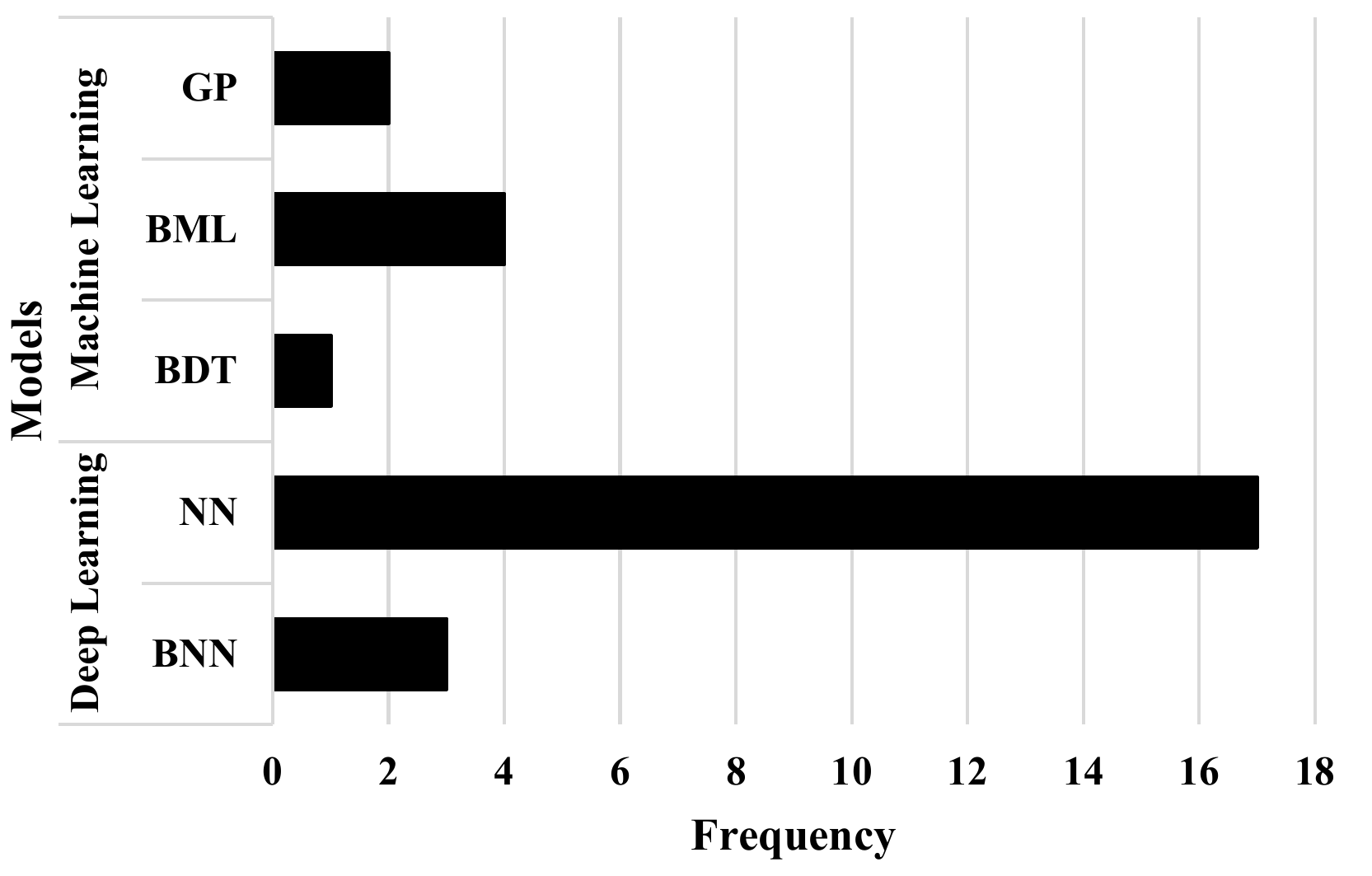}
\caption{\label{fig:modelsfreq}Frequency plot of the ML models used in the twenty-seven reviewed papers. The main models identified in the SLR are GP, BML, BDT, NN, and BNN. These models can be also classified into two major fields: ML and DL. There is an observable trend toward using DL models for the treatment of cosmological data.}
\end{figure}

For our SLR, the main ML models encountered were GP, BML, BDT, NN, and BNN. \footnote{\textbf{Taxonomy note:} Rather than a normative “traditional ML vs deep learning” dichotomy, we report results by \emph{model family} (GP, BML, BDT, NN, BNN) and—where relevant—by \emph{uncertainty handling} (deterministic vs probabilistic), which aligns with our corpus and the speed–calibration comparisons.} Some details about these models were explained in Section \ref{subsec:MachineLearning}. All of these models can be classified into two major fields: ML (including GP, BML, and BDT) and DL (comprising NN and BNN). The frequency of usage for each of these models is presented in Figure \ref{fig:modelsfreq}. Only models used specifically for analyzing cosmological data are considered in the count, i.e., models used solely for data generation or other tasks are excluded.

In detail, it can be observed that for a total of twenty-seven reviewed articles, the least-used ML model is BDT, mentioned in only one article ($3.7\%$) \cite{id_15}, followed by the GP model, which is used in two articles ($7.4\%$) \cite{id_21, id_11}. Next, the BNN model is employed in three articles ($11.1\%$) \cite{id_03,id_18,id_19}. The BML model was also observed in use in four articles ($14.8\%$) \cite{id_04,id_05,id_08,id_14}. Interestingly, there appears to be a limited usage of the BDT, BNN, and BML models, which are variations of classical models (like decision trees, NN, and other ML variants) used for Bayesian inference tasks. Finally, the most widely used model is NN, which appears in seventeen articles ($63.0\%$) \cite{id_01,id_02,id_07,id_09,id_10,id_12,id_13,id_16,id_17, id_20,id_22,id_23,id_24,id_25,id_26,id_06, id_27}. 

Overall, in twenty articles ($74.1\%$), the models used can be classified as DL, while in the remaining seven articles ($25.9\%$), the models used can be considered as ML models. This information clearly indicates a strong preference for DL models in handling the cosmological data mentioned in Section \ref{subsec:Databases}.

In the context of deriving observational constraints, the three most frequent families in our corpus offer complementary trade-offs: (i) deterministic neural networks (NN) scale well to high-dimensional inputs and deliver the largest computational speedups for repeated forward evaluations, but require external uncertainty handling (e.g., ensembling/MC dropout) and careful calibration; (ii) Bayesian neural networks (BNN) provide principled predictive uncertainty and can yield better-calibrated intervals, at the cost of higher training/inference complexity; and (iii) Gaussian processes (GP) are sample-efficient with analytic uncertainty and strong calibration in low–moderate dimensional settings, but suffer from $\mathcal{O}(n^3)$ scaling and reduced effectiveness as dimensionality grows. Given heterogeneous datasets and metrics across the included studies, we refrain from aggregate rankings and report these family-level trade-offs to contextualize the preferences observed in Fig.~\ref{fig:modelsfreq}. Consistent with these frequencies, the predominance of NN reflects pragmatic scaling: once trained, NN surrogates amortize repeated forward evaluations and leverage modern hardware/toolchains, whereas GP training scales as $\mathcal{O}(n^3)$ and is kernel-sensitive in higher dimensions, and BNNs incur additional inference cost and implementation complexity—factors that limit their routine use on large, high-dimensional data.

\subsection{\label{subsec:Aim} Research Objectives}
From the twenty-seven reviewed papers, we can conclude that it is possible to identify two main goals: 1) the enhancement of parameter estimation using ML techniques (hereafter referred to as 'improvement'), and 2) the application of such improved estimation methods to address cosmological problems (hereafter referred to as 'application'). In the first goal, the papers are focusing on the drawbacks of the parameter estimation, mainly focused on the problems associated with the MCMC analysis related to the computing time for big datasets and free parameters, as we mentioned in Section \ref{sec:Introduction}. For the second goal, the papers use a previously improved parameter estimation through a certain ML technique to tackle some specific physical cosmological problems, taking advantage of the improvement in the computing time (or other improvements) of the new parameter estimation technique.

\begin{figure}
\centering
\includegraphics[scale=0.30]{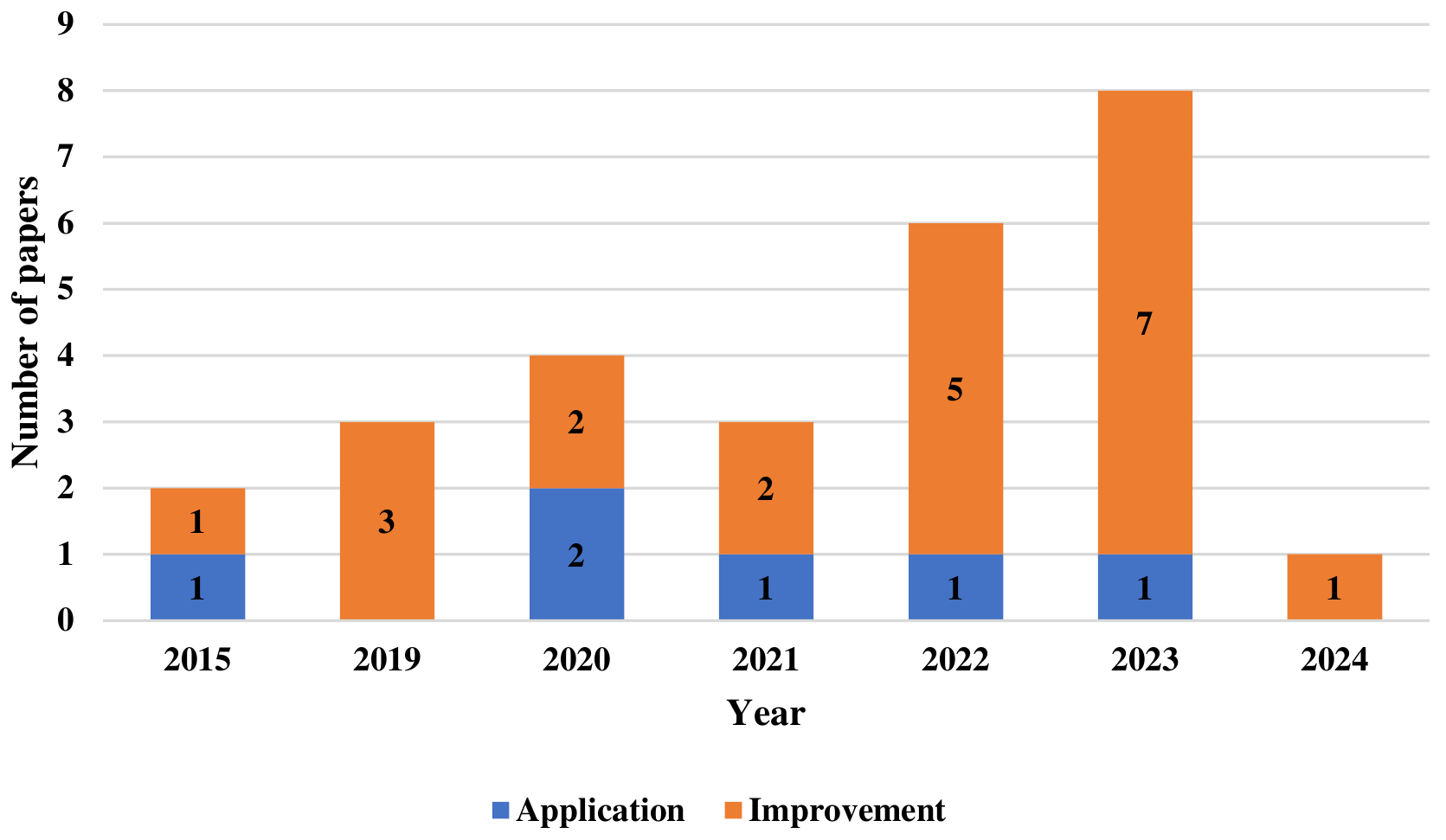}
\caption{\label{fig:yearsandapplication} Number of papers and the year of availability online for the twenty-seven reviewed papers. The articles are divided into two classifications based on their research aim: 1) the improvement of the parameter estimation through ML techniques (improvement) and 2) the application of an improved parameter estimation analysis based on a ML technique to face cosmological problems (application).}
\end{figure}

In Figure \ref{fig:yearsandapplication}, we depict the number of papers and the year of availability online for the twenty-seven reviewed papers. From the Figure, we can see that from 2014 to 2024 twenty-one papers ($77.8\%$) are focused on the improvement of the parameter estimation, meanwhile, six papers ($22.2\%$) are aimed at application, giving us insights that the main aim of these works is the improvement of the cosmological constraints. Considering the years of publication, we do not have reviewed papers in the years 2014, 2016, 2017, and 2018. In 2015, one paper ($50\%$) focuses on improvement \cite{id_15} and one paper ($50\%$) focuses on application \cite{id_14}. On the other hand, in 2019, three papers ($100\%$) are focused on the improvement of the parameter estimation \cite{id_01,id_09,id_18}. In 2020, two papers ($50\%$) are focused on improvement \cite{id_03,id_12} and two papers ($50\%$) are focused on application \cite{id_04,id_08}; while in 2021, two papers ($66.7\%$) focused their study in the improvement of the cosmological constraints \cite{id_13,id_16} and one paper ($33.3\%$) is focused in the application \cite{id_05}. In 2022, five papers ($83.3\%$) focused their aim on the improvement of the parameter estimation \cite{id_02,id_07,id_10,id_17,id_27} and only one paper ($16.7\%$) focused their aim in the application of an improvement parameter estimation \cite{id_19}. Similarly, in 2023, seven papers ($87.5\%$) are focused in improvement \cite{id_06,id_20,id_22,id_23,id_24,id_25,id_26}, while only one paper is focused in the application \cite{id_21}. Finally, in 2024, there is only one reviewed paper ($100\%$) and is focused on improvement \cite{id_11}. Note that in the last three years, thirteen of the twenty-seven reviewed papers ($48.1\%$) are focused on improvement, i.e., nearly half of the reviewed papers. This gives us insights into the relevance that is taking in the last years to obtain a more efficient parameter estimation for handling the incoming observational cosmological data. The most remarkable results for both improvement and application aims are listed below.
\\

\textbf{Improvement aim:}
\begin{itemize}[topsep = 0pt]
    \item[\textbf{1.}] Some improvements for the cosmological parameters estimation are studied in \cite{id_02}, with an enhancement of $20$ times faster to reproduce the correct contours compared with the MCMC case. Regarding the DL model, a densely connected neural network with three hidden layers, each consisting of 1024 neurons and using ReLU activation functions, was used. This architecture results in a number of trainable parameters on the order of millions, due to the full connectivity between layers. 
    \item[\textbf{2.}] In Ref. \cite{id_03}, an enhancement of $10^4$ times is reported in the computing time in comparison with classical methods for the parameter estimation, which accelerates the performance, but is less precise in comparison with the standard MCMC method. In the study, a BNN with the Visual Geometry Group (VGG) architecture was used, with a customized calibration method.
    \item[\textbf{3.}] A similar result as the above point was obtained in \cite{id_18}, but a BNN with different weight sampling methods is used to provide tighter constraints for the cosmological parameters. The findings of this paper serve as a guide for the models used in \cite{id_03}.
    \item[\textbf{4.}] In Ref. \cite{id_12}, the authors show a reduction in the computing time, giving an excellent performance in the parameter estimation compared with MCMC for the $\Lambda$CDM model. Also, a detailed explanation of the hyperparameters and the steps to train the model is given. In particular, a NN with three hidden layers (reducing the number of neurons in each layer) together with the ReLU activation function is used.
    \item[\textbf{5.}] The number of executions in the Einstein-Boltzmann Solvers for the CMB data are reduced in Ref. \cite{id_09} in comparison with the standard procedure, which saves computational resources, translating into faster computations, avoiding the bottleneck on the solvers in $\Lambda$CDM model with a massive neutrino model. From the ML point of view, three NNs are used which are made up of a combination of densely connected layers and convolutional layers. These layers are generally used in image classification tasks but, in this particular case, are used to reduce the number of neurons, instead of densely connecting the whole network. Along with this, the ReLU activation function is used together with Leaky ReLU, a version of ReLU that allows a small amount of negative data to be output.
    \item[\textbf{6.}] The authors in Ref. \cite{id_10} report that the estimation of parameters from MCMC is more efficient with the solutions provided by an ANN, improving the numerical integration in the $\Lambda$CDM model, the Chevallier-Polarski-Linder parametric dark energy, a quintessence model with exponential potential, and the Hu-Sawicki $f(R)$ model, estimating that the error is 1$\%$ in the region of the parameter space for a 95$\%$ of confidence for all models.
    \item[\textbf{7.}] A new method for the parameter estimation up to 8 times faster than the standard procedure is presented in Ref. \cite{id_06}.
    \item[\textbf{8.}] In Ref. \cite{id_13}, using NN techniques, the authors accelerate the estimation of the cosmological parameters, taking $10$ hours compared with the $5$ months required by the standard Boltzmann codes. Interestingly, the values of $\chi_\text{min}^2$ are similar to the standard compute for the $\Lambda$CDM model.
    \item[\textbf{9.}] In a similar way as in the above point, in Ref. \cite{id_23} the authors achieve high precision in the $\chi^2_\text{min}$ criteria, with a difference of $\Delta\chi^2\simeq 0.2$ compared with the results obtained by Cosmic Linear Anisotropy Solving System (CLASS), being up to $2$ times faster than this standard procedure.
    \item[\textbf{10.}] Finally, in Ref. \cite{id_22}, the authors show deviations for the parameters $H_0$, $\Omega_bh^2$, $\Omega_ch^2$, $\tau$, $A_s$, and $n_s$ of $0.013\sigma$, $0.020\sigma$, $0.010\sigma$, $0.073\sigma$, $0.094\sigma$, and $0.051\sigma$, respectively, between the MCMC method and the ML technique for the $\Lambda$CDM and $\omega$CDM models.
\end{itemize}
\bigskip

\textbf{Application aim:}
\begin{itemize}[topsep = 0pt]
    \item[\textbf{1.}] An improvement parameter estimation with ML techniques was applied to solve the $H_0$ tension in Ref. \cite{id_04}. In particular, through a BML method, the authors studied a Universe dominated by one fluid with a generalized equation of state.
    \item[\textbf{2.}] In a similar way as in the above point, the authors in Ref. \cite{id_05} apply a BML method in a model with cosmological constant, baryonic matter, and barotropic dark matter, and a model with barotropic dark energy, baryonic matter, and barotropic dark matter. 
    \item[\textbf{3.}] In Ref. \cite{id_19}, the authors show that the $H_0$ tension can be alleviate using BNN in a $f(t)$ modified gravity, specifically in a $f(t)$ exponential model.
    \item[\textbf{4.}] Finally, in Ref. \cite{id_08}, the authors prove the opacity of the Universe through BNN in the $\Lambda$CDM and $x$CDM models, obtaining that the Universe is not completely transparent, which also impacts the $H_0$ tension. Regarding the implementation of the ML techniques, all scenarios use PyMC3, a probabilistic Python framework that has the necessary tools to apply ML approach to probabilistic tasks.
\end{itemize}
\bigskip

\begin{figure}
\centering
\includegraphics[scale=1.80]{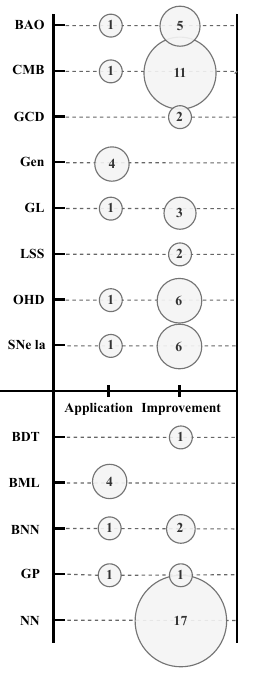}
\caption{\label{fig:samplesandapplication} Data sample for each cosmological data set (upper) and ML model (lower) used in the twenty-seven reviewed papers according to their research aim: 1) the improvement of the parameter estimation through ML techniques (improvement) and 2) the application of an improve parameter estimation analysis based on a ML technique to face cosmological problems (application).}
\end{figure}

In the upper part of Figure \ref{fig:samplesandapplication}, we present the data sample considered for each cosmological data set used in the twenty-seven reviewed papers according to the two main classifications of their research aim mentioned above, namely, improvement and application. As we can see, BAO was used in five papers ($83.3\%$) whose objective is to improve the parameter estimation \cite{id_09,id_10,id_12,id_20,id_24} and one paper ($16.7\%$) considers this data in the application of an improve parameter estimation thought a ML technique \cite{id_21}. On the other hand, eleven papers ($91.7\%$) used the CMB data in the improvement aim \cite{id_01,id_03,id_07,id_09,id_12,id_13,id_15,id_18,id_22,id_23,id_24}, while only one paper ($8.3\%$) considers this data in the application aim \cite{id_21}. This is an expected result and is a consequence of the conclusion obtained in Subsection \ref{subsec:Databases}, i.e., CMB data is the most expensive data at a computational level. Interestingly, two papers ($100\%$) use GCD \cite{id_25,id_27} and LSS \cite{id_02,id_09} data in the improvement of the cosmological constraints, while four papers ($100\%$) use generated data in the application of improved parameter estimation techniques \cite{id_04,id_05,id_08,id_14}. For the GL data, three papers ($75\%$) aim to improve the parameter estimation \cite{id_02,id_09,id_13} and one paper ($25\%$) aims to apply an improve parameter estimation \cite{id_19}. Finally, six papers ($85.7\%$) consider in the improvement aim the OHD \cite{id_06,id_10,id_11,id_16,id_20,id_26} and SNe Ia \cite{id_06,id_10,id_12,id_16,id_20,id_22} data, and one paper ($14.3\%$) considers in the application aim the OHD \cite{id_19} and SNe Ia \cite{id_21} data.

In the lower part of Figure \ref{fig:samplesandapplication}, we can observe the frequency of ML and DL model usage based on whether the paper’s objective is application or improvement. Specifically, the BDT model, used once, appears in an article \cite{id_15} ($100\%$) aiming to enhance the parameter estimation. The BML model is exclusively used in papers where the goal is to improve the cosmological constraints, specifically in four papers \cite{id_04, id_05, id_08, id_14} ($100\%$). The BNN model is used once \cite{id_19} ($33.3\%$) in an application article and twice (66.7\%) \cite{id_03, id_18} in improvement-focused papers. The GP model is used twice: once in an improvement paper ($50\%$) \cite{id_11} and once in an application paper ($50 \%$) \cite{id_21}.
Finally, the NN model is exclusively used in improvement-focused articles, particularly in seventeen cases ($100 \%$) \cite{id_01, id_02,id_06,id_07,id_09, id_10, id_12, id_13, id_16, id_17, id_20, id_22, id_23, id_24, id_25, id_26, id_27}. As it was mentioned in the previous Section, NN models offer higher representational capacity due to their deep architecture layers, making them particularly useful when dealing with large datasets.

\begin{figure}
\centering
\includegraphics[scale=0.42]{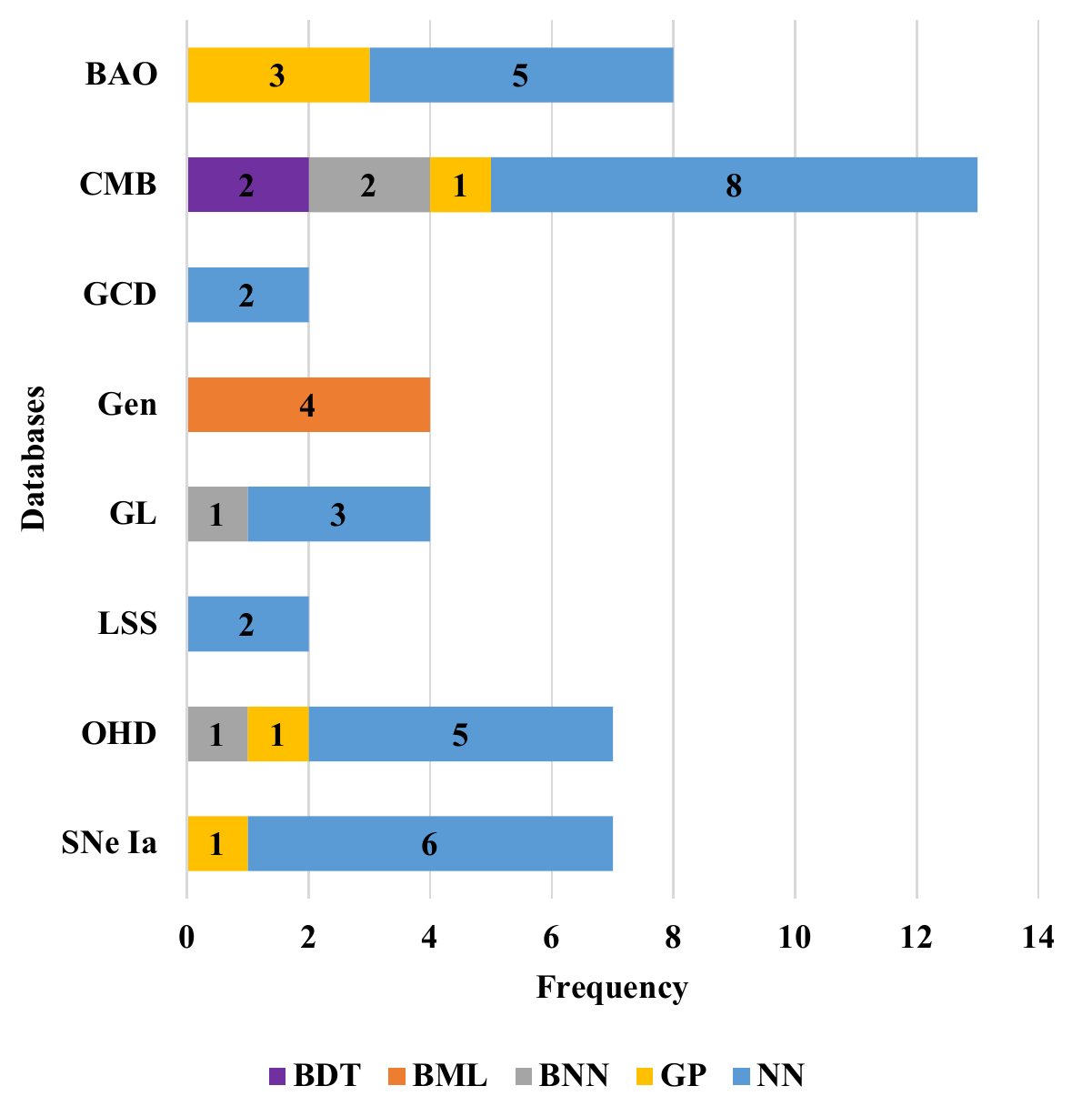}
\caption{\label{fig:modelsbydd}Frequency plot of the usage of the databases employed in the twenty-seven reviewed papers. Additionally, it specifies which ML/DL model was used for each database.}
\end{figure}

In Figure \ref{fig:modelsbydd}, we present a frequency plot of the usage of databases employed in the twenty-seven reviewed papers, specifying which ML/DL model was used for each database. Note that this plot is slightly different from Figure \ref{fig:samplesandapplication} because, in the latter, we present the number of papers that use a specific database while, in Figure \ref{fig:modelsbydd}, we present the ML/DL model used in each database which can be more than one. Following this line, the BAO database, is considered three times ($37.5\%$) the GP ML model and five times ($62.5\%$) the NN DL model. In this case, Ref. \cite{id_21} uses three different catalogs of BAO for the GP ML model. On the other hand, for the CMB database, it is considered one time ($7.7\%$) the GP ML model, two times ($15.4\%$) the BDT ML and BNN DL models, and eight ($61.5\%$) times the NN DL model. As before, Ref. \cite{id_15} uses Planck and WMAP catalogs for the BDT ML model. For the GCD and LSS databases, it is only considered the NN DL model ($100\%$); while for the GL database, it is considered once ($25\%$) the BNN DL model and three times the NN DL model ($75\%$). Following, in the OHD database, it is considered one ($14.3\%$) the BNN DL and GP ML models, and five times ($71.4\%$) the NN DL Model; while, for the SNe Ia database, it is considered one ($14.3\%$) the GP ML model and six times ($85.7\%$) the NN DL model. Finally, in the generated database, it is only considered the BML model ($100\%$). It is important to highlight that NN is the most used DL model, either to improve the parameter estimation or apply an improved cosmological constraint. In particular, for the CMB database, NN is considered eight times ($61.5\%$) which is an expected result, again due to the fact that CMB data is the most expensive data at a computational level which justifies NNs are used because they can obtain deeper representations of the input data. Also, for the generated database, all articles are focused on solving the same problem, which is the $H_0$ tension. For this purpose, the authors improve the ML technique applied through the papers.

\subsection{\label{subsec:Metadata}Year and Type of Publication}
In Figure \ref{fig:yearsandapplication}, we can see the number of articles per year selected in our SLR. As a reminder, we focus our review between the years 2014 to 2024. From a total of twenty-seven reviewed papers, two of them $(7.4\%)$ where available in 2015 \cite{id_15,id_14}, three of them ($11.1\%$) in 2019 \cite{id_01,id_09,id_18}, four ($14.8\%$) in 2020 \cite{id_03,id_12,id_04,id_08}, and three of them ($11.1\%$) in 2021 \cite{id_13,id_16,id_05}. Interestingly, in 2022 a large amount of six of the selected papers ($22.2\%$) were available online \cite{id_02,id_07,id_10,id_17,id_27,id_19}, while eight papers ($29.6\%$) were available in 2023 \cite{id_06,id_20,id_22,id_23,id_24,id_26,id_21}. Finally, one of the selected papers ($3.8\%$) where available online in 2024 \cite{id_11}. In general, the number of publications has been increasing in the last few years, especially in 2022 and 2023, amassing $51.8\%$ of the selected papers, being the 2023 year with the highest number of selected publications. This indicates the trend in the area of cosmology to search for new methods, in this case, artificial intelligence, to determine Bayesian inference.

On the other hand, in Table \ref{tab:journals}, the journals where the selected articles were published are displayed. As it is possible to see, six papers ($22.2\%$) were published in the Journal of Cosmology and Astroparticle Physics (JCAP), five papers ($18.5\%$) in Physical Review D (PRD), five papers ($18.5\%)$ in Monthly Notices of the Royal Astronomical Society (MNRAS), three papers ($11.1\%$) in The Astrophysical Journal Supplement Series (ApJS), and one paper ($3.7\%$) where published in Galaxies, Symmetry, The European Physical Journal C (EPJC) and Science Direct (SD). Four papers ($14.9\%$) were available in the preprint repository arXiv. This means that these papers are accessible online and can be cited but they have not gone through a peer review process. It is important to mention that all the journals, in which the selected papers were published, are Q1 and Q2 journals in the area of Physics, without the presence of journals in the area of computational science.

\begin{table}
\centering
\begin{tblr}{
  width = \linewidth,
  colspec = {Q[435]Q[204]Q[302]},
  hline{1-2,11} = {-}{},
}
\textbf{Journal} & \textbf{Number of papers} & \textbf{Refs}\\
Journal of Cosmology and Astroparticle
  Physics & 6 & \cite{id_07, id_09, id_15, id_17, id_21, id_23}\\
Physical Review D & 5 & \cite{id_04, id_06,id_10, id_18, id_26}\\
Monthly Notices of the Royal Astronomical
  Society & 5 & \cite{id_01, id_02, id_13, id_24, id_25}\\
  Preprint & 4 & \cite{id_03, id_05, id_20, id_27}\\
The Astrophysical Journal Supplement
  Series & 3 & \cite{id_12, id_16, id_22}\\
Galaxies & 1 & \cite{id_11}\\
Symmetry & 1 & \cite{id_14}\\
The European Physical Journal C & 1 & \cite{id_19} \\
Science Direct & 1 & \cite{id_08}
\end{tblr}
\caption{\label{tab:journals} Journals where the articles were published.}
\end{table}

\begin{figure}
\centering
\includegraphics[scale=0.35]{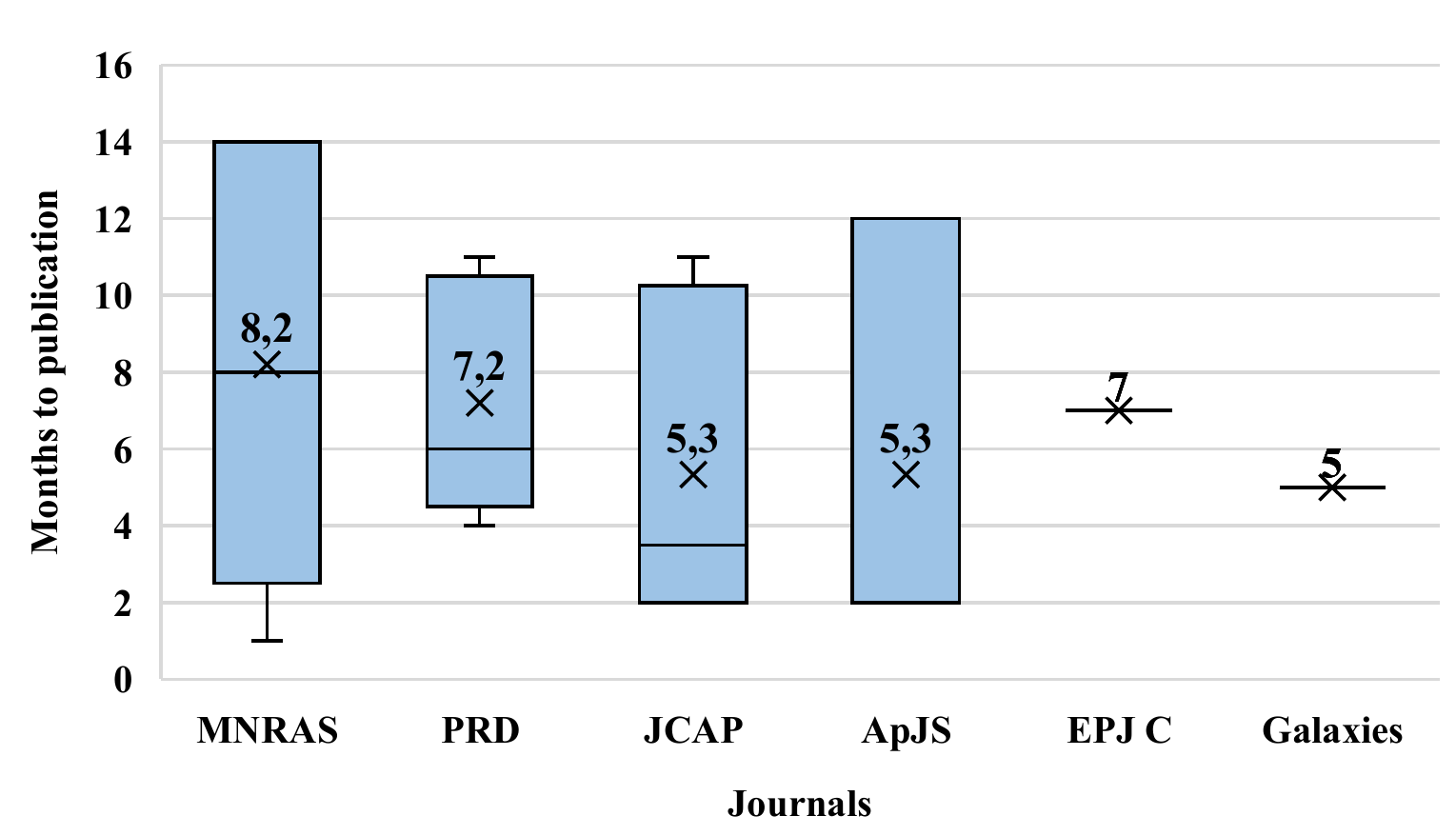}
\caption{\label{fig:arxivtojournal}Amount of months in which the reviewed papers were available in the online repository arXiv until their publication in a journal.}
\end{figure}

The preprint online repository arXiv is of interest, not only because nine of the selected papers are available in that repository and were not published in a journal, but also because the $100\%$ of the twenty-seven selected papers in our SLR are available in that repository. This highlights the importance of the arXiv preprint repository in the area of cosmology (and other physics areas). However, this leads to confusion in the years of publication in this SLR because we have a year in which the paper is available online in the arXiv repository and another year where the paper where published in a journal after a peer review process. Following this line, in Figure \ref{fig:arxivtojournal}, we present the number of months in which the reviewed papers were available in the online repository arXiv until their publication in a journal (if this is the case). As we can see, the selected papers published in MNRAS take a mean of $8.2$ months since the availability in arXiv to the publication in the journal, with a maximum of $14$ and a minimum of $1$ months. On the other hand, PRD takes a mean of $7.2$ months to be published since their availability in arXiv with a maximum of $11$ and a minimum of $4$ months. In the same line, for JCAP the mean time of publication since were available in arXiv to the publication is $5.3$ months, with a maximum of $11$ and a minimum of $2$ months. In the case of ApJS, it has the same range of MNRAS but with a mean of $5.3$ months, a maximum of $12$, and a minimum of $2$ months. Finally, for EPJC and Galaxies, the time of publication were $7$ and $5$ months, respectively. Something out of the ordinary is the case of Symmetry, which goes up to 6 years from its first appearance in arXiv until its publication in the journal. This appears to be a singular case and does not seem to reflect an issue with the journal itself. The long delay between the arXiv submission (2015) and the final publication in Symmetry (2021) may be due to substantial updates made to the manuscript over time, including the integration of the ML topic. However, as this is based on a single paper, no general conclusions should be drawn.

This metric contextualizes method adoption and prevents conflation of preprint and peer-reviewed timelines in our year-based analysis.

\section{\label{sec:Findings}Review findings and future research directions}
This Section presents an in-depth discussion of the review findings derived from the synthesis and analysis of the data. Additionally, it offers practical recommendations to address key research gaps, which could lead to potential future research opportunities. Finally, Section~\ref{sec:tech_benchmarking} provides a concise technique-level comparison and benchmarking considerations, clarifying why cross-paper benchmarking is not methodologically sound given heterogeneous tasks/datasets/metrics and linking these points to the expanded \textit{Performance} summaries in Appendix~A.

\subsection{\label{subsec:Outcomes}Main Outcomes}
\begin{figure*}
\centering
\includegraphics[width=\textwidth]{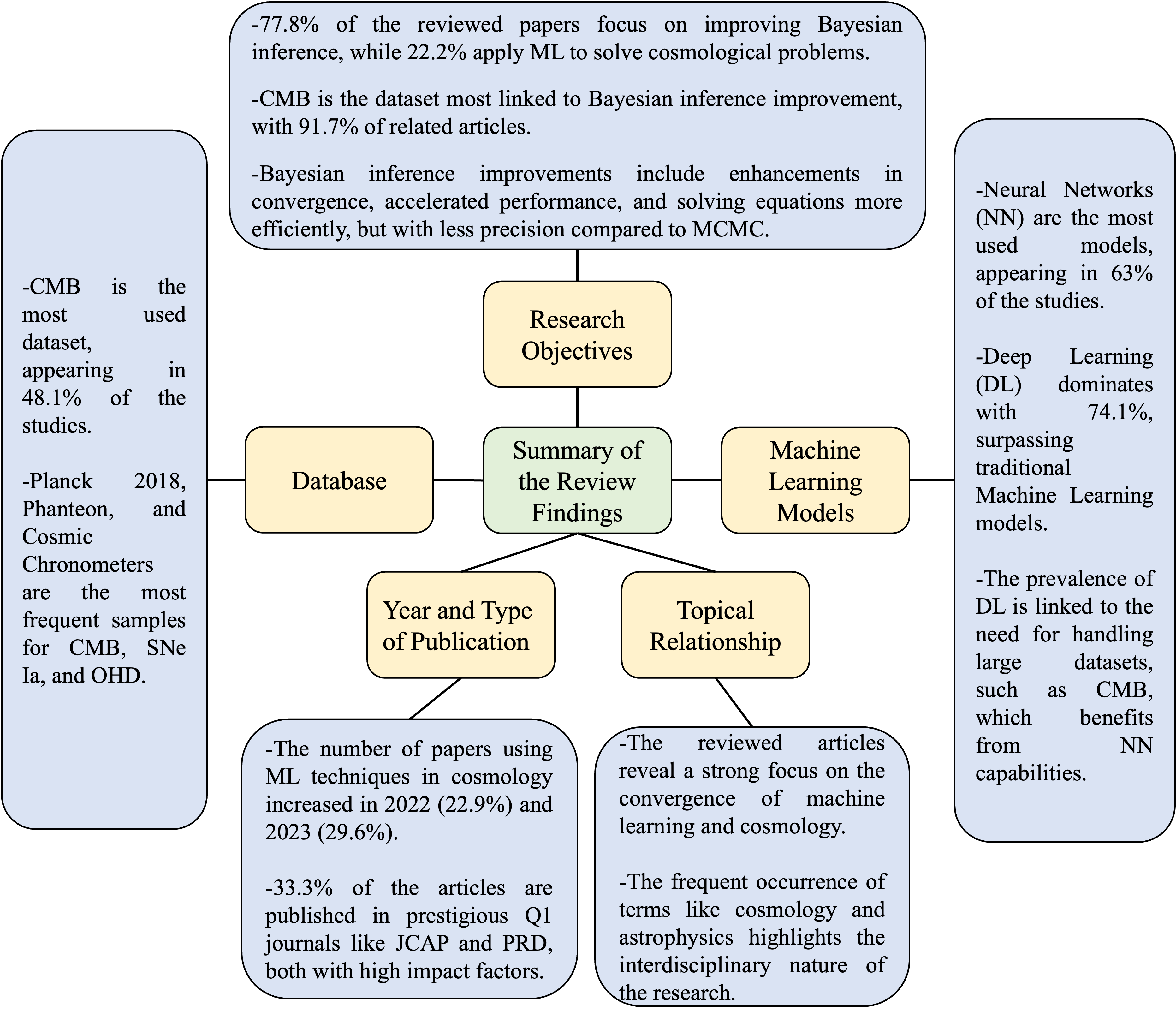}
\caption{\label{fig:Summary}Summary of the review findings.}
\end{figure*}

The findings are organized around five main themes presented in Subsection \ref{subsec:Executing}, with an overview of the review findings illustrated in Figure \ref{fig:Summary}.

\begin{itemize}
\item\textbf{Topical Relationship:} As it was presented in Subsection \ref{subsec:Relation}, the analysis of titles and keywords confirms a clear thematic alignment across the selected studies. The recurring presence of terms such as “cosmology,” “astrophysics,” “parameters,” “neural,” and “Bayesian” illustrates the strong focus on applying machine learning techniques to fundamental cosmological challenges. These challenges predominantly involve the estimation of free parameters and the analysis of nongalactic observational data. Rather than revealing a broad interdisciplinary diffusion, the thematic patterns suggest that the application of ML in cosmology remains largely grounded in the physics and astrophysics domains. This concentration highlights both the relevance and the early stage of this interdisciplinary field, where ML methods are still being explored and adapted to address domain-specific problems. The prominence of terms related to data analysis and inference further reinforces the conclusion that ML is primarily being employed to enhance the efficiency and accuracy of traditional model-fitting techniques within established cosmological frameworks.

\item\textbf{Databases:} Following Subsection \ref{subsec:Databases}, the main datasets used in the reviewed articles are: SNe Ia, OHD, BAO, CMB, LSS, GL, and GCD. For the twenty-seven articles considered in the revision, it can be noted that the most used databases are CMB with a $48.1\%$, BAO with a $29.6\%$, and SNe Ia and OHD with $25.9\%$. Meanwhile, GL, LSS, and GCD are less than $15\%$. It is interesting to note that CMB is the most used data set since it is the most expensive data at the computational level. From the data samples, it can be seen that Planck 2018 ($46.1\%$), Pantheon ($71.4\%$), and Cosmic Chronometers ($57.1\%$) are the most used in CMB, SNe Ia, and OHD, respectively.
\item\textbf{Machine Learning Models:} In general, for both ML and DL models, it was found in Subsection \ref {subsec:MlDlModels} that the majority of the reviewed articles uses NN ($63\%$). The other encountered models do not exceed $15\%$ usage. Moreover, when considering the technique type alone, DL significantly surpasses ML ($74.1\%$ DL compared to $25.9\%$ ML). This trend is probably due to the large amount of data and parameters to be processed, where DL techniques are often better suited due to their capacity to learn patterns at greater depth. This observation is further reinforced by the discovery that CMB, a rather large and complex database, is primarily handled using NNs.
\item\textbf{Research Objectives:} A remarkable result of our SLR is presented in Subsection \ref{subsec:Aim}, in which a $77.8\%$ of the papers are focused on improvement the parameter estimation and $22.2\%$ in the applications of an improved cosmological constraints through ML techniques to solve cosmological problems. In this line, the most studied problem is the $H_0$ tension in different cosmological models. On the other hand, in the improvement there are more varied results, such as enhancement in the convergence, accelerating the performance of the inference, and solving equations more efficiently with ML techniques, but always with less precision compared to the MCMC method. Lower values of precision are achieved for the $\chi^2$ \cite{id_23} in $\Lambda$CDM and little deviations in $\sigma$ for some cosmological parameters \cite{id_22}. In $\Lambda$CDM and $\omega$CDM. Concerning the focus on different databases, it is important to consider that from CMB the $91.7\%$ of the articles are related to the improvement of the parameter estimation, which is expected as CMB is the most expensive data at the computational level. In the case of OHD and SNe IA, both with $85.7\%$, are focused on improvement, and also for BAO with $83.3\%$ of the papers. On the other hand, it is interesting that NN is only used in the improvement of the cosmological constraints and BML is only used in application. In this line, NN is widely used in the reviewed papers for all the databases with $61.5\%$ for the CMB, $62.5\%$ for BAO, $71.5\%$ for OHD, $100\%$ for GCD and LSS, which results in the use of different catalogs $31$ times for the databases mentioned before. Meanwhile, the second technique more used is GP with only $6$ times applied for all the databases.
\item\textbf{Year and Type of Publication:} Finally, from Subsection \ref{subsec:Metadata}, we can see that the last years show an increase in the number of articles available to cite using ML techniques to improve the parameter estimation in cosmology or applied in cosmological problems, such as in 2022 with a $22.9\%$ of the reviewed papers and 2023 with a $29.6\%$. On the other hand, $33.3\%$ of the articles are published in prestigious journals such as JCAP and PRD, which are Q1 journals in the area of physics with a hight impact factor. In this line, it is important to note that the $100\%$ of the reviewed papers are available in the arXiv online repository, giving insights into the useful of this repository in the area of cosmology. We also report the time (in months) between first availability on arXiv and publication in a peer-reviewed venue, as a proxy for the pace at which results are vetted; this helps contextualize the corpus’s reliance on preprints, the maturity of the literature, and the appropriate level of caution when interpreting findings. The updated analysis shows that the mean time is $5.3$ months for JCAP, $7.2$ months for PRD, and $8.2$ months for MNRAS. Other journals such as ApJS show a comparable mean time to JCAP, also at $5.3$ months, while EPJC and Galaxies take 7 and 5 months, respectively. An outlier in the dataset is Symmetry, where one article took 6 years to transition from arXiv to formal publication—likely due to substantial post-submission modifications and topic shifts, rather than delays inherent to the journal itself.
\end{itemize}

\subsection{\label{subsec:Gaps}Research Gaps and Recommendations}
\begin{figure*}
\centering
\includegraphics[width=\textwidth]{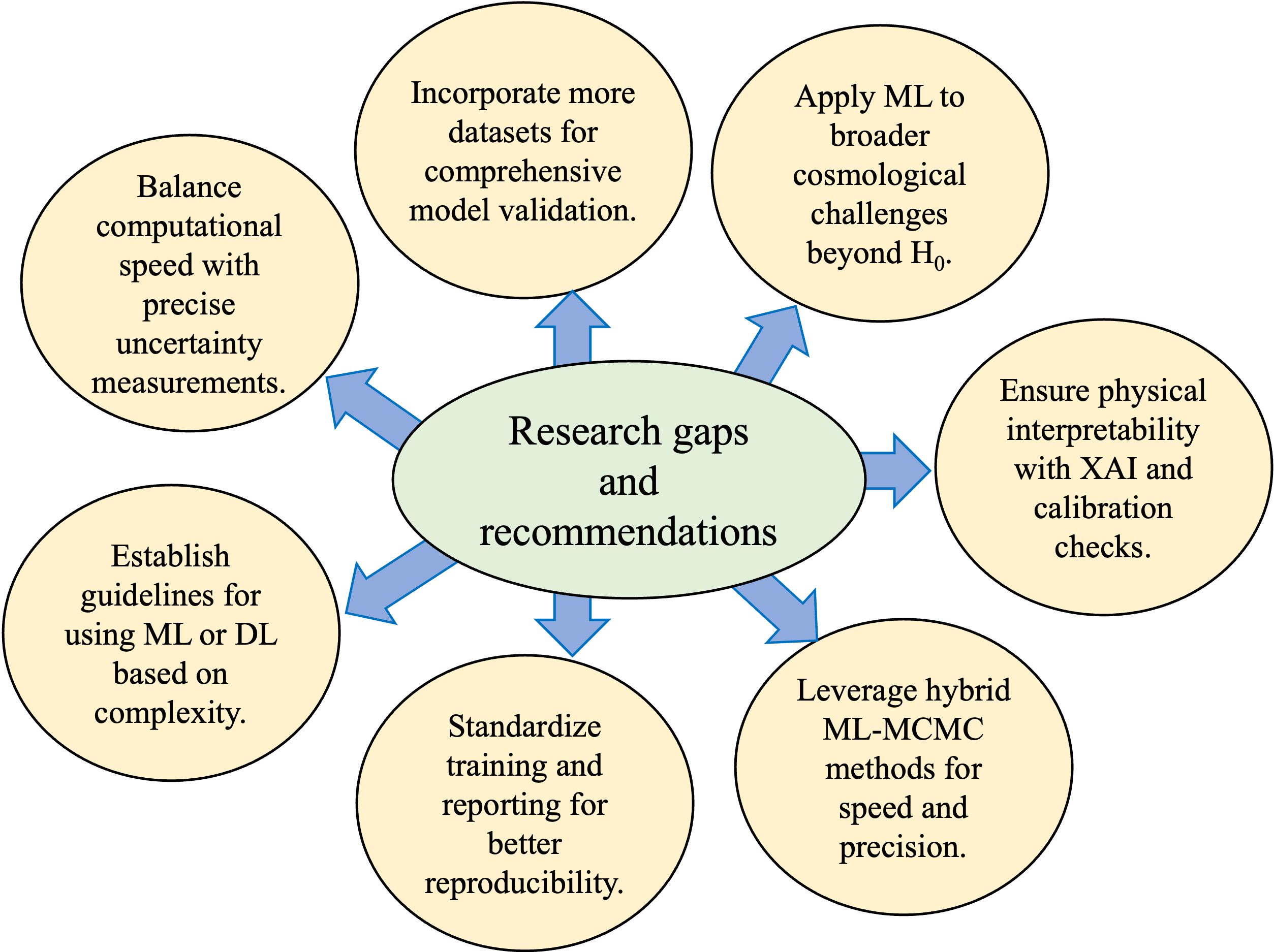}
\caption{\label{fig:Researchgaps}Summary of the research gaps and recommendations identified in the SLR.}
\end{figure*}

According to the main findings of our SLR, we organize the gaps and recommendations into three themes—\textbf{Methodological}, \textbf{Application}, and \textbf{Data-related}—summarized in Figure~\ref{fig:Researchgaps}.

\noindent\textbf{Methodological gaps and recommendations}
\begin{itemize}[leftmargin=1.2em]

\item[(M1)] \textbf{Improvement vs. Precision for ML Techniques}

Various ML techniques, including BNN, GP, and DL models, have been applied to cosmological data analysis, showing their potential to improve parameter estimation and to handle large, complex datasets. However, a central tension emerges between computational acceleration and the fidelity/calibration of the resulting constraints. In our corpus, methods that deliver the largest speedups sometimes exhibit wider credible regions or miscalibrated posteriors relative to classical baselines, highlighting that efficiency gains do not automatically guarantee precision. The cosmological consequences of using faster but less precise surrogates are concrete: posteriors may be biased, credible regions under/over-estimated, and thus key scientific claims can be distorted—e.g., tensions (such as $H_0$) artificially inflated or masked, Bayes factors and model selection misreported, and cross-probe consistency (CMB/BAO/SNe/LSS) mischaracterized.

There are, nevertheless, regimes where prioritizing speed is justified: when repeated forward evaluations dominate wall-time (e.g., emulating Einstein–Boltzmann pipelines inside MCMC), during rapid exploratory scans in high-dimensional spaces, or for triage/operational tasks. By contrast, precision must take precedence for final parameter estimation intended for publication, tension quantification across probes, and model comparison—settings where coverage and bias directly affect scientific validity. In practice, speed-first surrogates (often NN-based) are valuable for amortizing computation, provided they are accompanied by explicit uncertainty calibration and validation against classical pipelines; BNN/GP approaches, while costlier, tend to offer stronger uncertainty calibration when their assumptions hold.

\emph{Recommendations.} We recommend reporting both efficiency (speedup~$\times$, wall-time, ESS/s) \emph{and} calibration/accuracy (coverage, bias, posterior width, posterior predictive checks/PIT) to make the trade-off explicit; validating surrogates against Einstein–Boltzmann solvers or exact likelihoods at checkpoints; adopting physics-aware inductive biases where possible (e.g., spherical/equivariant layers, operator-learning surrogates); and using hybrid pipelines that combine speed-first emulation with precision-first verification (e.g., periodic exact re-evaluation, proposal preconditioning, stress tests under distribution shift). These practices mitigate the risk that acceleration comes at the expense of reliable cosmological constraints.

\item[(M2)] \textbf{Hybrid Approaches: Combining Machine Learning and MCMC Methods}

Hybrid methodologies emerge from the reviewed papers that combine ML techniques with the traditional MCMC method. One combination is the use of ML techniques to solve differential equations. For example, in Ref. \cite{id_06,id_10}, NNs are employed to solve the cosmological background equations, while in Ref. \cite{id_24}, NNs are used to solve the Einstein-Boltzmann equations. In both cases, the solutions are used as input for the MCMC method, accelerating the computation time and maintaining the precision. These examples demonstrate the versatility of neural networks in efficiently handling computationally expensive components of cosmological modeling, paving the way for their integration into traditional inference workflows. This highlights the importance of not only focusing on improving the MCMC method, but also on investing resources in optimizing the most time-consuming aspects of the procedure involving CMB data, i.e., the development of more efficient Einstein–Boltzmann solvers.

Another combination is using the results of ML techniques as a prior as in Refs. \cite{id_03, id_18}. In these works, The results obtained from BNN are used as prior information/input for the MCMC method, accelerating the computations with similar precision as in the classical implementation of MCMC. This approach is particularly effective in reducing the dimensionality of the parameter space, allowing MCMC methods to focus on fine-tuning within a more constrained region. Such integration not only reduces computational overhead but also enhances the stability of the inference process in high-dimensional scenarios.

Despite these advances, the adoption of hybrid methodologies is not without challenges. Ensuring compatibility between ML-generated outputs and MCMC implementations requires careful validation, especially when physical constraints must be preserved. Additionally, the interpretability of ML-based priors remains an area of concern, as it can obscure the underlying assumptions driving the parameter inference. Addressing these challenges will be critical to ensuring the robustness and reliability of hybrid approaches.

Based on the above, a recommendation is to explore these hybrid methodologies since they accelerate the computations and give us uncertainties similar to the traditional MCMC method. Future research should focus on standardizing frameworks for integrating ML techniques with MCMC, establishing benchmarks to compare hybrid and traditional methods, and exploring the potential of emerging ML techniques such as physics-informed neural networks (PINNs) to further optimize cosmological computations. These efforts will help realize the full potential of hybrid methods in advancing precision cosmology.

Hybrid methodologies that combine ML techniques with MCMC have demonstrated their potential to enhance cosmological computations by balancing precision and efficiency. For instance, NNs can be used to solve cosmological equations or reduce the dimensionality of parameter spaces before applying MCMC, as described in (M2). These approaches leverage the strengths of DL, discussed in (M4), particularly in handling high-dimensional data and learning intricate patterns in cosmological datasets. By integrating the scalability and adaptability of DL models into hybrid methodologies, researchers could achieve significant gains in both computational performance and accuracy. These efforts will help realize the full potential of hybrid methods in advancing precision cosmology.

These hybrid methods, which balance precision and efficiency, complement the strengths of DL discussed in (M4), particularly in analyzing high-dimensional datasets and addressing complex cosmological problems.

\item[(M3)] \textbf{Inconsistent Reporting Standards in Model Training} 

    The evidence gathered from the training phases of different models lacks a consistent standard across studies, leading to variability in reporting. Some papers focus on the theoretical aspects, detailing architectural choices and modifications \cite{id_20, id_21, id_22, id_07}, while others provide comprehensive descriptions of experimental setups, such as the libraries used, programming languages, and environmental contexts \cite{id_03, id_04, id_05, id_08, id_14}. A subset of studies presents detailed visualizations of model architectures, including layer interconnections and input-output flows \cite{id_01, id_02, id_09, id_17}, whereas others offer step-by-step guidelines for training procedures \cite{id_12}.

Despite these contributions, there remains a notable absence of unified guidelines for documenting the training process, leading to substantial heterogeneity in the level of detail provided. For instance, while some studies excel in presenting architectural visualizations or procedural guidelines, critical aspects such as dataset distribution and preprocessing techniques are often underreported. Without these details, it is difficult to evaluate the representativeness and generalizability of the models. Moreover, studies rarely specify training duration, making it challenging to assess computational efficiency or scalability.

However, significant gaps remain in the reporting of critical elements such as dataset distribution (how data is divided into subsets for training, validation, and testing), training duration (the total time spent training a model), computing environments (e.g., the hardware or cloud infrastructure used), and hyperparameters (configurations that control the training process, such as the learning rate, which determines how much the model adjusts its parameters in each iteration, and activation functions, which define how signals are passed between layers in neural networks). Equally crucial are other factors that enhance comparability, such as the precise evaluation metrics used, a thorough description of the training pipeline, and the rationale behind specific parameter choices. The absence of these elements creates significant obstacles for meta-analysis, as variability in reporting undermines the comparability of results across studies. These omissions hinder the comparability of results and the identification of optimal configurations, complicating efforts to determine the most effective models or hyperparameter settings.

Furthermore, the lack of comprehensive context makes it challenging to replicate experiments under identical conditions. For example, some studies fail to describe the computing hardware used (e.g., GPUs, CPUs, or cloud infrastructure), which significantly impacts training performance and cost-effectiveness. In fields like cosmology, where data sizes and computational demands are substantial, these details are critical for assessing the feasibility of deploying similar models in real-world scenarios. This variability raises questions about the reliability and reproducibility of findings, ultimately limiting their utility in advancing the field.

As a result, this lack of standardization highlights the urgent need for a more structured approach to documenting the training phases of machine learning models. Developing and adopting reporting frameworks that ensure transparency and completeness, akin to PRISMA guidelines for systematic reviews, would greatly enhance reproducibility, reliability, and the cumulative progress of machine learning in cosmology.

\item[(M4)] \textbf{Deep Learning vs. Traditional Machine Learning in Cosmological Applications}

Our review confirms that DL models generally outperform traditional ML methods, particularly when processing large datasets with numerous variables. NNs excel in learning intricate patterns through their deep architectures, offering extensive customization to suit various cosmological applications. The findings reveal a marked preference for DL algorithms across the reviewed studies, suggesting that ML models, while useful for preliminary analysis or feature selection, may struggle with the complexities of cosmological data. Specifically, DL methods such as CNNs and RNNs demonstrate superior performance in tasks requiring spatial and temporal pattern recognition, such as analyzing sky surveys and time-series data from telescopes. These strengths make DL particularly advantageous for applications like detecting gravitational waves, mapping dark matter distributions, and estimating cosmological parameters.

Additionally, the flexibility of NNs facilitates essential adaptations for cosmological inference, enabling precise parameter estimation and more effective handling of high-dimensional observational data. For example, custom loss functions tailored to cosmological objectives (e.g., minimizing deviations in parameter estimation) and the integration of physical constraints within NN architectures allow for more accurate modeling of astrophysical phenomena. Moreover, the scalability of DL models enables them to handle the exponential growth of data from next-generation surveys, such as the Vera Rubin Observatory and the Euclid mission, where traditional ML methods often falter.

Across the surveyed studies, architectural choices are largely adapted from computer science (e.g., CNN/U-Net variants for map-like data; MLPs for emulators), with only limited explicit encoding of cosmological inductive biases. Notable exceptions include spherical/equivariant convolutions for CMB/weak-lensing maps and operator-learning surrogates for Einstein–Boltzmann pipelines. This pattern helps explain why the largest speedups sometimes coexist with calibration issues: acceleration is prioritized, while symmetry constraints and uncertainty modeling are not always built in. A practical direction is to combine physics-aware architectures with explicit calibration checks (coverage/PIT) and comparisons to classical baselines, so that efficiency gains do not compromise constraint reliability.

However, the increased computational demands and the risk of overfitting in DL models highlight the importance of establishing clear evaluation metrics and benchmarks to assess their efficiency and reliability in cosmological contexts. Studies should also explore hybrid approaches that combine the strengths of ML and DL—for instance, using ML for feature extraction and DL for parameter inference—to optimize resource utilization while maintaining accuracy.

Future research could benefit from clearly defined guidelines for when ML methods are appropriate versus when the added complexity of DL is justified. These guidelines should consider not only the size and complexity of the datasets but also factors such as computational resources, interpretability needs, and the specific objectives of the study. Developing such a framework would enable researchers to make informed decisions, maximizing the scientific impact of their analyses while minimizing resource expenditure.  

\item[(M5)] \textbf{Interpretability and physical faithfulness (XAI)}

A recurring concern in precision cosmology is trust in “black-box” models. Beyond aggregate performance, we need evidence that models learn physics-relevant structure rather than survey- or instrument-specific artifacts. In the reviewed literature, XAI is unevenly reported. Useful practices include: (i) attribution and saliency analyses on map-like inputs (e.g., integrated gradients, Grad-CAM) with checks that highlighted regions align with physically meaningful features; (ii) stability tests of explanations under small input perturbations and across instruments/surveys to detect spurious correlations; (iii) counterfactual “injection” and ablation tests using simulations (turning on/off specific effects) to verify causal sensitivity to the intended signal; (iv) enforcing inductive biases via symmetry-aware architectures (e.g., spherical/equivariant layers) and operator-learning surrogates; and (v) posterior diagnostics (coverage, posterior predictive checks, simulation-based calibration) to ensure that uncertainty is not only reported but also calibrated.

\emph{Recommendations.} Report at least one attribution method with stability checks; include cross-survey/domain-shift tests; provide physics “unit tests” via controlled simulation injections; prefer architectures that encode known symmetries; and release code to reproduce explanations and diagnostics alongside classical-baseline comparisons.

\end{itemize}

\noindent\textbf{Application gaps}
\begin{itemize}[leftmargin=1.2em]

\item[(A1)] \textbf{ML Technique Applied to Cosmological Problems}

The reviewed papers considered two scenarios: the improvement in the algorithm and the application to a cosmological problem. In the first case, an exploration of the algorithm was focused on refining the parameter estimation processes through ML techniques. In the second case, the ML technique was directly toward addressing the $H_0$ tension. This tension refers to the discrepancy between the observational technique (such as CMB) and the local measurements of the Hubble constant $H_0$. The study of this problem was in the context of different cosmological scenarios to $\Lambda$CDM such as a Universe dominated by only one fluid with a general barotropic equation \cite{id_04} or a Universe with barotropic dark energy and dark matter \cite{id_05}. In the same line, was studied to probe the opacity of the Universe with a $x$CDM cosmological model giving as a result that impact in the $H_0$ tension problem \cite{id_08}.

While the reviewed papers focus primarily on this specific problem, there exists a significant gap in exploring different cosmological issues that could benefit from these ML techniques. For instance, other cosmological problems could be treated with the ML techniques used in the reviewed articles, such as the cosmological constant problem \cite{Weinberg:1988cp} concerning the large discrepancy between the theoretical value of the cosmological constant and the observed value (which is related to dark energy). The ML techniques explored in the reviewed articles could help to test alternative models to $\Lambda$CDM that might solve this problem. On the other hand, the coincidence problem \cite{Velten:2014nra}, related to the fact that energy densities for dark matter and dark energy are of the same order of magnitude at the current time, which can be seen as a fine-tuning problem, could also be explored with ML techniques to test alternatives models to explain this issue.

Another cosmological issue is the possibility of a warm dark matter component in the Universe \cite{Newton:2020cog}, an alternative to the cold dark matter considered in $\Lambda$CDM, this could leave imprints in the LSS and the CMB background that could be analyzed with the ML techniques and be tested in different cosmological scenarios. Finally, to mention a few, the phantom dark energy \cite{Rest:2013mwz} which accelerates the expansion of the Universe and its origin is unknown, could also be tested in different cosmological scenarios with the use of ML techniques.

It is important to emphasize that all the aforementioned problems benefit from the use of ML techniques for parameter estimation, thereby helping to alleviate the tensions within the $\Lambda$CDM model. Likewise, other problems may also benefit from these parameter estimation results

\end{itemize}

\noindent\textbf{Data-related gaps}
\begin{itemize}[leftmargin=1.2em]

\item[(D1)] \textbf{Cosmological Data Used in ML Techniques}

Traditional cosmological methods, such as those using SNe Ia, BAO, and CMB data, continue to be fundamental tools in cosmology. These provide insights into the expansion, the energy densities, and the parameters of a specific cosmological model. The incorporation of ML techniques enhanced their robustness and efficiency, through their abilities to analyze a large and complex dataset, having the potential to optimize the parameter estimation of alternative cosmological models to $\Lambda$CDM. Nevertheless, we identify that a large number of the reviewed works do not use the majority of the available databases and focus their studies on a relatively narrow subset of the available dataset. For instance, several studies use particular datasets, like only CMB, or combine SNe Ia with other datasets such as OHD without incorporating the total available dataset.

The use of some of the datasets mentioned before brings some valuable results to cosmology but limits the scope of the analysis and estimation of the cosmological parameters for certain cosmological models. Moreover, this could may cause biases because each dataset has its own uncertainties and observational limitations. In this sense, we identify a gap because any alternative cosmological model, to be considered viable, must be able to describe the total background cosmological data in the same or better way as the $\Lambda$CDM model. In the case of $\Lambda$CDM, it has been extensively tested and validated with a wide range of observational data, and any alternative model must be consistent, at least at the same level, with the current observational data. This includes not only using a single dataset but also accounting to join all of them for a full analysis. Therefore, we recommend focusing future studies considering to use of larger numbers of datasets such as gravitational waves, CMB, LSS, OHD, and SNe Ia, among others in order to be fully consistent in the analysis of alternative cosmological models. Nevertheless, we would like to emphasize that this is not a straightforward task due to the complexity of the cosmological model and the specificities of its analysis.

\end{itemize}

\subsection{\label{sec:tech_benchmarking}Technique-level comparison and benchmarking considerations}

\paragraph{Why cross-paper benchmarking is not methodologically sound.}
The 27 papers in our corpus differ substantially in (i) \textit{task framing} (e.g., emulation of Boltzmann solvers, likelihood-free simulation-based inference, or direct regression), (ii) \textit{cosmological probes and datasets} (CMB, BAO, SNe\,Ia, OHD, LSS, gravitational lensing), (iii) \textit{pipelines} (likelihoods, priors, simulators, samplers, train/validation splits), and—crucially—(iv) \textit{reported metrics} and reporting conventions. Because there is no common baseline across data, simulators, priors, calibration diagnostics, or compute budgets, constructing an ``apples-to-apples'' ranking of NN, BNN, and GP from published results would be misleading. A genuinely comparable benchmark would require re-executing all methods under a harmonized protocol (shared datasets/splits, identical likelihoods or simulators, agreed-upon diagnostics, fixed compute budgets), which lies beyond the scope of a systematic review and is closer to a dedicated community benchmarking effort.

\paragraph{When each family of techniques tends to be preferable.}
Our synthesis reveals a consistent trade-off between \textit{efficiency} and \textit{uncertainty quality}:
\begin{itemize}
    \item \textbf{NN-based accelerators} (emulators/surrogates, accelerated profile-likelihood computations) are preferable when \textit{wall-time} and throughput dominate (e.g., repeated model evaluations), provided that calibration is checked against standard pipelines.
    \item \textbf{BNN/GP approaches} tend to be preferable when \textit{calibrated posteriors} and well-characterized uncertainty are the priority, accepting higher computational cost when their assumptions hold.
    \item \textbf{Hybrid schemes} (e.g., NN emulators within MCMC, or normalizing flows in simulation-based inference) are attractive when both speed and calibration matter, using ML to accelerate expensive steps while retaining principled posterior checks.
\end{itemize}

\paragraph{Minimal conditions for a future community benchmark.}
To enable fair, objective comparisons, we encourage convergence on: (i) common public datasets/simulators with predefined splits; (ii) standard \textit{efficiency metrics} (e.g., speedup against a reference pipeline; wall-time per effective sample); (iii) \textit{calibration diagnostics} (credible-interval coverage, PIT/SBC, posterior-width ratios); (iv) \textit{accuracy criteria} (NLL/$\chi^2$ on held-out data; parameter bias); (v) transparent \textit{compute-budget reporting} (hardware, runtime caps, seeds); and (vi) \textit{out-of-domain} tests to quantify robustness under domain shift.

\paragraph{Practical implication for this review.}
Until such conditions are met, cross-paper ``performance tables'' should be interpreted as \textit{as-reported summaries} rather than direct rankings. Consistent with this principle, Appendix~A also includes a \textit{Performance} column that consolidates, for each study, the outcomes reported by the authors and explicitly flags their non-comparability across heterogeneous setups.

\section{\label{sec:Threats}Threats to Validity}
In the course of conducting this SLR on the application of ML techniques for observational constraints in cosmology, several potential threats to validity have been identified. These threats may impact the generalizability, reliability, and interpretation of the findings. The following Section categorizes and discusses these threats.

\subsection{\label{subsec:External}External Validity}
This review focuses on the use of machine learning (ML) techniques in cosmology, particularly in improving parameter estimation through Bayesian inference and related methods. While the findings may offer valuable insights, their applicability to other domains, or even different contexts within cosmology itself, is limited. The cosmological datasets reviewed, such as SNe Ia data, CMB measurements, and LSS surveys, have unique characteristics that may not generalize well to other fields. For example, ML models trained on cosmological data often deal with high-dimensional and noisy datasets, which may behave differently in non-cosmological domains. In fact, it is usually the opposite, as it is cosmologists who apply ML techniques developed in other fields (such as computer science) to adapt them to the specific needs of cosmological data. Therefore, researchers applying these techniques outside of cosmology should be cautious when attempting to extrapolate conclusions directly.

On the other hand, the generalization across different cosmological datasets can be challenging. Data from diverse cosmological surveys, such as the LSST or Euclid, may have varying levels of noise, resolution, and biases. These differences can impact the performance of ML models in ways that limit their generalizability. The findings in this review, which focus on specific datasets, may not hold across all future surveys. 

In addition, the clear thematic alignment we highlight may partly reflect our inclusion criteria (which prioritize works that explicitly derive observational constraints) together with the modest size of the corpus (27 papers, 2014–2024). This naturally limits thematic diversity at present; accordingly, this survey should be regarded as an initial baseline to be updated as the literature expands and reporting standards mature.

Finally, the results are highly time-specific. The ML techniques reviewed in this paper rely on the data and methods currently available. As new cosmological data emerge and ML models evolve, the conclusions drawn from the present review may become outdated or require significant adjustment.

\subsection{\label{subsec:Construct}Construct Validity}
Construct validity refers to how well the ML techniques and evaluation methods used in the studies reviewed actually measure the intended cosmological parameters. One potential threat to construct validity in this review is the diversity of evaluation metrics used across different studies. Some studies may focus on metrics such as accuracy or mean squared error, while others emphasize Bayesian inference performance or uncertainty estimation. This variability in evaluation criteria complicates efforts to compare and synthesize results across studies, potentially leading to inconsistent conclusions about the effectiveness of ML techniques in cosmology.

Another concern is the complexity and uniqueness of cosmological data. Cosmological datasets often involve high noise levels, incomplete information, and non-linear relationships between variables. In some cases, ML models may overfit to these unique characteristics rather than generalize to broader cosmological phenomena. This could create an illusion of strong model performance on the training data, but such models might underperform when applied to new or unseen cosmological observations.

\subsection{\label{subsec:Internal}Internal Validity}
Internal validity concerns whether the conclusions drawn in the reviewed studies are truly reflective of the methods used, rather than external biases or influences. One prominent threat to internal validity in this review is the possibility of publication bias. Studies that report positive or statistically significant results may be more likely to be published, leading to a potential overestimation of the effectiveness of ML techniques in cosmology. Negative or inconclusive findings might remain unpublished, thus skewing the overall picture of the field.

Additionally, there is a risk of researcher bias during the review process. Although rigorous inclusion and exclusion criteria were employed, the subjective interpretation of the relevance and quality of the studies reviewed may introduce biases. The decisions made regarding which studies to include in the review could influence the overall findings, particularly if certain datasets or methods were preferentially selected.

Finally, the variability in the quality and characteristics of the cosmological datasets used in the reviewed studies also poses a risk to internal validity. Some datasets may be more amenable to ML techniques due to higher data quality or fewer inherent biases, leading to potentially skewed conclusions regarding the effectiveness of the techniques across the field.

\subsection{\label{subsec:Conclusion}Conclusion Validity}
Conclusion validity refers to the strength and reliability of the conclusions drawn from the reviewed studies. A key threat to conclusion validity in this review lies in the interpretation of the findings. Different researchers may prioritize different aspects of ML applications, such as computational efficiency, accuracy, or interpretability. This variation in emphasis can affect the conclusions drawn from the review, as certain aspects of the findings may be given more weight than others depending on the researcher’s focus.

Moreover, inconsistency in the results across studies presents another challenge. Some ML techniques may perform well under certain cosmological conditions but fail to generalize to others. This inconsistency makes it difficult to draw broad, definitive conclusions about the overall utility of ML for cosmological parameter estimation. It is essential to recognize that while certain methods may show promise, the overall effectiveness of ML in cosmology remains context-dependent, with substantial variability in performance depending on the dataset and problem at hand.

\section{\label{sec:Conclusions}Conclusions}
This study presents a systematic review exploring ML techniques applied to derive observational constraints in cosmology, offering a comprehensive synthesis of the most relevant methodologies at this interdisciplinary intersection. Through the analysis of twenty-seven articles, we identified that NNs and DL approaches dominate the field due to their ability to handle large datasets and extract complex patterns efficiently.

The results of this review indicate that, while ML techniques have shown significant improvements compared to traditional methods such as MCMC, particularly in efficiency (with precision maintained when surrogate models are properly validated), challenges remain regarding their full integration into conventional cosmological practices. The observed preference for hybrid models and the use of simulated data also suggests that the scientific community is in a transitional phase, experimenting with approaches that combine the robustness of classical methods with the flexibility of ML. Specifically, we highlight the potential of DL to address computational bottlenecks in processing CMB data and free cosmological parameters. However, the lack of standardization in evaluation criteria and the limited adoption of certain models, such as BNNs, reveal areas for improvement and the need for future research.

On the applied side, progress has been made in resolving critical issues such as the Hubble constant ($H_0$) tension using Bayesian techniques, but further studies are needed to consolidate these methodologies in broader scenarios. The increasing availability of data through upcoming projects, such as LSST and Euclid, positions ML as an essential tool to tackle emerging challenges in precision cosmology.

In alignment with our objectives (RQ1–RQ4): NN are most frequent, with BNN and GP less common (RQ1); NN surrogates typically provide the largest speedups for repeated forward evaluations—enabling practical MCMC—while maintaining precision when validated against Einstein–Boltzmann solvers, whereas BNN and GP generally offer stronger uncertainty calibration under their assumptions (RQ2); head-to-head comparisons are scarce and reporting remains heterogeneous, with limited cross-survey validation (RQ3); and the common mix of observational and simulated data calls for explicit calibration checks and stress tests (RQ4). In practice, we recommend NN for high-dimensional, speed-critical emulation (with uncertainty calibration checks), BNN when well-calibrated posteriors are paramount and data are moderate, and GP for small/medium datasets in low–moderate dimensional settings.

Finally, we propose that future research focus on developing methodological frameworks that integrate common standards for model validation and on creating interpretable approaches to foster trust within the cosmological community. Interdisciplinary collaboration between cosmologists and ML experts will be crucial to maximizing the impact of these techniques and ensuring their sustainable adoption in scientific practice.

\vspace{6pt} 





\authorcontributions{Conceptualization, L.R. and E.G.; methodology, L.R., S.E., E.G., C.M. and F.L; software, L.R. and S.E.; validation, L.R., S.E., E.G., C.M. and F.L; formal analysis, L.R., S.E., E.G., C.M. and F.L; investigation, L.R., S.E., E.G. and C.M.; data curation, L.R. and S.E.; writing---original draft preparation, L.R. and E.G.; writing---review and editing, L.R., S.E., E.G., C.M. and F.L; visualization, L.R., S.E., E.G. and C.M.; supervision, L.R. and E.G. All authors have read and agreed to the published version of the manuscript.}

\dataavailability{No new data were created for this work.}

\acknowledgments{E.G. acknowledges the scientific support of Núcleo de Investigación No. 7 UCN-VRIDT 076/2020, Núcleo de Modelación y Simulación Científica (NMSC).}

\conflictsofinterest{The authors declare no conflicts of interest.} 



\abbreviations{Abbreviations}{
The following abbreviations are used in this manuscript:\\

\noindent 
\begin{tabular}{@{}ll}
MCMC &  Markov chain Monte Carlo \\
ML &  Machine Learning \\
SLR & Systematic Literature Review \\
DL & Deep Learning \\
CDM & Cold Dark Matter \\ 
SNe Ia & Type Ia Supernovae \\
SNe & Supernovae \\
OHD & Observational Hubble Parameter Data \\
BAO & Baryon Acoustic Oscillations \\
CMB & Cosmic Microwave Background \\
WMAP &  Wilkinson Microwave Anisotropy Probe \\
LSS & Large Scale Structure \\
2dFGRS & 2-degree Field Galaxy Redshift Survey \\
SDSS & Sloan Digital Sky Survey \\
GL & Gravitational Lensing \\
H0LiCOW & $H_{0}$ Lenses in COSMOGRAIL’s Wellspring \\
LSST & Legacy Survey of Space and Time \\
SPHEREx & Spectro-Photometer for the History of the Universe, Epoch of Reionization, and Ices Explorer \\
NGRST & Nancy G. Roman Space Telescope \\
DESI & Dark Energy Spectroscopic Instrument \\
PFS6 & Prime Focus Spectrograph \\
NN &  Neural Network \\
RNN & Recurrent Neural Network \\
BML & Bayesian Machine Learning \\
GP & Gaussian Processes \\
BDT & Bayesian Decision Trees \\
BNN & Bayesian Neural Network \\
CNN &  Convolutional Neural Network \\
GCD & Galaxy Clustering Data \\
WFIRST & Wide Field Infrared Survey Telescope \\
JACP & Journal of Cosmology and Astroparticle Physics \\ 
PRD & Physical Review D \\
MNRAS & Monthly Notices of the Royal Astronomical Society \\
ApJS & The Astrophysical Journal Supplement Series \\
EPJC & The European Physical Journal C \\
PINN & Physics-Informed Neural Network
\end{tabular}
}

\newgeometry{left=6cm, right=6cm, top=5mm, bottom=10mm,}
\appendixtitles{no} 
\appendixstart
\appendix
\section[\appendixname~\thesection]{List of selected papers for review and Extracted Data }\label{apendix:selectedPapers}

\DefTblrTemplate{contfoot-text}{default}{}
\DefTblrTemplate{conthead-text}{default}{}
\DefTblrTemplate{caption}{default}{}
\DefTblrTemplate{conthead}{default}{}
\DefTblrTemplate{capcont}{default}{}
\begin{table}[H]
\begin{adjustwidth}{-\extralength}{0cm}
\scriptsize

\newcolumntype{L}[1]{>{\RaggedRight\arraybackslash\hsize=#1\hsize}X}

\begin{tabularx}{\fulllength}{L{2.0} L{1.6} L{0.4} L{0.8} L{0.2} L{1.0} c c}
\toprule
\textbf{Title\footnotemark} & \textbf{Author(s)} & \textbf{Databases} & \textbf{Catalogs} & \textbf{ML Models} & \textbf{Research Aim} & \textbf{Ref} & \textbf{Year} \\
\midrule
Accelerated Bayesian inference using deep learning & Moss, Adam & CMB & Planck 2015 & NN & Improvement & \cite{id_01} & 2020\\
Accelerating cosmological inference with Gaussian processes and neural networks – an application to LSST Y1 weak lensing and galaxy clustering & Boruah, Supranta S; Eifler, Tim; Miranda, Vivian; Krishanth, P M Sai & Weak Lensing & LSST & NN & Improvement & \cite{id_02} & 2022\\
Accelerating MCMC algorithms through Bayesian Deep Networks & Hortua, Hector J.; Volpi, Riccardo; Marinelli, Dimitri; Malago, Luigi & CMB & Simulated & BNN & Improvement & \cite{id_03} & 2020\\
An analysis of the H 0 tension problem in the Universe with viscous dark fluid & Elizalde, Emilio; Khurshudyan, Martiros; Odintsov, Sergei D.; Myrzakulov, Ratbay & Generated & --- & BML & Application & \cite{id_04} & 2020\\
An approach to cold dark matter deviation and the $H_{0}$ tension problem by using machine learning & Elizalde, Emilio; Gluza, Janusz; Khurshudyan, Martiros & Generated & --- & BML & Application & \cite{id_05} & 2021\\
Faster Bayesian inference with neural network bundles and new results for $f(R)$ models & Chantada, Augusto T. and Landau, Susana J. and Protopapas, Pavlos and Scóccola, Claudia G. and Garraffo, Cecilia & SNe Ia, OHD & WMAP, N/S, N/S, SDSS, 2dF survey & NN & Improvement & \cite{id_06} & 2023\\
CONNECT: a neural network based framework for emulating cosmological observables and cosmological parameter inference & Nygaard, Andreas; Holm, Emil Brinch; Hannestad, Steen; Tram, Thomas & CMB & Planck 2018 & NN & Improvement & \cite{id_07} & 2023\\
Constraints on Cosmic Opacity from Bayesian Machine Learning: The hidden side of the $H_{0}$ tension problem & Elizalde, Emilio; Khurshudyan, Martiros & Generated & --- & BML & Application & \cite{id_08} & 2020\\
CosmicNet. Part I. Physics-driven implementation of neural networks within Einstein-Boltzmann Solvers & Albers, Jasper; Fidler, Christian; Lesgourgues, Julien; Schöneberg, Nils; Torrado, Jesus & CMB, Lensing, BAO & Planck 2018, N/S, N/S & NN & Improvement & \cite{id_09} & 2019\\
Cosmology-informed neural networks to solve the background dynamics of the Universe & Chantada, Augusto T.; Landau, Susana J.; Protopapas, Pavlos; Scóccola, Claudia G.; Garraffo, Cecilia & SNe Ia, OHD, BAO & Pantheon, Cosmic Chronometers, N/S & NN & Improvement & \cite{id_10} & 2022\\
Constraints on prospective deviations from the cold dark matter model using a Gaussian process & Khurshudyan, Martiros; Elizalde, Emilio & OHD & WMAP, SDSS, 2dF survey & GP & Improvement & \cite{id_11} & 2024\\
ECoPANN: A Framework for Estimating Cosmological Parameters Using Artificial Neural Networks & Wang, Guo-Jian; Li, Si-Yao; Xia, Jun-Qing & CMB, SNe Ia, BAO & Simulated, Simulated, Simulated & NN & Improvement & \cite{id_12} & 2020\\
KiDS-1000 cosmology: machine learning – accelerated constraints on interacting dark energy with CosmoPower & Spurio Mancini, A; Pourtsidou, A & CMB, Weak Lensing & Planck 2018, KiDS-1000 & NN & Improvement & \cite{id_13} & 2022\\
Late Time Attractors of Some Varying Chaplygin Gas Cosmological Models & Khurshudyan, Martiros; Myrzakulov, Ratbay & Generated & --- & BML & Application & \cite{id_14} & 2021\\
Learn-as-you-go acceleration of cosmological parameter estimates & Aslanyan, Grigor; Easther, Richard; Price, Layne C. & CMB & Planck, WMAP & BDT & Improvement & \cite{id_15} & 2015\\
Likelihood-free Cosmological Constraints with Artificial Neural Networks: An Application on Hubble Parameters and SNe Ia & Wang, Yu-Chen; Xie, Yuan-Bo; Zhang, Tong-Jie; Huang, Hui-Chao; Zhang, Tingting; Liu, Kun & SNe Ia, OHD & Pantheon, N/S & NN & Improvement & \cite{id_16} & 2021\\
LINNA: Likelihood Inference Neural Network Accelerator & To, Chun-Hao; Rozo, Eduardo; Krause, Elisabeth; Wu, Hao-Yi; Wechsler, Risa H.; Salcedo, Andrés N. & Dark Energy Survey (DES) & DES (year 1) & NN & Improvement & \cite{id_17} & 2023\\
Parameter estimation for the cosmic microwave background with Bayesian neural networks & Hortúa, Héctor J.; Volpi, Riccardo; Marinelli, Dimitri; Malagò, Luigi & CMB & Simulated & BNN & Improvement & \cite{id_18} & 2020\\
Solving the $H_{0}$ tension in f(T) gravity through Bayesian machine learning & Aljaf, Muhsin; Elizalde, Emilio; Khurshudyan, Martiros; Myrzakulov, Kairat; Zhadyranova, Aliya & Strong Lensing, OHD & H0LiCOW, Cosmic Chronometers & BNN & Application & \cite{id_19} & 2022\\
A semi-model-independent approach to describe a cosmological database & Mehrabi, Ahmad & SNe Ia, OHD, BAO & Pantheon, Cosmic Chronometers, N/S & NN & Improvement & \cite{id_20} & 2023\\
A thorough investigation of the prospects of eLISA in addressing the Hubble tension: Fisher forecast, MCMC and Machine Learning & Shah, Rahul; Bhaumik, Arko; Mukherjee, Purba; Pal, Supratik & CMB, BAO, SNe Ia & Planck 2018, 6dFGS, SDSS MGS, BOSS DR12, Pantheon & GP & Application & \cite{id_21} & 2023\\
CoLFI: Cosmological Likelihood-free Inference with Neural Density Estimators & Wang, Guo-Jian; Cheng, Cheng; Ma, Yin-Zhe; Xia, Jun-Qing; Abebe, Amare; Beesham, Aroonkumar & CMB, SNe Ia & Planck 2015, Pantheon & NN & Improvement & \cite{id_22} & 2023\\
Fast and effortless computation of profile likelihoods using CONNECT & Nygaard, Andreas; Holm, Emil Brinch; Hannestad, Steen; Tram, Thomas & CMB & Planck 2018 & NN & Improvement & \cite{id_23} & 2023\\
High-accuracy emulators for observables in $\Lambda$CDM, $N_{eff}$, $\Sigma m_{v}$, and $w$ cosmologies & Bolliet, Boris; Mancini, Alessio Spurio; Hill, J. Colin; Madhavacheril, Mathew; Jense, Hidde T.; Calabrese, Erminia; Dunkley, Jo & CMB, LSS, BAO & Planck 2018, DES (year 1), BOSS DR12 & NN & Improvement & \cite{id_24} & 2023\\
NAUTILUS: boosting Bayesian importance nested sampling with deep learning & Lange, Johannes U. & Galaxy Clustering data & Halo Connection & NN & Improvement & \cite{id_25} & 2023\\
Test of artificial neural networks in likelihood-free cosmological constraints: A comparison of information maximizing neural networks and denoising autoencoder & Chen, Jie-Feng; Wang, Yu-Chen; Zhang, Tingting; Zhang, Tong-Jie & OHD & N/S & NN & Improvement & \cite{id_26} & 2023\\
Estimating Cosmological Constraints from Galaxy Cluster Abundance using Simulation-Based Inference & Reza, Moonzarin; Zhang, Yuanyuan; Nord, Brian; Poh, Jason; Ciprijanovic, Aleksandra; Strigari, Louis & Galaxy Clustering data & Pantheon+, Cosmic Chronometers & NN & Improvement & \cite{id_27} & 2022\\
\bottomrule
\end{tabularx}
\end{adjustwidth}
\end{table}

\begin{table}[H]
\begin{adjustwidth}{-\extralength}{0cm}
\scriptsize
\centering
\begin{tblr}{
  width = \linewidth,
  colspec = {Q[467]Q[475]},
  hline{1,29} = {-}{0.08em},
  hline{2} = {-}{0.05em},
}
\textbf{Title} & \textbf{Performance}\footnotemark\\

Accelerated Bayesian inference using deep learning
  & The model accelerates MCMC convergence, achieving independent samples in just $\approx 10$ likelihood evaluations. \\

Accelerating cosmological inference with Gaussian processes and neural networks – an application to LSST Y1 weak lensing and galaxy clustering
  & The emulator achieves full MCMC–level accuracy while reducing inference time by over two orders of magnitude. \\

Accelerating MCMC algorithms through Bayesian Deep Networks
  & The Bayesian neural network accelerates MCMC inference by $\sim10{,}000\times$ at the cost of slightly increased uncertainties. \\

An analysis of the H$_0$ tension problem in the Universe with viscous dark fluid
  & The Bayesian model puts very tight constraints on the parameters ($\sigma<0.15$) and successfully resolves the cosmological H$_0$ tension. \\

An approach to cold dark matter deviation and the $H_0$ tension problem by using machine learning
  & Bayesian ML achieved $\sim0.2\ \mathrm{km\,s^{-1}\,Mpc^{-1}}$ precision on $H_0$ estimates and detected dark matter deviations with $5\sigma$ significance. \\

Faster Bayesian inference with neural network bundles and new results for $f(R)$ models
  & The neural network bundle delivers under $2\%$ error while accelerating Bayesian inference by up to $90\times$. \\

CONNECT: a neural network based framework for emulating cosmological observables and cosmological parameter inference
  & CONNECT emulates CLASS with neural networks, achieving model evaluations in milliseconds and parameter deviations below $0.1\sigma$. \\

Constraints on Cosmic Opacity from Bayesian Machine Learning: The hidden side of the $H_0$ tension problem
  & Bayesian ML analysis yields tight cosmic opacity constraints—with uncertainties as low as $\sim0.2\%$ for $H_0$—shedding light on the $H_0$ tension. \\

CosmicNet. Part I. Physics-driven implementation of neural networks within Einstein-Boltzmann Solvers
  & The neural-network–accelerated perturbation module in CLASS achieves comparable cosmological accuracy while speeding up computations by nearly $28\times$. \\

Cosmology-informed neural networks to solve the background dynamics of the Universe
  & The cosmology-informed neural network achieves sub-percent error ($1\%$) across the parameter space with high evaluation speed. \\

Constraints on prospective deviations from the cold dark matter model using a Gaussian process
  & The GP model estimates the Hubble constant at $\sim71\ \mathrm{km\,s^{-1}\,Mpc^{-1}}$ with about $5\%$ uncertainty. \\

ECoPANN: A Framework for Estimating Cosmological Parameters Using Artificial Neural Networks
  & ECoPANN delivers cosmological parameter estimates as accurately as MCMC but in seconds instead of hours. \\

KiDS-1000 cosmology: machine learning – accelerated constraints on interacting dark energy with CosmoPower
  & The CosmoPower neural emulator delivers CLASS-level accuracy while running $\sim400\times$ faster. \\

Late Time Attractors of Some Varying Chaplygin Gas Cosmological Models
  & The Bayesian machine learning model infers cosmological parameters with $\sim0.5\%$ uncertainty in $H_0$ but fails to resolve the $H_0$ tension or fit high-redshift $H(z)$ data. \\

Learn-as-you-go acceleration of cosmological parameter estimates
  & Learn-as-you-go emulation speeds up cosmological parameter estimation by about $6\times$ without sacrificing accuracy. \\

Likelihood-free Cosmological Constraints with Artificial Neural Networks: An Application on Hubble Parameters and SNe Ia
  & The likelihood‐free DAE+MAF method estimates cosmological parameters as accurately as traditional MCMC without needing an explicit likelihood. \\

LINNA: Likelihood Inference Neural Network Accelerator
  & LINNA achieves cosmological parameter inference with $0.2\sigma$ bias and $\sim60\times$ speedup over brute-force methods. \\

Parameter estimation for the cosmic microwave background with Bayesian neural networks
  & VGG with Flipout delivers the most accurate and fastest CMB parameter estimation. \\

Solving the $H_0$ tension in $f(T)$ gravity through Bayesian machine learning
  & Bayesian machine learning precisely constrained cosmological parameters and demonstrated that exponential $f(T)$ models resolve the $H_0$ tension. \\

A semi-model-independent approach to describe a cosmological database
  & Reduce el estadístico $\chi^2$ en $\approx14$ puntos para datos de $H(z)$ y en $\approx22$ puntos para el conjunto Pantheon de SNIa. \\

A thorough investigation of the prospects of eLISA in addressing the Hubble tension: Fisher forecast, MCMC and Machine Learning
  & The machine learning approach (GP) significantly outperforms Fisher and MCMC methods, reducing the Hubble tension by an additional $1.24\sigma$ down to $3.74\sigma$. \\

CoLFI: Cosmological Likelihood-free Inference with Neural Density Estimators
  & CoLFI’s Mixture Neural Network matches MCMC precision with $100\times$ fewer simulations. \\

Fast and effortless computation of profile likelihoods using CONNECT
  & CONNECT achieves sub-$0.1\%$ error in $C_\ell$ emulation and speeds up profile likelihood computations by $10^3$–$10^{12}\times$. \\

High-accuracy emulators for observables in $\Lambda$CDM, $N_{\mathrm{eff}}$, $\Sigma m_\nu$, and $w$ cosmologies
  & The emulators deliver sub-percent accuracy across cosmological observables with a $\sim1000\times$ speedup over traditional Boltzmann codes. \\

NAUTILUS: boosting Bayesian importance nested sampling with deep learning
  & Nautilus cuts likelihood evaluations by up to $100\times$ while maintaining over $99\%$ accuracy in Bayesian evidence estimates. \\

Test of artificial neural networks in likelihood-free cosmological constraints: A comparison of information maximizing neural networks and denoising autoencoder
  & MAF-DAE yields tighter cosmological constraints with minimal information loss compared to MAF-IMNN. \\

Estimating Cosmological Constraints from Galaxy Cluster Abundance using Simulation-Based Inference
  & The SBI inference accurately recovers cosmological parameters with uncertainties comparable to MCMC. \\

\end{tblr}
\end{adjustwidth}
\end{table}

\footnotetext{All studies included in this review adhered to the predefined inclusion criteria.}

\footnotetext{The “Performance” values reported in this table are drawn solely from the conclusions and results presented in each paper. Since each study employs different metrics, evaluation methodologies, and experimental conditions, these figures should not be interpreted as directly comparable; they merely reflect what each author was able to quantify and emphasize in their work.}

\restoregeometry
\isPreprints{}{
} 

\reftitle{References}


\bibliography{bibliography}

\begin{thebibliography}{999}

\bibitem[Gelman et~al.(1995)Gelman, Carlin, Stern, and
  Rubin]{gelman1995bayesian}
Gelman, A.; Carlin, J.B.; Stern, H.S.; Rubin, D.B.
\newblock {\em Bayesian data analysis}; Chapman and Hall/CRC,  1995.

\bibitem[Lewis and Bridle(2002)]{lewis2002cosmological}
Lewis, A.; Bridle, S.
\newblock {Cosmological parameters from CMB and other data: A Monte Carlo
  approach}.
\newblock {\em Physical Review D} {\bf 2002}, {\em 66},~103511.

\bibitem[Ntampaka et~al.(2015)Ntampaka, Trac, Sutherland, Battaglia,
  P{\'o}czos, and Schneider]{ntampaka2015machine}
Ntampaka, M.; Trac, H.; Sutherland, D.J.; Battaglia, N.; P{\'o}czos, B.;
  Schneider, J.
\newblock A machine learning approach for dynamical mass measurements of galaxy
  clusters.
\newblock {\em The Astrophysical Journal} {\bf 2015}, {\em 803},~50.

\bibitem[Jpt(2008)]{jpt2008cochrane}
Jpt, H.
\newblock Cochrane handbook for systematic reviews of interventions.
\newblock {\em http://www. cochrane-handbook. org} {\bf 2008}.

\bibitem[L{\'o}pez-S{\'a}nchez et~al.(2023)L{\'o}pez-S{\'a}nchez,
  Hern{\'a}ndez-Oca{\~n}a, Ch{\'a}vez-Bosquez, and
  Hern{\'a}ndez-Torruco]{lopez2023supervised}
L{\'o}pez-S{\'a}nchez, M.; Hern{\'a}ndez-Oca{\~n}a, B.; Ch{\'a}vez-Bosquez, O.;
  Hern{\'a}ndez-Torruco, J.
\newblock Supervised Deep Learning Techniques for Image Description: A
  Systematic Review.
\newblock {\em Entropy} {\bf 2023}, {\em 25},~553.

\bibitem[Bozkurt and Sharma(2020)]{bozkurt2020emergency}
Bozkurt, A.; Sharma, R.C.
\newblock Emergency remote teaching in a time of global crisis due to
  CoronaVirus pandemic.
\newblock {\em Asian journal of distance education} {\bf 2020}, {\em
  15},~i--vi.

\bibitem[Lochner et~al.(2016)Lochner, McEwen, Peiris, Lahav, and
  Winter]{lochner2016photometric}
Lochner, M.; McEwen, J.D.; Peiris, H.V.; Lahav, O.; Winter, M.K.
\newblock Photometric supernova classification with machine learning.
\newblock {\em The Astrophysical Journal Supplement Series} {\bf 2016}, {\em
  225},~31.

\bibitem[Dieleman et~al.(2015)Dieleman, Willett, and
  Dambre]{dieleman2015rotation}
Dieleman, S.; Willett, K.W.; Dambre, J.
\newblock Rotation-invariant convolutional neural networks for galaxy
  morphology prediction.
\newblock {\em Monthly notices of the royal astronomical society} {\bf 2015},
  {\em 450},~1441--1459.

\bibitem[Mukhanov(2005)]{Mukhanov:2005sc}
Mukhanov, V.
\newblock {\em {Physical Foundations of Cosmology}}; Cambridge University
  Press: Oxford,  2005.
\newblock {\url{https://doi.org/10.1017/CBO9780511790553}}.

\bibitem[Bertone and Hooper(2018)]{Bertone:2016nfn}
Bertone, G.; Hooper, D.
\newblock {History of dark matter}.
\newblock {\em Rev. Mod. Phys.} {\bf 2018}, {\em 90},~045002,
  \href{http://arxiv.org/abs/1605.04909}{{\normalfont
  [arXiv:astro-ph.CO/1605.04909]}}.
\newblock {\url{https://doi.org/10.1103/RevModPhys.90.045002}}.

\bibitem[Salucci(2018)]{Salucci:2018eie}
Salucci, P.
\newblock {Dark Matter in Galaxies: evidences and challenges}.
\newblock {\em Found. Phys.} {\bf 2018}, {\em 48},~1517--1537,
  \href{http://arxiv.org/abs/1807.08541}{{\normalfont
  [arXiv:astro-ph.GA/1807.08541]}}.
\newblock {\url{https://doi.org/10.1007/s10701-018-0209-5}}.

\bibitem[Riess et~al.(1998)]{SupernovaSearchTeam:1998fmf}
Riess, A.G.;  et~al.
\newblock {Observational evidence from supernovae for an accelerating universe
  and a cosmological constant}.
\newblock {\em Astron. J.} {\bf 1998}, {\em 116},~1009--1038,
  \href{http://arxiv.org/abs/astro-ph/9805201}{{\normalfont
  [astro-ph/9805201]}}.
\newblock {\url{https://doi.org/10.1086/300499}}.

\bibitem[Perlmutter et~al.(1999)]{SupernovaCosmologyProject:1998vns}
Perlmutter, S.;  et~al.
\newblock {Measurements of $\Omega$ and $\Lambda$ from 42 High Redshift
  Supernovae}.
\newblock {\em Astrophys. J.} {\bf 1999}, {\em 517},~565--586,
  \href{http://arxiv.org/abs/astro-ph/9812133}{{\normalfont
  [astro-ph/9812133]}}.
\newblock {\url{https://doi.org/10.1086/307221}}.

\bibitem[Fischer(2018)]{Fischer:2018zkr}
Fischer, A.E.
\newblock {Friedmann\textquoteright{}s equation and the creation of the
  universe}.
\newblock {\em Int. J. Mod. Phys. D} {\bf 2018}, {\em 27},~1847013.
\newblock {\url{https://doi.org/10.1142/S0218271818470132}}.

\bibitem[Workman et~al.(2022)]{ParticleDataGroup:2022pth}
Workman, R.L.;  et~al.
\newblock {Review of Particle Physics}.
\newblock {\em PTEP} {\bf 2022}, {\em 2022},~083C01.
\newblock {\url{https://doi.org/10.1093/ptep/ptac097}}.

\bibitem[Velten et~al.(2014)Velten, vom Marttens, and Zimdahl]{Velten:2014nra}
Velten, H.E.S.; vom Marttens, R.F.; Zimdahl, W.
\newblock {Aspects of the cosmological \textquotedblleft{}coincidence
  problem\textquotedblright{}}.
\newblock {\em Eur. Phys. J. C} {\bf 2014}, {\em 74},~3160,
  \href{http://arxiv.org/abs/1410.2509}{{\normalfont
  [arXiv:astro-ph.CO/1410.2509]}}.
\newblock {\url{https://doi.org/10.1140/epjc/s10052-014-3160-4}}.

\bibitem[Aghanim et~al.(2020)]{Planck:2018vyg}
Aghanim, N.;  et~al.
\newblock {Planck 2018 results. VI. Cosmological parameters}.
\newblock {\em Astron. Astrophys.} {\bf 2020}, {\em 641},~A6,
  \href{http://arxiv.org/abs/1807.06209}{{\normalfont
  [arXiv:astro-ph.CO/1807.06209]}}.
\newblock [Erratum: Astron.Astrophys. 652, C4 (2021)],
  {\url{https://doi.org/10.1051/0004-6361/201833910}}.

\bibitem[Liu et~al.(2023)Liu, Roepke, and Han]{Liu:2023qmw}
Liu, Z.W.; Roepke, F.K.; Han, Z.
\newblock {Type Ia Supernova Explosions in Binary Systems: A Review}.
\newblock {\em Res. Astron. Astrophys.} {\bf 2023}, {\em 23},~082001,
  \href{http://arxiv.org/abs/2305.13305}{{\normalfont
  [arXiv:astro-ph.HE/2305.13305]}}.
\newblock {\url{https://doi.org/10.1088/1674-4527/acd89e}}.

\bibitem[Moresco et~al.(2016)Moresco, Pozzetti, Cimatti, Jimenez, Maraston,
  Verde, Thomas, Citro, Tojeiro, and Wilkinson]{Moresco:2016mzx}
Moresco, M.; Pozzetti, L.; Cimatti, A.; Jimenez, R.; Maraston, C.; Verde, L.;
  Thomas, D.; Citro, A.; Tojeiro, R.; Wilkinson, D.
\newblock {A 6\% measurement of the Hubble parameter at $z\sim0.45$: direct
  evidence of the epoch of cosmic re-acceleration}.
\newblock {\em JCAP} {\bf 2016}, {\em 05},~014,
  \href{http://arxiv.org/abs/1601.01701}{{\normalfont
  [arXiv:astro-ph.CO/1601.01701]}}.
\newblock {\url{https://doi.org/10.1088/1475-7516/2016/05/014}}.

\bibitem[Scolnic et~al.(2018)]{Pan-STARRS1:2017jku}
Scolnic, D.M.;  et~al.
\newblock {The Complete Light-curve Sample of Spectroscopically Confirmed SNe
  Ia from Pan-STARRS1 and Cosmological Constraints from the Combined Pantheon
  Sample}.
\newblock {\em Astrophys. J.} {\bf 2018}, {\em 859},~101,
  \href{http://arxiv.org/abs/1710.00845}{{\normalfont
  [arXiv:astro-ph.CO/1710.00845]}}.
\newblock {\url{https://doi.org/10.3847/1538-4357/aab9bb}}.

\bibitem[Brout et~al.(2022)]{Brout:2022vxf}
Brout, D.;  et~al.
\newblock {The Pantheon+ Analysis: Cosmological Constraints}.
\newblock {\em Astrophys. J.} {\bf 2022}, {\em 938},~110,
  \href{http://arxiv.org/abs/2202.04077}{{\normalfont
  [arXiv:astro-ph.CO/2202.04077]}}.
\newblock {\url{https://doi.org/10.3847/1538-4357/ac8e04}}.

\bibitem[Magana et~al.(2018)Magana, Amante, Garcia-Aspeitia, and
  Motta]{Magana:2017nfs}
Magana, J.; Amante, M.H.; Garcia-Aspeitia, M.A.; Motta, V.
\newblock {The Cardassian expansion revisited: constraints from updated Hubble
  parameter measurements and type Ia supernova data}.
\newblock {\em Mon. Not. Roy. Astron. Soc.} {\bf 2018}, {\em 476},~1036--1049,
  \href{http://arxiv.org/abs/1706.09848}{{\normalfont
  [arXiv:astro-ph.CO/1706.09848]}}.
\newblock {\url{https://doi.org/10.1093/mnras/sty260}}.

\bibitem[Jimenez and Loeb(2002)]{Jimenez:2001gg}
Jimenez, R.; Loeb, A.
\newblock {Constraining cosmological parameters based on relative galaxy ages}.
\newblock {\em Astrophys. J.} {\bf 2002}, {\em 573},~37--42,
  \href{http://arxiv.org/abs/astro-ph/0106145}{{\normalfont
  [astro-ph/0106145]}}.
\newblock {\url{https://doi.org/10.1086/340549}}.

\bibitem[Abdul~Karim et~al.(2025)]{DESI:2025zgx}
Abdul~Karim, M.;  et~al.
\newblock {DESI DR2 Results II: Measurements of Baryon Acoustic Oscillations
  and Cosmological Constraints},  2025,
  \href{http://arxiv.org/abs/2503.14738}{{\normalfont
  [arXiv:astro-ph.CO/2503.14738]}}.

\bibitem[Peebles and Yu(1970)]{Peebles:1970ag}
Peebles, P.J.E.; Yu, J.T.
\newblock {Primeval adiabatic perturbation in an expanding universe}.
\newblock {\em Astrophys. J.} {\bf 1970}, {\em 162},~815--836.
\newblock {\url{https://doi.org/10.1086/150713}}.

\bibitem[Eisenstein and Hu(1998)]{Eisenstein:1997ik}
Eisenstein, D.J.; Hu, W.
\newblock {Baryonic features in the matter transfer function}.
\newblock {\em Astrophys. J.} {\bf 1998}, {\em 496},~605,
  \href{http://arxiv.org/abs/astro-ph/9709112}{{\normalfont
  [astro-ph/9709112]}}.
\newblock {\url{https://doi.org/10.1086/305424}}.

\bibitem[Beutler et~al.(2011)]{Beutler_2011}
Beutler, F.;  et~al.
\newblock The 6dF Galaxy Survey: baryon acoustic oscillations and the local
  Hubble constant: 6dFGS: BAOs and the local Hubble constant.
\newblock {\em Mon. Not. Roy. Astron. Soc.} {\bf 2011}, {\em 416},~3017–3032.
\newblock {\url{https://doi.org/10.1111/j.1365-2966.2011.19250.x}}.

\bibitem[Ross et~al.(2015)]{Ross:2014qpa}
Ross, A.J.;  et~al.
\newblock {The clustering of the SDSS DR7 main Galaxy sample \textendash{} I. A
  4 per cent distance measure at $z = 0.15$}.
\newblock {\em Mon. Not. Roy. Astron. Soc.} {\bf 2015}, {\em 449},~835--847,
  \href{http://arxiv.org/abs/1409.3242}{{\normalfont
  [arXiv:astro-ph.CO/1409.3242]}}.
\newblock {\url{https://doi.org/10.1093/mnras/stv154}}.

\bibitem[Alam et~al.(2017)]{BOSS:2016wmc}
Alam, S.;  et~al.
\newblock {The clustering of galaxies in the completed SDSS-III Baryon
  Oscillation Spectroscopic Survey: cosmological analysis of the DR12 galaxy
  sample}.
\newblock {\em Mon. Not. Roy. Astron. Soc.} {\bf 2017}, {\em 470},~2617--2652,
  \href{http://arxiv.org/abs/1607.03155}{{\normalfont
  [arXiv:astro-ph.CO/1607.03155]}}.
\newblock {\url{https://doi.org/10.1093/mnras/stx721}}.

\bibitem[Aghamousa et~al.(2016)]{DESI:2016fyo}
Aghamousa, A.;  et~al.
\newblock {The DESI Experiment Part I: Science,Targeting, and Survey Design},
  2016,  \href{http://arxiv.org/abs/1611.00036}{{\normalfont
  [arXiv:astro-ph.IM/1611.00036]}}.

\bibitem[{Penzias} and {Wilson}(1965)]{1965ApJ...142..419P}
{Penzias}, A.A.; {Wilson}, R.W.
\newblock {A Measurement of Excess Antenna Temperature at 4080 Mc/s.}
\newblock {\em Astrophys. J.} {\bf 1965}, {\em 142},~419--421.
\newblock {\url{https://doi.org/10.1086/148307}}.

\bibitem[Tegmark et~al.(2004)]{SDSS:2003eyi}
Tegmark, M.;  et~al.
\newblock {Cosmological parameters from SDSS and WMAP}.
\newblock {\em Phys. Rev. D} {\bf 2004}, {\em 69},~103501,
  \href{http://arxiv.org/abs/astro-ph/0310723}{{\normalfont
  [astro-ph/0310723]}}.
\newblock {\url{https://doi.org/10.1103/PhysRevD.69.103501}}.

\bibitem[Wang and Mukherjee(2006)]{Wang:2006ts}
Wang, Y.; Mukherjee, P.
\newblock {Robust dark energy constraints from supernovae, galaxy clustering,
  and three-year wilkinson microwave anisotropy probe observations}.
\newblock {\em Astrophys. J.} {\bf 2006}, {\em 650},~1--6,
  \href{http://arxiv.org/abs/astro-ph/0604051}{{\normalfont
  [astro-ph/0604051]}}.
\newblock {\url{https://doi.org/10.1086/507091}}.

\bibitem[{Hinshaw} et~al.(2013)]{2013ApJS..208...19H}
{Hinshaw}, G.;  et~al.
\newblock {Nine-year Wilkinson Microwave Anisotropy Probe (WMAP) Observations:
  Cosmological Parameter Results}.
\newblock {\em Astrophys. J. Suppl. Series} {\bf 2013}, {\em 208},~19,
  \href{http://arxiv.org/abs/1212.5226}{{\normalfont
  [arXiv:astro-ph.CO/1212.5226]}}.
\newblock {\url{https://doi.org/10.1088/0067-0049/208/2/19}}.

\bibitem[Aghanim et~al.(2020)]{Planck:2018nkj}
Aghanim, N.;  et~al.
\newblock {Planck 2018 results. I. Overview and the cosmological legacy of
  Planck}.
\newblock {\em Astron. Astrophys.} {\bf 2020}, {\em 641},~A1,
  \href{http://arxiv.org/abs/1807.06205}{{\normalfont
  [arXiv:astro-ph.CO/1807.06205]}}.
\newblock {\url{https://doi.org/10.1051/0004-6361/201833880}}.

\bibitem[Springel et~al.(2006)Springel, Frenk, and White]{Springel:2006vs}
Springel, V.; Frenk, C.S.; White, S.D.M.
\newblock {The large-scale structure of the Universe}.
\newblock {\em Nature} {\bf 2006}, {\em 440},~1137,
  \href{http://arxiv.org/abs/astro-ph/0604561}{{\normalfont
  [astro-ph/0604561]}}.
\newblock {\url{https://doi.org/10.1038/nature04805}}.

\bibitem[Colless et~al.(2001)]{2DFGRS:2001zay}
Colless, M.;  et~al.
\newblock {The 2dF Galaxy Redshift Survey: Spectra and redshifts}.
\newblock {\em Mon. Not. Roy. Astron. Soc.} {\bf 2001}, {\em 328},~1039,
  \href{http://arxiv.org/abs/astro-ph/0106498}{{\normalfont
  [astro-ph/0106498]}}.
\newblock {\url{https://doi.org/10.1046/j.1365-8711.2001.04902.x}}.

\bibitem[York et~al.(2000)]{SDSS:2000hjo}
York, D.G.;  et~al.
\newblock {The Sloan Digital Sky Survey: Technical Summary}.
\newblock {\em Astron. J.} {\bf 2000}, {\em 120},~1579--1587,
  \href{http://arxiv.org/abs/astro-ph/0006396}{{\normalfont
  [astro-ph/0006396]}}.
\newblock {\url{https://doi.org/10.1086/301513}}.

\bibitem[Wong et~al.(2020)]{Wong:2019kwg}
Wong, K.C.;  et~al.
\newblock {H0LiCOW \textendash{} XIII. A 2.4 per cent measurement of H0 from
  lensed quasars: 5.3\ensuremath{\sigma} tension between early- and
  late-Universe probes}.
\newblock {\em Mon. Not. Roy. Astron. Soc.} {\bf 2020}, {\em 498},~1420--1439,
  \href{http://arxiv.org/abs/1907.04869}{{\normalfont
  [arXiv:astro-ph.CO/1907.04869]}}.
\newblock {\url{https://doi.org/10.1093/mnras/stz3094}}.

\bibitem[Turner(2022)]{Turner:2022gvw}
Turner, M.S.
\newblock {The Road to Precision Cosmology},  2022,
  \href{http://arxiv.org/abs/2201.04741}{{\normalfont
  [arXiv:astro-ph.CO/2201.04741]}}.
\newblock {\url{https://doi.org/10.1146/annurev-nucl-111119-041046}}.

\bibitem[Abdalla et~al.(2022)]{Abdalla:2022yfr}
Abdalla, E.;  et~al.
\newblock {Cosmology intertwined: A review of the particle physics,
  astrophysics, and cosmology associated with the cosmological tensions and
  anomalies}.
\newblock {\em JHEAp} {\bf 2022}, {\em 34},~49--211,
  \href{http://arxiv.org/abs/2203.06142}{{\normalfont
  [arXiv:astro-ph.CO/2203.06142]}}.
\newblock {\url{https://doi.org/10.1016/j.jheap.2022.04.002}}.

\bibitem[Riess et~al.(2022)]{Riess:2021jrx}
Riess, A.G.;  et~al.
\newblock {A Comprehensive Measurement of the Local Value of the Hubble
  Constant with 1 km s$^{−1}$ Mpc$^{−1}$ Uncertainty from the Hubble Space
  Telescope and the SH0ES Team}.
\newblock {\em Astrophys. J. Lett.} {\bf 2022}, {\em 934},~L7,
  \href{http://arxiv.org/abs/2112.04510}{{\normalfont
  [arXiv:astro-ph.CO/2112.04510]}}.
\newblock {\url{https://doi.org/10.3847/2041-8213/ac5c5b}}.

\bibitem[Hogg and Foreman-Mackey(2018)]{Hogg:2017akh}
Hogg, D.W.; Foreman-Mackey, D.
\newblock {Data analysis recipes: Using Markov Chain Monte Carlo}.
\newblock {\em Astrophys. J. Suppl.} {\bf 2018}, {\em 236},~11,
  \href{http://arxiv.org/abs/1710.06068}{{\normalfont
  [arXiv:astro-ph.IM/1710.06068]}}.
\newblock {\url{https://doi.org/10.3847/1538-4365/aab76e}}.

\bibitem[Goodman and Weare(2010)]{Goodman_Ensemble_2010}
Goodman, J.; Weare, J.
\newblock Ensemble samplers with affine invariance.
\newblock {\em Communications in applied mathematics and computational science}
  {\bf 2010}, {\em 5},~65--80.

\bibitem[Foreman-Mackey et~al.(2013)Foreman-Mackey, Hogg, Lang, and
  Goodman]{Foreman-Mackey:2012any}
Foreman-Mackey, D.; Hogg, D.W.; Lang, D.; Goodman, J.
\newblock {emcee: The MCMC Hammer}.
\newblock {\em Publ. Astron. Soc. Pac.} {\bf 2013}, {\em 125},~306--312,
  \href{http://arxiv.org/abs/1202.3665}{{\normalfont
  [arXiv:astro-ph.IM/1202.3665]}}.
\newblock {\url{https://doi.org/10.1086/670067}}.

\bibitem[Hajian(2007)]{Hajian:2006mt}
Hajian, A.
\newblock {Efficient Cosmological Parameter Estimation with Hamiltonian Monte
  Carlo}.
\newblock {\em Phys. Rev. D} {\bf 2007}, {\em 75},~083525,
  \href{http://arxiv.org/abs/astro-ph/0608679}{{\normalfont
  [astro-ph/0608679]}}.
\newblock {\url{https://doi.org/10.1103/PhysRevD.75.083525}}.

\bibitem[Ivezi\'c et~al.(2019)]{LSST:2008ijt}
Ivezi\'c, v.;  et~al.
\newblock {LSST: from Science Drivers to Reference Design and Anticipated Data
  Products}.
\newblock {\em Astrophys. J.} {\bf 2019}, {\em 873},~111,
  \href{http://arxiv.org/abs/0805.2366}{{\normalfont
  [arXiv:astro-ph/0805.2366]}}.
\newblock {\url{https://doi.org/10.3847/1538-4357/ab042c}}.

\bibitem[Laureijs et~al.(2011)]{laureijs2011euclid}
Laureijs, R.;  et~al.
\newblock Euclid Definition Study Report,  2011,
  \href{http://arxiv.org/abs/1110.3193}{{\normalfont
  [arXiv:astro-ph.CO/1110.3193]}}.

\bibitem[Dor\'e et~al.(2014)]{SPHEREx:2014bgr}
Dor\'e, O.;  et~al.
\newblock {Cosmology with the SPHEREX All-Sky Spectral Survey},  2014,
  \href{http://arxiv.org/abs/1412.4872}{{\normalfont
  [arXiv:astro-ph.CO/1412.4872]}}.

\bibitem[Spergel et~al.(2015)]{Spergel:2015sza}
Spergel, D.;  et~al.
\newblock {Wide-Field InfrarRed Survey Telescope-Astrophysics Focused Telescope
  Assets WFIRST-AFTA 2015 Report},  2015,
  \href{http://arxiv.org/abs/1503.03757}{{\normalfont
  [arXiv:astro-ph.IM/1503.03757]}}.

\bibitem[Takada et~al.(2014)]{Takada_2014}
Takada, M.;  et~al.
\newblock Extragalactic science, cosmology, and Galactic archaeology with the
  Subaru Prime Focus Spectrograph.
\newblock {\em Publications of the Astronomical Society of Japan} {\bf 2014},
  {\em 66}.
\newblock {\url{https://doi.org/10.1093/pasj/pst019}}.

\bibitem[Janiesch et~al.(2021)Janiesch, Zschech, and
  Heinrich]{janiesch_machine_2021}
Janiesch, C.; Zschech, P.; Heinrich, K.
\newblock Machine learning and deep learning.
\newblock {\em Electronic Markets} {\bf 2021}, {\em 31},~685--695.
\newblock {\url{https://doi.org/10.1007/s12525-021-00475-2}}.

\bibitem[Rosenblatt(1958)]{rosenblatt_perceptron_1958}
Rosenblatt, F.
\newblock The perceptron: a probabilistic model for information storage and
  organization in the brain.
\newblock {\em Psychological Review} {\bf 1958}, {\em 65},~386.

\bibitem[Minsky and Papert(1969)]{minsky_introduction_1969}
Minsky, M.; Papert, S.
\newblock An introduction to computational geometry.
\newblock {\em Cambridge tiass., HIT} {\bf 1969}, {\em 479},~480.

\bibitem[Chollet(2021)]{chollet2021deep}
Chollet, F.
\newblock {\em Deep learning with Python}; Simon and Schuster,  2021.

\bibitem[Aggarwal(2018)]{aggarwal_introduction_2018}
Aggarwal, C.C.
\newblock {\em An {Introduction} to {Neural} {Networks}}; Springer
  International Publishing,  2018.
\newblock Publication Title: Neural Networks and Deep Learning: A Textbook,
  {\url{https://doi.org/10.1007/978-3-319-94463-0_1}}.

\bibitem[Bharadiya(2023)]{bharadiya_review_2023}
Bharadiya, J.P.
\newblock A {Review} of {Bayesian} {Machine} {Learning} {Principles},
  {Methods}, and {Applications},  2023.

\bibitem[Seeger(2004)]{seeger_gaussian_2004}
Seeger, M.
\newblock {GAUSSIAN} {PROCESSES} {FOR} {MACHINE} {LEARNING}.
\newblock {\em International Journal of Neural Systems} {\bf 2004}, {\em
  14},~69--106.
\newblock {\url{https://doi.org/10.1142/S0129065704001899}}.

\bibitem[Nuti et~al.(2019)Nuti, Rugama, and Cross]{nuti_bayesian_2019}
Nuti, G.; Rugama, L.A.J.; Cross, A.I.
\newblock A {Bayesian} {Decision} {Tree} {Algorithm},  2019.
\newblock Publisher: [object Object] Version Number: 3,
  {\url{https://doi.org/10.48550/ARXIV.1901.03214}}.

\bibitem[Dension(1998)]{dension_bayesian_1998}
Dension, D.
\newblock A {Bayesian} {CART} algorithm.
\newblock {\em Biometrika} {\bf 1998}, {\em 85},~363--377.
\newblock {\url{https://doi.org/10.1093/biomet/85.2.363}}.

\bibitem[Blundell et~al.(2015)Blundell, Cornebise, Kavukcuoglu, and
  Wierstra]{blundell_weight_2015}
Blundell, C.; Cornebise, J.; Kavukcuoglu, K.; Wierstra, D.
\newblock Weight {Uncertainty} in {Neural} {Networks},  2015.
\newblock arXiv:1505.05424 [cs, stat].

\bibitem[de~Dios Rojas~Olvera et~al.(2022)de~Dios Rojas~Olvera,
  G{\'o}mez-Vargas, and V{\'a}zquez]{de2022observational}
de~Dios Rojas~Olvera, J.; G{\'o}mez-Vargas, I.; V{\'a}zquez, J.A.
\newblock Observational cosmology with artificial neural networks.
\newblock {\em Universe} {\bf 2022}, {\em 8},~120.

\bibitem[Moriwaki et~al.(2023)Moriwaki, Nishimichi, and
  Yoshida]{moriwaki2023machine}
Moriwaki, K.; Nishimichi, T.; Yoshida, N.
\newblock Machine learning for observational cosmology.
\newblock {\em Reports on Progress in Physics} {\bf 2023}, {\em 86},~076901.

\bibitem[Lahav(2023)]{lahav2023deep}
Lahav, O.
\newblock Deep Machine Learning in Cosmology: Evolution or Revolution?
\newblock {\em arXiv preprint arXiv:2302.04324} {\bf 2023}.

\bibitem[Dvorkin et~al.(2022)Dvorkin, Mishra-Sharma, Nord, Villar, Avestruz,
  Bechtol, {\'C}iprijanovi{\'c}, Connolly, Garrison, Narayan,
  et~al.]{dvorkin2022machine}
Dvorkin, C.; Mishra-Sharma, S.; Nord, B.; Villar, V.A.; Avestruz, C.; Bechtol,
  K.; {\'C}iprijanovi{\'c}, A.; Connolly, A.J.; Garrison, L.H.; Narayan, G.;
  et~al.
\newblock Machine learning and cosmology.
\newblock {\em arXiv preprint arXiv:2203.08056} {\bf 2022}.

\bibitem[Han et~al.(2015)Han, Ding, Zhang, and Zhao]{han2015improving}
Han, B.; Ding, H.; Zhang, Y.; Zhao, Y.
\newblock Improving accuracy of Quasars' photometric redshift estimation by
  integration of KNN and SVM.
\newblock {\em Proceedings of the International Astronomical Union} {\bf 2015},
  {\em 11},~209--209.

\bibitem[Di~Valentino et~al.(2025)Di~Valentino, Levi~Said, Riess, Pollo,
  Poulin, and {CosmoVerse Network}]{CosmoVerse2025}
Di~Valentino, E.; Levi~Said, J.; Riess, A.G.; Pollo, A.; Poulin, V.;
  {CosmoVerse Network}.
\newblock The CosmoVerse White Paper: Addressing observational tensions in
  cosmology with systematics and fundamental physics.
\newblock {\em Physics of the Dark Universe} {\bf 2025}, {\em 49},~101965.
\newblock {\url{https://doi.org/10.1016/j.dark.2025.101965}}.

\bibitem[Spurio~Mancini et~al.(2022)Spurio~Mancini, Piras, Alsing, Joachimi,
  and Hobson]{SpurioMancini2022CosmoPower}
Spurio~Mancini, A.; Piras, D.; Alsing, J.; Joachimi, B.; Hobson, M.P.
\newblock CosmoPower: emulating cosmological power spectra for accelerated
  Bayesian inference from next-generation surveys.
\newblock {\em Monthly Notices of the Royal Astronomical Society} {\bf 2022},
  {\em 511},~1771--1788.
\newblock {\url{https://doi.org/10.1093/mnras/stac064}}.

\bibitem[Mootoovaloo et~al.(2025)Mootoovaloo, Garc{\'i}a-Garc{\'i}a, Alonso,
  and Ruiz-Zapatero]{Mootoovaloo2025Emuflow}
Mootoovaloo, A.; Garc{\'i}a-Garc{\'i}a, C.; Alonso, D.; Ruiz-Zapatero, J.
\newblock emuflow: normalizing flows for joint cosmological analysis.
\newblock {\em Monthly Notices of the Royal Astronomical Society} {\bf 2025},
  {\em 536},~190--202.
\newblock {\url{https://doi.org/10.1093/mnras/stae2604}}.

\bibitem[Kitchenham and Charters(2007)]{kitchenham2007guidelines}
Kitchenham, B.; Charters, S.
\newblock Guidelines for Performing Systematic Literature Reviews in Software
  Engineering.
\newblock Technical report, EBSE Technical Report, Version 2.3, EBSE-2007-01.
  Keele University and Durham University, UK,  2007.

\bibitem[Kitchenham(2004)]{kitchenham2004procedures}
Kitchenham, B.
\newblock Procedures for performing systematic reviews.
\newblock {\em Keele, UK, Keele University} {\bf 2004}, {\em 33},~1--26.

\bibitem[Rojas et~al.(2025)Rojas, Espinoza, González, Maldonado, and
  Luo]{SLRProtocol2025}
Rojas, L.; Espinoza, S.; González, E.; Maldonado, C.; Luo, F.
\newblock Protocol for the Systematic Literature Review (PRISMA 2020): ML for
  Observational Constraints in Cosmology,  2025.
\newblock Accessed: 10 August 2025,
  {\url{https://doi.org/10.5281/zenodo.16899506}}.

\bibitem[Lange(2023)]{id_25}
Lange, J.U.
\newblock {nautilus: boosting Bayesian importance nested sampling with deep
  learning}.
\newblock {\em Mon. Not. Roy. Astron. Soc.} {\bf 2023}, {\em 525},~3181--3194,
  \href{http://arxiv.org/abs/2306.16923}{{\normalfont
  [arXiv:astro-ph.IM/2306.16923]}}.
\newblock {\url{https://doi.org/10.1093/mnras/stad2441}}.

\bibitem[Reza et~al.(2022)Reza, Zhang, Nord, Poh, Ciprijanovic, and
  Strigari]{id_27}
Reza, M.; Zhang, Y.; Nord, B.; Poh, J.; Ciprijanovic, A.; Strigari, L.
\newblock {Estimating Cosmological Constraints from Galaxy Cluster Abundance
  using Simulation-Based Inference}.
\newblock In Proceedings of the {39th International Conference on Machine
  Learning Conference},  7 2022,
  \href{http://arxiv.org/abs/2208.00134}{{\normalfont
  [arXiv:astro-ph.CO/2208.00134]}}.

\bibitem[Aljaf et~al.(2022)Aljaf, Elizalde, Khurshudyan, Myrzakulov, and
  Zhadyranova]{id_19}
Aljaf, M.; Elizalde, E.; Khurshudyan, M.; Myrzakulov, K.; Zhadyranova, A.
\newblock {Solving the $H_{0}$ tension in f(T) gravity through Bayesian machine
  learning}.
\newblock {\em Eur. Phys. J. C} {\bf 2022}, {\em 82},~1130,
  \href{http://arxiv.org/abs/2205.06252}{{\normalfont
  [arXiv:astro-ph.CO/2205.06252]}}.
\newblock {\url{https://doi.org/10.1140/epjc/s10052-022-11109-y}}.

\bibitem[Mancini~Spurio and Pourtsidou(2022)]{id_13}
Mancini~Spurio, A.; Pourtsidou, A.
\newblock {KiDS-1000 cosmology: machine learning \textendash{} accelerated
  constraints on interacting dark energy with CosmoPower}.
\newblock {\em Mon. Not. Roy. Astron. Soc.} {\bf 2022}, {\em 512},~L44--L48,
  \href{http://arxiv.org/abs/2110.07587}{{\normalfont
  [arXiv:astro-ph.CO/2110.07587]}}.
\newblock {\url{https://doi.org/10.1093/mnrasl/slac019}}.

\bibitem[Boruah et~al.(2022)Boruah, Eifler, Miranda, and M]{id_02}
Boruah, S.S.; Eifler, T.; Miranda, V.; M, S.K.P.
\newblock {Accelerating cosmological inference with Gaussian processes and
  neural networks \textendash{} an application to LSST Y1 weak lensing and
  galaxy clustering}.
\newblock {\em Mon. Not. Roy. Astron. Soc.} {\bf 2022}, {\em 518},~4818--4831,
  \href{http://arxiv.org/abs/2203.06124}{{\normalfont
  [arXiv:astro-ph.CO/2203.06124]}}.
\newblock {\url{https://doi.org/10.1093/mnras/stac3417}}.

\bibitem[Albers et~al.(2019)Albers, Fidler, Lesgourgues, Sch\"oneberg, and
  Torrado]{id_09}
Albers, J.; Fidler, C.; Lesgourgues, J.; Sch\"oneberg, N.; Torrado, J.
\newblock {CosmicNet. Part I. Physics-driven implementation of neural networks
  within Einstein-Boltzmann Solvers}.
\newblock {\em JCAP} {\bf 2019}, {\em 09},~028,
  \href{http://arxiv.org/abs/1907.05764}{{\normalfont
  [arXiv:astro-ph.CO/1907.05764]}}.
\newblock {\url{https://doi.org/10.1088/1475-7516/2019/09/028}}.

\bibitem[Chantada et~al.(2023)Chantada, Landau, Protopapas, Sc\'occola, and
  Garraffo]{id_10}
Chantada, A.T.; Landau, S.J.; Protopapas, P.; Sc\'occola, C.G.; Garraffo, C.
\newblock {Cosmology-informed neural networks to solve the background dynamics
  of the Universe}.
\newblock {\em Phys. Rev. D} {\bf 2023}, {\em 107},~063523,
  \href{http://arxiv.org/abs/2205.02945}{{\normalfont
  [arXiv:astro-ph.CO/2205.02945]}}.
\newblock {\url{https://doi.org/10.1103/PhysRevD.107.063523}}.

\bibitem[Wang et~al.(2021)Wang, Xie, Zhang, Huang, Zhang, and Liu]{id_16}
Wang, Y.C.; Xie, Y.B.; Zhang, T.J.; Huang, H.C.; Zhang, T.; Liu, K.
\newblock {Likelihood-free Cosmological Constraints with Artificial Neural
  Networks: An Application on Hubble Parameters and SNe Ia}.
\newblock {\em Astrophys. J. Supp.} {\bf 2021}, {\em 254},~43,
  \href{http://arxiv.org/abs/2005.10628}{{\normalfont
  [arXiv:astro-ph.CO/2005.10628]}}.
\newblock {\url{https://doi.org/10.3847/1538-4365/abf8aa}}.

\bibitem[Mehrabi(2023)]{id_20}
Mehrabi, A.
\newblock {A semi-model-independent approach to describe a cosmological
  database},  2023,  \href{http://arxiv.org/abs/2301.07369}{{\normalfont
  [arXiv:astro-ph.CO/2301.07369]}}.

\bibitem[Shah et~al.(2023)Shah, Bhaumik, Mukherjee, and Pal]{id_21}
Shah, R.; Bhaumik, A.; Mukherjee, P.; Pal, S.
\newblock {A thorough investigation of the prospects of eLISA in addressing the
  Hubble tension: Fisher forecast, MCMC and Machine Learning}.
\newblock {\em JCAP} {\bf 2023}, {\em 06},~038,
  \href{http://arxiv.org/abs/2301.12708}{{\normalfont
  [arXiv:astro-ph.CO/2301.12708]}}.
\newblock {\url{https://doi.org/10.1088/1475-7516/2023/06/038}}.

\bibitem[Wang et~al.(2023)Wang, Cheng, Ma, Xia, Abebe, and Beesham]{id_22}
Wang, G.J.; Cheng, C.; Ma, Y.Z.; Xia, J.Q.; Abebe, A.; Beesham, A.
\newblock {CoLFI: Cosmological Likelihood-free Inference with Neural Density
  Estimators}.
\newblock {\em Astrophys. J. Suppl.} {\bf 2023}, {\em 268},~7,
  \href{http://arxiv.org/abs/2306.11102}{{\normalfont
  [arXiv:astro-ph.CO/2306.11102]}}.
\newblock {\url{https://doi.org/10.3847/1538-4365/ace113}}.

\bibitem[Chantada et~al.(2024)Chantada, Landau, Protopapas, Sc\'occola, and
  Garraffo]{id_06}
Chantada, A.T.; Landau, S.J.; Protopapas, P.; Sc\'occola, C.G.; Garraffo, C.
\newblock {Faster Bayesian inference with neural network bundles and new
  results for f(R) models}.
\newblock {\em Phys. Rev. D} {\bf 2024}, {\em 109},~123514,
  \href{http://arxiv.org/abs/2311.15955}{{\normalfont
  [arXiv:astro-ph.CO/2311.15955]}}.
\newblock {\url{https://doi.org/10.1103/PhysRevD.109.123514}}.

\bibitem[Wang et~al.(2020)Wang, Li, and Xia]{id_12}
Wang, G.J.; Li, S.Y.; Xia, J.Q.
\newblock {ECoPANN: A Framework for Estimating Cosmological Parameters using
  Artificial Neural Networks}.
\newblock {\em Astrophys. J. Suppl.} {\bf 2020}, {\em 249},~25,
  \href{http://arxiv.org/abs/2005.07089}{{\normalfont
  [arXiv:astro-ph.CO/2005.07089]}}.
\newblock {\url{https://doi.org/10.3847/1538-4365/aba190}}.

\bibitem[Khurshudyan and Elizalde(2024)]{id_11}
Khurshudyan, M.; Elizalde, E.
\newblock {Constraints on Prospective Deviations from the Cold Dark Matter
  Model Using a Gaussian Process}.
\newblock {\em Galaxies} {\bf 2024}, {\em 12},~31,
  \href{http://arxiv.org/abs/2402.08630}{{\normalfont
  [arXiv:gr-qc/2402.08630]}}.
\newblock {\url{https://doi.org/10.3390/galaxies12040031}}.

\bibitem[Chen et~al.(2023)Chen, Chen, Wang, Wang, Zhang, Zhang, and
  Zhang]{id_26}
Chen, J.F.; Chen, J.; Wang, Y.C.; Wang, Y.; Zhang, T.; Zhang, T.J.; Zhang, T.
\newblock {Test of artificial neural networks in likelihood-free cosmological
  constraints: A comparison of information maximizing neural networks and
  denoising autoencoder}.
\newblock {\em Phys. Rev. D} {\bf 2023}, {\em 107},~063517,
  \href{http://arxiv.org/abs/2211.05064}{{\normalfont
  [arXiv:astro-ph.CO/2211.05064]}}.
\newblock {\url{https://doi.org/10.1103/PhysRevD.107.063517}}.

\bibitem[To et~al.(2023)To, Rozo, Krause, Wu, Wechsler, and Salcedo]{id_17}
To, C.H.; Rozo, E.; Krause, E.; Wu, H.Y.; Wechsler, R.H.; Salcedo, A.N.
\newblock {LINNA: Likelihood Inference Neural Network Accelerator}.
\newblock {\em JCAP} {\bf 2023}, {\em 01},~016,
  \href{http://arxiv.org/abs/2203.05583}{{\normalfont
  [arXiv:astro-ph.CO/2203.05583]}}.
\newblock {\url{https://doi.org/10.1088/1475-7516/2023/01/016}}.

\bibitem[Bolliet et~al.(2024)Bolliet, Spurio~Mancini, Hill, Madhavacheril,
  Jense, Calabrese, and Dunkley]{id_24}
Bolliet, B.; Spurio~Mancini, A.; Hill, J.C.; Madhavacheril, M.; Jense, H.T.;
  Calabrese, E.; Dunkley, J.
\newblock {High-accuracy emulators for observables in \ensuremath{\Lambda}CDM,
  Neff, \ensuremath{\Sigma}m\ensuremath{\nu}, and w cosmologies}.
\newblock {\em Mon. Not. Roy. Astron. Soc.} {\bf 2024}, {\em 531},~1351--1370,
  \href{http://arxiv.org/abs/2303.01591}{{\normalfont
  [arXiv:astro-ph.CO/2303.01591]}}.
\newblock {\url{https://doi.org/10.1093/mnras/stae1201}}.

\bibitem[Nygaard et~al.(2023{\natexlab{a}})Nygaard, Holm, Hannestad, and
  Tram]{id_07}
Nygaard, A.; Holm, E.B.; Hannestad, S.; Tram, T.
\newblock {CONNECT: a neural network based framework for emulating cosmological
  observables and cosmological parameter inference}.
\newblock {\em JCAP} {\bf 2023}, {\em 05},~025,
  \href{http://arxiv.org/abs/2205.15726}{{\normalfont
  [arXiv:astro-ph.IM/2205.15726]}}.
\newblock {\url{https://doi.org/10.1088/1475-7516/2023/05/025}}.

\bibitem[Nygaard et~al.(2023{\natexlab{b}})Nygaard, Holm, Hannestad, and
  Tram]{id_23}
Nygaard, A.; Holm, E.B.; Hannestad, S.; Tram, T.
\newblock {Fast and effortless computation of profile likelihoods using
  CONNECT}.
\newblock {\em JCAP} {\bf 2023}, {\em 11},~064,
  \href{http://arxiv.org/abs/2308.06379}{{\normalfont
  [arXiv:astro-ph.CO/2308.06379]}}.
\newblock {\url{https://doi.org/10.1088/1475-7516/2023/11/064}}.

\bibitem[Moss(2020)]{id_01}
Moss, A.
\newblock {Accelerated Bayesian inference using deep learning}.
\newblock {\em Mon. Not. Roy. Astron. Soc.} {\bf 2020}, {\em 496},~328--338,
  \href{http://arxiv.org/abs/1903.10860}{{\normalfont
  [arXiv:astro-ph.CO/1903.10860]}}.
\newblock {\url{https://doi.org/10.1093/mnras/staa1469}}.

\bibitem[Aslanyan et~al.(2015)Aslanyan, Easther, and Price]{id_15}
Aslanyan, G.; Easther, R.; Price, L.C.
\newblock Learn-as-you-go acceleration of cosmological parameter estimates.
\newblock {\em Journal of Cosmology and Astroparticle Physics} {\bf 2015}, {\em
  2015},~005–005.
\newblock {\url{https://doi.org/10.1088/1475-7516/2015/09/005}}.

\bibitem[Hortua et~al.(2020{\natexlab{a}})Hortua, Volpi, Marinelli, and
  Malago]{id_03}
Hortua, H.J.; Volpi, R.; Marinelli, D.; Malago, L.
\newblock {Accelerating MCMC algorithms through Bayesian Deep Networks}.
\newblock In Proceedings of the {34th Conference on Neural Information
  Processing Systems},  11 2020,
  \href{http://arxiv.org/abs/2011.14276}{{\normalfont
  [arXiv:astro-ph.CO/2011.14276]}}.

\bibitem[Hortua et~al.(2020{\natexlab{b}})Hortua, Volpi, Marinelli, and
  Malag\`o]{id_18}
Hortua, H.J.; Volpi, R.; Marinelli, D.; Malag\`o, L.
\newblock {Parameter estimation for the cosmic microwave background with
  Bayesian neural networks}.
\newblock {\em Phys. Rev. D} {\bf 2020}, {\em 102},~103509,
  \href{http://arxiv.org/abs/1911.08508}{{\normalfont
  [arXiv:astro-ph.IM/1911.08508]}}.
\newblock {\url{https://doi.org/10.1103/PhysRevD.102.103509}}.

\bibitem[Elizalde et~al.(2020)Elizalde, Khurshudyan, Odintsov, and
  Myrzakulov]{id_04}
Elizalde, E.; Khurshudyan, M.; Odintsov, S.D.; Myrzakulov, R.
\newblock {Analysis of the $H_0$ tension problem in the Universe with viscous
  dark fluid}.
\newblock {\em Phys. Rev. D} {\bf 2020}, {\em 102},~123501,
  \href{http://arxiv.org/abs/2006.01879}{{\normalfont
  [arXiv:gr-qc/2006.01879]}}.
\newblock {\url{https://doi.org/10.1103/PhysRevD.102.123501}}.

\bibitem[Elizalde et~al.(2021)Elizalde, Gluza, and Khurshudyan]{id_05}
Elizalde, E.; Gluza, J.; Khurshudyan, M.
\newblock {An approach to cold dark matter deviation and the $H_{0}$ tension
  problem by using machine learning},  2021,
  \href{http://arxiv.org/abs/2104.01077}{{\normalfont
  [arXiv:astro-ph.CO/2104.01077]}}.

\bibitem[Elizalde and Khurshudyan(2022)]{id_08}
Elizalde, E.; Khurshudyan, M.
\newblock {Constraints on cosmic opacity from Bayesian machine learning: The
  hidden side of the H0 tension problem}.
\newblock {\em Phys. Dark Univ.} {\bf 2022}, {\em 37},~101114,
  \href{http://arxiv.org/abs/2006.12913}{{\normalfont
  [arXiv:astro-ph.CO/2006.12913]}}.
\newblock {\url{https://doi.org/10.1016/j.dark.2022.101114}}.

\bibitem[Khurshudyan and Myrzakulov(2021)]{id_14}
Khurshudyan, M.; Myrzakulov, R.
\newblock {Late time attractors of some varying Chaplygin gas cosmological
  models}.
\newblock {\em Symmetry} {\bf 2021}, {\em 13},~769,
  \href{http://arxiv.org/abs/1509.07357}{{\normalfont
  [arXiv:gr-qc/1509.07357]}}.
\newblock {\url{https://doi.org/10.3390/sym13050769}}.

\bibitem[Weinberg(1989)]{Weinberg:1988cp}
Weinberg, S.
\newblock {The Cosmological Constant Problem}.
\newblock {\em Rev. Mod. Phys.} {\bf 1989}, {\em 61},~1--23.
\newblock {\url{https://doi.org/10.1103/RevModPhys.61.1}}.

\bibitem[Newton et~al.(2021)Newton, Leo, Cautun, Jenkins, Frenk, Lovell, Helly,
  Benson, and Cole]{Newton:2020cog}
Newton, O.; Leo, M.; Cautun, M.; Jenkins, A.; Frenk, C.S.; Lovell, M.R.; Helly,
  J.C.; Benson, A.J.; Cole, S.
\newblock {Constraints on the properties of warm dark matter using the
  satellite galaxies of the Milky Way}.
\newblock {\em JCAP} {\bf 2021}, {\em 08},~062,
  \href{http://arxiv.org/abs/2011.08865}{{\normalfont
  [arXiv:astro-ph.CO/2011.08865]}}.
\newblock {\url{https://doi.org/10.1088/1475-7516/2021/08/062}}.

\bibitem[Rest et~al.(2014)]{Rest:2013mwz}
Rest, A.;  et~al.
\newblock {Cosmological Constraints from Measurements of Type Ia Supernovae
  discovered during the first 1.5 yr of the Pan-STARRS1 Survey}.
\newblock {\em Astrophys. J.} {\bf 2014}, {\em 795},~44,
  \href{http://arxiv.org/abs/1310.3828}{{\normalfont
  [arXiv:astro-ph.CO/1310.3828]}}.
\newblock {\url{https://doi.org/10.1088/0004-637X/795/1/44}}.

\end{thebibliography}

%


\PublishersNote{}
\isPreprints{}{
} 
\end{document}